%% file: draft0.tex
\documentclass[useAMS,usenatbib]{mn2e/mn2e}
\usepackage{graphicx}
\usepackage{hyperref}
\usepackage{longtable}
\usepackage{graphicx}
\usepackage{amsmath}
\usepackage[dvips,usenames]{color}

\input{mn2e/journalDefs}
\input{mn2e/emulateMNRAS}
\newcommand{\alphaMLT}{\mbox{$\alpha_{\rm MLT
}$}}

\newcommand{\diff}{\mbox{${\rm d}$}}

\newcommand{\jks}{\mbox{$J\!-\!K_{\rm s}$}}

\newcommand{\ks}{\mbox{$K_{\rm s}$}}

\newcommand{\dmo}{\mbox{$(m\!-\!M)_{0}$}}

\newcommand{\feh}{\mbox{\rm [{\rm Fe}/{\rm H}]}}

\newcommand{\Msun}{\mbox{$M_{\odot}$}}
\newcommand{\Rsun}{\mbox{$R_{\odot}$}}
\newcommand{\Teff}{\mbox{$T_{\rm eff}$}}
\newcommand{\logT}{\mbox{$\log\Teff$}}

\newcommand{\logg}{\mbox{$\log g$}}

\newcommand{\Ttau}{\mbox{$T$--\,$\tau$}}
\newcommand{\beq}{\begin{equation}}
\newcommand{\eeq}{\end{equation}}
\newcommand{\beqa}{\begin{eqnarray}}
\newcommand{\eeqa}{\end{eqnarray}}

\title[Very-low mass stars]{Improving PARSEC models for very low mass stars}

\author[Chen et al.]{Yang Chen$^{1,2}$, L\'eo Girardi$^3$, Alessandro Bressan$^1$, Paola Marigo$^4$, \newauthor Mauro Barbieri$^4$, Xu Kong$^{2,5}$
  \\
  $^{1}$SISSA, via Bonomea 265, I-34136 Trieste, Italy \\
  $^{2}$Department of Astronomy, University of Science and Technology of China, Hefei, Anhui 230026, China\\
  $^{3}$Osservatorio Astronomico di Padova -- INAF,
  Vicolo dell'Osservatorio 5, I-35122 Padova, Italy \\
  $^{4}$Dipartimento di Fisica e Astronomia, Universit\`a di Padova, Vicolo
  dell'Osservatorio 2, I-35122 Padova, Italy\\
  $^{5}$Key Laboratory for Research in Galaxies and Cosmology, USTC, Chinese Academy of Sciences, China}

\begin{document}

\date{Accepted 2014 August 6. Received 2014 August 6; in original form 2014 May 26}

\pagerange{\pageref{firstpage}--\pageref{lastpage}} \pubyear{2014}

\maketitle

\label{firstpage}

\begin{abstract}
Many stellar models present difficulties in
reproducing basic observational relations of very low mass stars (VLMS), including
the mass--radius relation and the optical colour--magnitudes of cool dwarfs.
Here, we improve PARSEC models on these points.
We implement the $\Ttau$ relations from PHOENIX BT-Settl model atmospheres as the
outer boundary conditions in the PARSEC code, finding that this change alone
reduces the discrepancy in the mass--radius relation from 8 to 5 per cent.
We compare the models with multi--band photometry of clusters Praesepe and M\,67,
showing that the use of $\Ttau$ relations clearly improves the description of the
optical colours and magnitudes. But anyway, using both Kurucz and PHOENIX model
spectra, model colours are still systematically fainter and bluer than the
observations. We then apply a shift to the above $\Ttau$ relations, increasing from $0$ at \Teff\,=\,4730\,K to $\sim$14\% at \Teff\,=\,3160\,K, 
to reproduce the observed mass--radius radius relation of dwarf stars.
Taking this experiment as a calibration of the \Ttau\ relations, we can reproduce the optical and near infrared CMDs of low mass stars in the old metal--poor globular clusters NGC\,6397 and 47\,Tuc, and in the intermediate--age and young solar--metallicity open clusters M\,67 and Praesepe. 
Thus, we extend PARSEC models using this calibration, providing VLMS models more suitable for the lower main
sequence stars over a wide range of metallicities and wavelengths. 
Both sets of models are available on PARSEC webpage.
\end{abstract}


%%%%%%%%%%%%%%%%%%%%%%%%%%%%%%%%%%%%%%%%%%%%%%%%%%%%%%%%%%%%%%%%%%%%%%
\section{Introduction}
\label{sec_intro}

Very low mass stars (VLMS; $M\la0.6~M_\odot$) are by far the most
numerous stars in the Galaxy. For a \citet{Kroupa01} or
\citet{Chabrier01} initial mass function (IMF), they constitute about 1/3
of the formed stars. Contrarily to the more massive stars, they remain
burning hydrogen during the entire Hubble time, being observable at
about the same luminosities from the moment they settle on the main
sequence (MS) up to very old ages. At the near--solar metallicities
that characterize the Solar Neighbourhood, they appear mostly as M
dwarfs, with their spectral energy distribution (SED) peaking at
near--infrared wavelengths, and marked by numerous molecular bands of
TiO, VO, water vapor, etc. \citep[see e.g.][]{AllardHauschildt95,
Allard_etal97}. At
the lower metallicities typical of the thick disk and halo, they
also appear as K dwarfs, with their SEDs peaking at red wavelengths
($R$ and $I$--bands).

VLMS appear copious in {\em any} deep imaging survey of the Galaxy
(such as Sloan Digital Sky Survey (SDSS) and Dark Energy Survey (DES)), and even more in infrared imaging campaigns
such as 2MASS \citep{2masspaper}, UKIRT Infrared Deep Sky Survey (UKIDSS) \citep{Lawrence_etal07},
European Southern Observatory (ESO)/UKIRT Infrared Deep Sky Survey (UKIDSS) public surveys \citep{vistapublicsurveys}, 
and {\sl WISE}
\citep{wise}.  Suffice it to mention that almost half of the 2MASS
point sources \citep{2mass} concentrate at $\jks\simeq0.85$, in a sort
of {\em vertical finger} in near--infrared colour--magnitude diagrams
\citep[e.g.][]{NikolaevWeinberg00}; this finger is dominated by M
dwarfs at magnitudes $\ks\ga14$, except at very low galactic latitudes
\citep{Zasowski_etal13}. In the optical, instead, VLMS appear along
well--defined colour--magnitude relations, as indicated by stars in open
clusters, that made them amenable for the distance derivations via
photometric parallaxes, and hence valuable probes of the Milky Way
structure \citep{Siegel_etal02,Juric_etal08,Ivezic_etal12}.

VLMS stars are also frequent among the targets of planet searches.
Indeed, almost the totality of {\em Kepler} planet candidates
\citep[95\% cf.][]{Borucki2011} are found around dwarfs with
masses below 1.2 \Msun, with the best chances of finding
Earth-mass planets being around the targets of even smaller masses \citep[e.g.][]{Quintana2014}.
In the case of transit detection, the presence of a well-defined mass--radius
relation \citep{Torres2010} allows the easy derivation of planetary properties.

Despite the great importance of the mass--radius
relation of VLMS, it has been poorly predicted and badly matched in
present grids of stellar models, with models tending to systematically
underestimate the stellar luminosity/radii for a given mass
\citep{Torres2010}. Significant clarifications have been recently
provided by \citet{Feiden2012} and \citet{Spada_etal13}, who identify the surface
boundary conditions in the VLMS models as a critical factor for
improving the data--model agreement. Anyway, even for the best models and
data, a discrepancy of about 3\% remains in the observed radii \citep{Spada_etal13}. Another recurrent 
discrepancy is in the colour--magnitude relations of VLMS: indeed, models that fairly well reproduce the 
near-infrared colours of VLMS in star clusters \citep[as in][]{Sarajedini2009}, tend to have optical 
colours which are far too blue at the bottom of the MS, as indicated in \citet[][]{An_etal08}, and as 
we will show in the following. A similar discrepancy also appears in low-metallicity globular clusters 
\citep[e.g.][]{Campos_etal13}. These disagreements imply that present isochrones cannot be safely used to 
estimate the absolute magnitudes -- and hence distances -- of field dwarfs, once their optical colours and 
apparent magnitudes are measured. Instead, empirical luminosity--colour relations have been preferred for 
this \citep[e.g.][]{Juric_etal08, Green_etal14}.

In this paper, we will revise the PAdova-TRieste Stellar Evolution
Code \citep[PARSEC;][]{parsec} seeking for a significant improvement of
their VLMS models. The way devised to do so is centred on the
revision of the $\Ttau$ relation used as the outer boundary condition
in stellar models, as will be described in
Sect.~\ref{sec_models}. The revised VLMS models will be transformed
into isochrones and compared to some key observations in
Sect.~\ref{sec_data}. The improvement in the models is clear, as
summarized in Sect.~\ref{sec_final}, however, an additional {\em ad hoc} correction to the $\Ttau$ relation 
is needed to bring models and data into agreement. A subsequent paper will be
devoted to a more thorough discussion of the available model
atmospheres and synthetic spectra for M dwarfs.

%%%%%%%%%%%%%%%%%%%%%%%%%%%%%%%%%%%%%%%%%%%%%%%%%%%%%%%%%%%%%%%%%%
\section{Models}
\label{sec_models}

\subsection{The stellar evolution code}

PARSEC is an extended and updated version of the code previously used
by \citet{Bressan_etal81, Girardi_etal00, Bertelli_etal08}, as
thoroughly described by \citet{parsec}. The main updates regard:
\begin{itemize}
\item full consideration of pre-main sequence phases;
\item the equation of state from FreeEOS v2.2.1 by Alan W. Irwin
  \footnote{\url{http://freeeos.sourceforge.net/}};
\item revised opacities from AESOPUS
  \citep{MarigoAringer09}\footnote{\url{http://stev.oapd.inaf.it/cgi-bin/aesopus}}
  and the OPAL group
  \citep{IglesiasRogers96}\footnote{\url{http://opalopacity.llnl.gov/}};
\item adoption of the revised \citet{Caffau_etal08, Caffau_etal09}
  solar chemical abundances;
\item extended nuclear networks and the reaction rates recommended in
  the updated JINA REACLIB Database
  \citep{Cyburt_etal10}\footnote{\url{http://groups.nscl.msu.edu/jina/reaclib/db/}};
\item microscopic diffusion is allowed to operate in low-mass stars;
\item the temperature gradient is described by the mixing length
  theory \citep{BV1958}, with the parameter $\alphaMLT=1.74$
  being calibrated by means of a Solar Model which is tested against
  the helioseismologic constraints \citep{Basuetal2009}.
\end{itemize}

The code as described in \citet{parsec} still makes use of the gray
atmosphere approximation \citep{Mihalas1978} as the external boundary
condition, i.e. the relation between the temperature and Rosseland mean optical
depth $\tau$ across the atmosphere, $\Ttau$, is given by
\begin{equation}
T^4(\tau)= \frac{3}{4} \Teff^4 \left[ \tau+q(\tau) \right]
\label{eq_gray}
\end{equation}
where $q(\tau)\approx2/3$ is the Hopf function.

The PARSEC v1.1 release of tracks contains stellar evolutionary models
in a wide range of masses down to 0.1~\Msun, with ages from 0
to 15 Gyr, and for several values of metal content going from
$Z=0.0001$ to $Z=0.06$.

These are the VLMS tracks we are going to revise. For guidance, Table~\ref{tab_notation} presents a summary 
of the main characteristics of these models, and of the other models that will be introduced later in this paper.

\begin{table*}
  \centering
  \caption{Summary of the PARSEC models for VLMS discussed in this paper.
  }
  \begin{tabular}{|lllp{0.2\textwidth}|p{0.35\textwidth}|}
    Model  & BC  & \Ttau & BC tables used for VLMS & basic description\\
    \hline
    v1.1  & OBC & Gray atmosphere     & \citet{CastelliKurucz03} and \citet{Allard_etal00} & previous version of PARSEC models and BC tables\\
    v1.1  & NBC & Gray atmosphere     & \citet{CastelliKurucz03} + PHOENIX BT-Settl & PARSEC v1.1 models interpolated with our new BC tables\\
    v1.2  & NBC & PHOENIX BT-Settl     & \citet{CastelliKurucz03} + PHOENIX BT-Settl & new VLMS models with $\Ttau$ relation from PHOENIX BT-Settl\\
    v1.2S & NBC & calibrated PHOENIX BT-Settl & \citet{CastelliKurucz03} + PHOENIX BT-Settl & new VLMS models with calibrated \Ttau\ relation with respect to PHOENIX BT-Settl\\
    \hline
  \end{tabular}
  \label{tab_notation}
\end{table*}

\subsection{The model atmospheres}
\label{sec_atmos}

Following the indications from various authors \citep[e.g.][]{VandenBerg_etal08, Feiden2012, Spada_etal13}, we replace the approximation in Eq.~\ref{eq_gray} by the $\Ttau$ relations
provided by real model atmospheres. We use the large library of model atmospheres from
PHOENIX~\citep[][and references therein]{Allard2012}\footnote{\url{http://perso.ens-lyon.fr/france.allard/index.html}.}, 
and in particular the set of BT-Settl models computed with the \citet{AGSS2009} chemical composition, which contains the most 
complete coverage in stellar parameters (temperature, gravity and metallicity) for both stellar spectra and atmosphere structures.

\begin{figure*}
\includegraphics[scale=0.8]{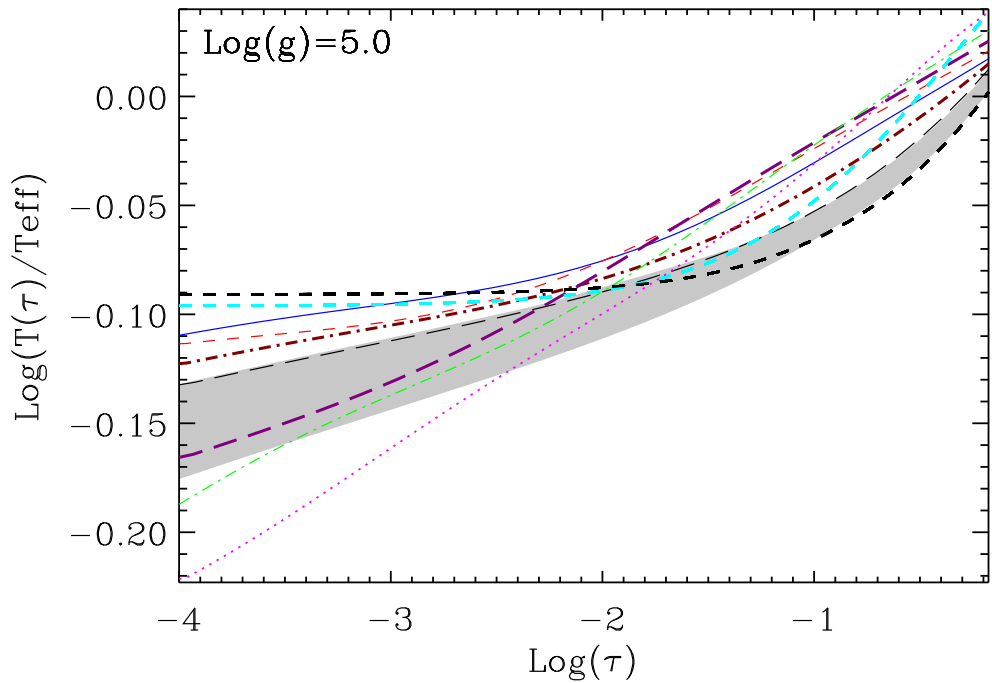}
\caption{A family of polynomial fits to the $\Ttau$ relations from PHOENIX (BT-Settl), for $\feh=0$ and $\logg=5.0$, in the region from $\tau=10^{-4}$ to $\tau=2/3$. 
All $\Ttau$ curves have been divided by \Teff\ so as to reduce the vertical scale in the plot.
 The magenta dot, green dash dot, purple thick long dash, red dash, blue solid, brown thick dash dot and black long dash curves are for 
 $ \Teff/{\rm K}=2600, 2800, 3000, 3500, 4000, 4300 \,{\rm and} \,4700$,  respectively.
 For $4700<\Teff/{\rm K} \leq 10000$, we show only the area occupied by the models (gray shaded). 
Other relations useful for the discussion in this paper are also presented: the \citet{KS1966} relation (cyan thick dash line) and the gray 
atmosphere approximation as in Eq.~\ref{eq_gray} (black thick dash line).}
\label{T_tau_nocorr}
\end{figure*}

The $\Ttau$ relations in PHOENIX (BT-Settl) cover the range of $2600 < \Teff/{\rm K} < 700000$ and $−0.5<\logg<6$ (with $g$ in cm\,s$^{-2}$),
for metallicities $Z$ between $\sim$0.04 and 0.000003. 
They are well-behaved and generally cover
the entire interval from $\tau = 0$ to $\tau \geq 100$. Fig.~\ref{T_tau_nocorr} shows some selected polynomial fits performed to the atmosphere models, 
from $\tau=10^{-4}$ to the boundary at $\tau=2/3$. 
They provide an excellent representation of the $\Ttau$ data. The polynomial fits are obtained for each metallicity, 
\Teff\ and \logg\ in the database, and later interpolated among these three parameters.

PARSEC solves the stellar structure at each time step via the \citet{Henyey} method as
described in \citet{Hofmeister_etal64} and \citet{Kippenhahn_book}. In the atmosphere integration, the family of $\Ttau$ relations -- written as a 
function of \Teff, \logg, and [Fe/H] -- replace the simple surface boundary condition formerly represented by Eq.~\ref{eq_gray} 
in the following way: The boundary is set at $\tau'=2/3$.
The pressure $P$ is integrated starting from the radiative pressure value at $\tau=0$ up to
its value at $\tau'$ via $\diff\tau/\diff P = \kappa R^2/GM$, where $\kappa$ is the Rosseland mean opacity, $R$ and $M$ are 
the stellar radius and total mass, and $G$ is the gravitational constant.

\subsection{The role of bolometric corrections}
\label{sec_bcs}

PHOENIX BT-Settl atmosphere models provide not only the $\Ttau$ relations to be used
as the external boundary conditions, but also an extended grid of synthetic spectral energy
distributions (SED) from which we can compute bolometric correction (${\rm BC}_\lambda$) tables. We have 
also implemented the new BC tables to convert the basic output of stellar models into the absolute magnitudes in
several passbands $M_\lambda$, with
\begin{equation}
M_\lambda = M_{\rm bol} - {\rm BC}_\lambda
\end{equation}
where $M_{\rm bol}=-2.5\log(L/L_\odot) - 4.7554$.
The formalism to compute ${\rm BC}_\lambda$ is thoroughly described in \citet{Girardi_etal02}, and will not be repeated here. 
Suffice it to recall that it depends primarily on $\Teff$, and to a lesser extent also on surface gravity and metallicity.

These tables of  ${\rm BC}_\lambda$ will be used later in this paper, for the \Teff\ interval between 2600 and 6000~K, 
as an alternative to the previous tables used in PARSEC -- which were based on \citet{CastelliKurucz03} and \citet{Allard_etal00} 
model atmospheres. In the following, new BC tables will be referred to as NBC, while the previous ones as OBC (see Table~\ref{tab_notation}).
As we will will discuss later, the new BC have a significant role mainly on the near-infrared colours.

\subsection{The new VLMS models}
\label{sec_vlms}

We follow the same procedure as described in \citet{parsec} to calibrate the solar model using the new $\Ttau$ relations. 
The new solar model has a $\alphaMLT=1.77$, which is slightly higher than the previous one used for PARSEC v1.1 \citep[namely $\alphaMLT=1.74$, see][]{parsec}. %This value will be assumed for PARSEC models from now on.
We have recomputed VLMS models using the new $\Ttau$ relations and
$\alphaMLT=1.77$, for all compositions contained in the previous
PARSEC v1.1 release, giving origin to PARSEC v1.2 tracks.
They start at the birth line
defined by a central temperature of $5\times10^4$~K, evolve through
the pre-main sequence where the main stages of D and $^3$He burning
occur, and finally settle on the zero-age main sequence
(ZAMS). Evolution in the main sequence is quite slow and takes longer
than 15 Gyr for all masses smaller than about 1~\Msun.

It is evident from the mass--radius relation of Fig.~\ref{fig_massradius1} and the $\log L$ versus $\log\Teff$ panels of 
Figs.~\ref{fig_Praesepecmds}, \ref{fig_m67cmds}, \ref{fig_47Tuc} and \ref{fig_NGC6397} (and discussions later in Sect.~\ref{sec_data}) 
that the use of the new $\Ttau$ relation has a significant impact in the stellar radii and on the shape of the lower main sequence, 
with the new ZAMS models becoming significantly larger, cooler (by up to $\Delta\Teff\simeq200$~K) and slightly less luminous, for 
model stars of the same mass (look at the difference between black and blue lines in Fig.~\ref{fig_massradius1}, and between blue 
and green curves in Figs.~\ref{fig_Praesepecmds}, \ref{fig_m67cmds}, \ref{fig_47Tuc} and \ref{fig_NGC6397}). In the HR diagrams, 
however, the lower ZAMS does never become as straight as the upper main sequence. As clarified long ago by \citet{Copeland_etal70}, 
the curved shape of the lower main sequence is mainly due to the changes in the equation of state (EOS), which  enters into a range 
of significantly higher densities for VLMS. Particularly important are the roles of Coulomb interactions and the formation of the H$_2$ 
molecule, which causes a strong reduction of the adiabatic temperature
gradient, $\nabla_{\rm ad}=(\partial \log T/\partial \log P)_S$, from
$0.4$ to about 0.1, in the outer layers of the coolest dwarfs.

%%%%%%%%%%%%%%%%%%%%%%%%%%%%%%%%%%%%%%%%%%%%%%%%%%%%%%%%%

%%%%%%%%%%%%%%%%%%%%%%%%%%%%%%%%%%%%%%%%%%%%%%%%%%%%%%%%%%%%%%%%%%
\section{Comparison with the data}
\label{sec_data}

In this section we compare the models with a few, selected observational datasets. We start with the mass--radius relation because 
it is a fundamental relation that does not involve the stellar SEDs. Then, we discuss the observed CMDs of two open clusters 
(Praesepe and M67, representing solar metallicity) and two globular clusters (47\,Tuc and NGC\,6397, representing metal poor environments).

%%%%%%%%%%%%%%%
\subsection{The mass--radius relation}
\label{sec_massradius}

\begin{figure*}
\includegraphics[scale=0.8]{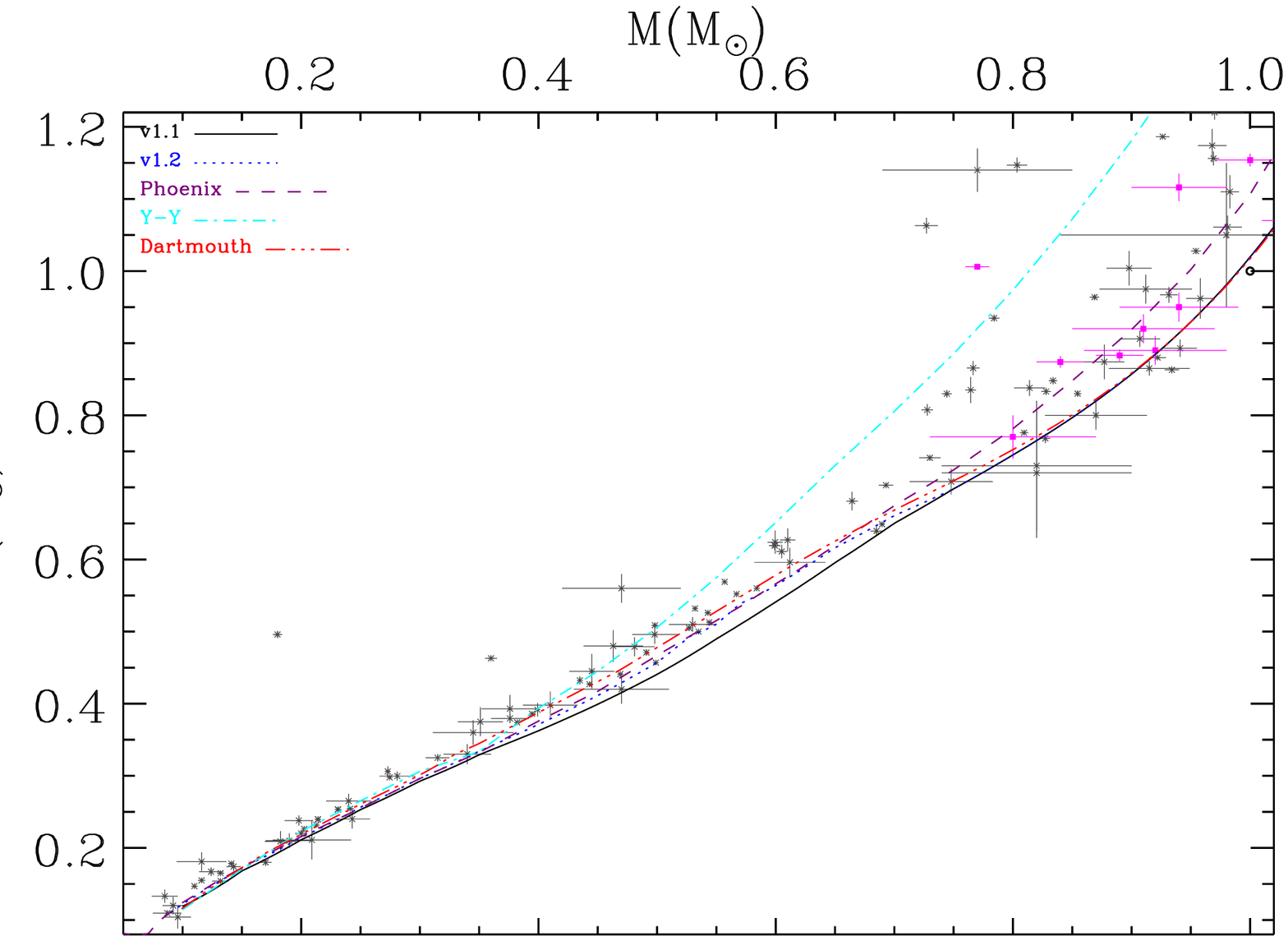}
\includegraphics[scale=0.8]{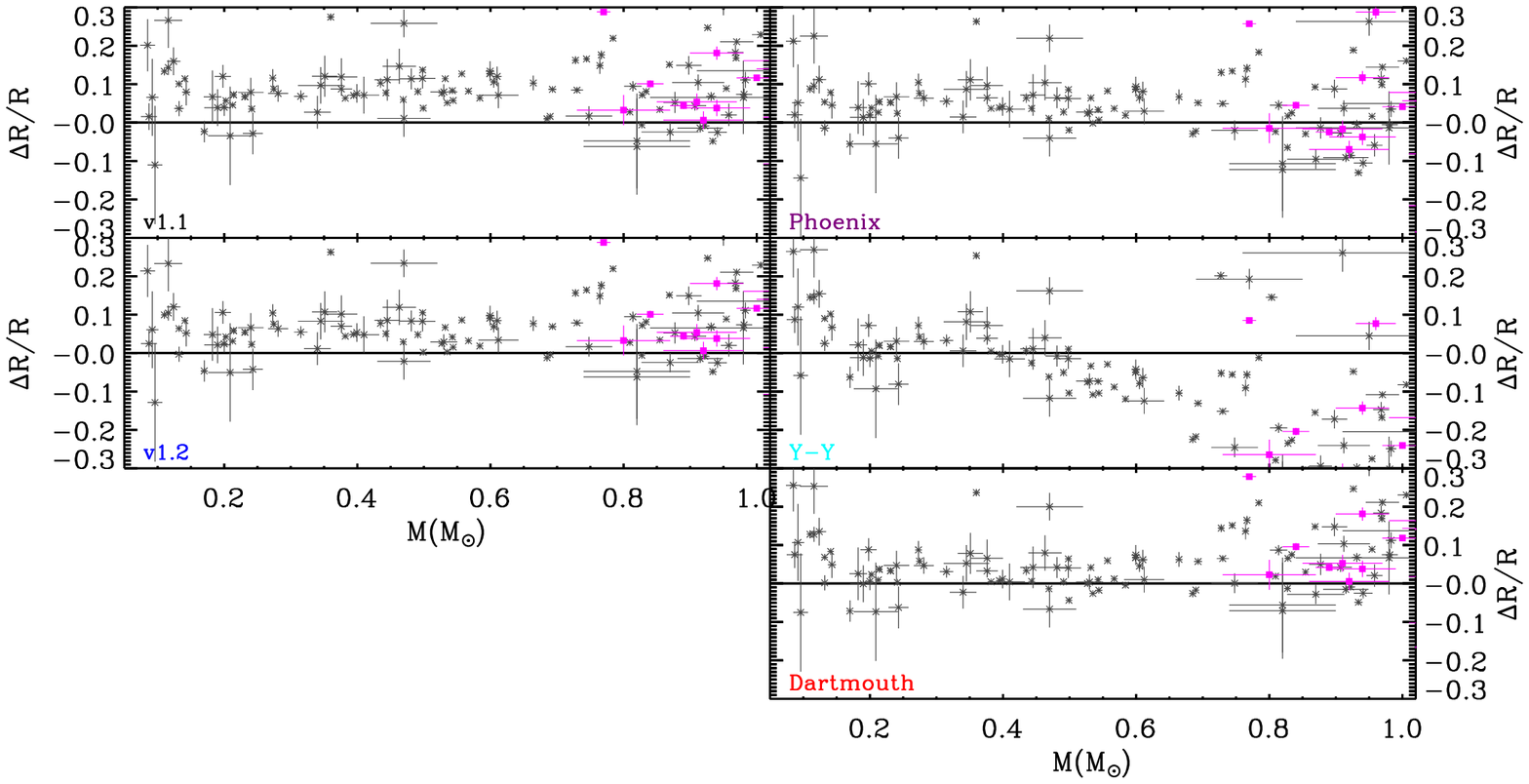}\hspace{-1.25cm}
\caption[]{
The empirical mass--radius relation for low mass stars in the solar neighbourhood using compiled data listed in Table~\ref{mass_radius_tab}.
Black asterisks are binary stars; Magenta squares are single stars. On the upper-left panel, isochrones of 5\,Gyr are overplotted for comparison for different models.
PARSEC v1.1 isochrones are shown in black while our updated isochrones (v1.2, Z=0.02) are in blue. PHOENIX (Z=0.02), Y-Y (Z=0.01631) 
and Dartmouth (Z=0.01885) models are in purple, cyan and red respectively. The other panels are the differences compared to the 
observation for different models as the labels shown. Y axis of $\Delta R/R$ is defined as $(R_{\rm obs}-R_{\rm mod,5Gyr})/R_{\rm obs}$.
}
\label{fig_massradius1}
\end{figure*}

Fig.~\ref{fig_massradius1} shows the mass-radius relation derived
from our tracks, as compared with recent observational data for nearby
stars, obtained either via asteroseismic techniques, or via
eclipsing binaries\footnote{
Stars measured via interferometric techniques are discarded, since their masses are derived using either empirical or theoretical mass-luminosity relations.}.
The full data set is presented in Table~\ref{mass_radius_tab}.
Among these observations, the most direct ones are
likely those from the eclipsing binaries (see \citealt{Torres2010},
and \citealt{Kraus2011}), since they do not depend on any
suspicious assumption or uncertain empirical calibration between the
photometry and $\Teff$. The same happens for the few eclipsing binaries in which the primary
is a white dwarf, taken from \citet{Parsons2012a, Parsons2012b},
in which the M dwarf masses and radii are particularly well
constrained.\footnote{The situation essentially does not change if we adopt the recent compilation of masses and radii from \citet{Eker2014}, 
which however is less complete for masses smaller than $0.4$~\Msun, and does not contain any star below 0.18~\Msun. While in this 
paper we pay more attention to the lower masses, we decide to just use our own collected data.}

It is obvious that the PARSEC v1.1 mass--radius relation is systematically below the empirical
data, with a typical deficit of 8~\% in the radius for a given mass, over the entire interval between 0.1 and 0.7~\Msun. For masses 
higher than $\sim0.7$~\Msun, the comparison between model and observed radii is not very significant since the radii increase with 
the stellar age, so that both models and observations tend to occupy a wider range in this parameter.

This mismatch in the stellar radii is very significant,
and has already been noticed by a number of authors \citep[e.g.][and
references therein]{Casagrande_etal08, Kraus2011}. It has inspired
a few alternative explanations, for instance an additional growth
in radius caused by rotation~\citep{Kraus2011,Irwin2011}, magnetic fields~\citep[e.g.][and
references therein]{Spruit1986,Feiden2012, Feiden2013, MacDonald2013, Jackson2014}. These mechanisms may indicate that eclipsing 
binaries follow a different mass--radius relation than single field stars, although \citet{Boyajian_etal12} and \citet{Spada_etal13} 
find that their radii are indistinguishable.

The lower-left panel of Fig.~\ref{fig_massradius1} shows that in PARSEC v1.2 models, this mismatch is reduced down to $\sim5$~\%.

We verified that it is really hard to eliminate this discrepancy in radii.
For a few tracks, we have explored a change in the EOS, testing for instance
the use of our previous \citet{Mihalas1990}
EOS for temperatures lower than $10^7$~K,
and the OPAL EOS \citep{opaleos} for higher temperatures. The effect on the radii was close to null.
Moreover, we note that our adopted FreeEOS is a modern EOS that includes all major
effects of relevance here. We note in particular that our
models reproduce the velocity of sound across the Sun to within
0.6~\%, which is well within the error bars. Therefore, the situation
is not easily remediable: Significant changes in the equation of
state, apart from not being motivated, would probably ruin the agreement
with the Standard Solar Model.

Changes in the mixing-length parameter $\alphaMLT$ also revealed to be
non-influential: models with $\alphaMLT$ as low as 0.1 have their radii
increased by only $\sim2$~\%. The use of very different metallicities
and helium contents does not change the situation either.

Fig.~\ref{fig_massradius1} also compares the mass--radius data with three other recent sets of models.
\begin{enumerate}
\item ``PHOENIX'', which are essentially the \cite{Baraffe1997, Baraffe_etal98}
theoretical isochrones transformed with their synthetic colour tables. They have not implemented 
their $\Ttau$ relation into their isochrones, but only the colours. 
\item Yale-Yonsei \citep[Y--Y; ][]{Spada_etal13} which used PHOENIX (BT-Settl) $\Ttau$ relation to improve their previous 
Y--Y models. $\alphaMLT=1.743$ was used. They demonstrate the large improvement compared to their previous models for masses below $0.6 \Msun$. 
\item Dartmouth \citep{Dotter2008} which used PHOENIX model atmospheres (both $\Ttau$ and the synthetic colour-$T_{\rm eff}$ transformations) and $\alphaMLT=1.938$.
\end{enumerate}
All models are computed for metallicities close to solar. It is easy to notice that PHOENIX models present almost the same discrepancy as ours, 
for the entire mass range of VLMS. Dartmouth models present more or less the same pattern, but with reduced discrepancies in the interval 
between 0.2 and 0.6~\Msun. Y-Y models, instead, present overestimated radii only for masses below 0.45~\Msun; they turn out to be underestimated 
instead for all higher masses. It is hard to trace back the origin of these differences.

Before exploring other possible changes to our models, we decided to look at the sequences of VLMS stars in open and globular clusters.

%%%%%%%%%%%%%

\subsection{The lower main sequence in Praesepe}

Praesepe is the perfect cluster to study the shape of the lower main sequence:
it is reasonably well populated, it has an age high enough to have all VLMS
already settled on the main sequence, and in addition it has excellent
(and uncontroversial) trigonometric parallaxes from Hipparcos. \citet{van_Leeuwen2009} finds $\dmo=6.30$~mag, $\log({\rm age/yr})$=8.90 
($\sim$0.8 Gyr), and $E(B\!-\!V)=0.01$~mag. In addition,
the cluster has been recently and deeply observed by the Panoramic Survey Telescope \& Rapid Response System (Pan-STARRS); \citet{WangPF2014} provides a comprehensive catalogue containing hundreds of VLMS, with memberships provided
by the combination of Pan-STARRS and 2MASS photometry, and PPMXL proper motions.

\begin{figure*}
\includegraphics[width=0.5\textwidth]{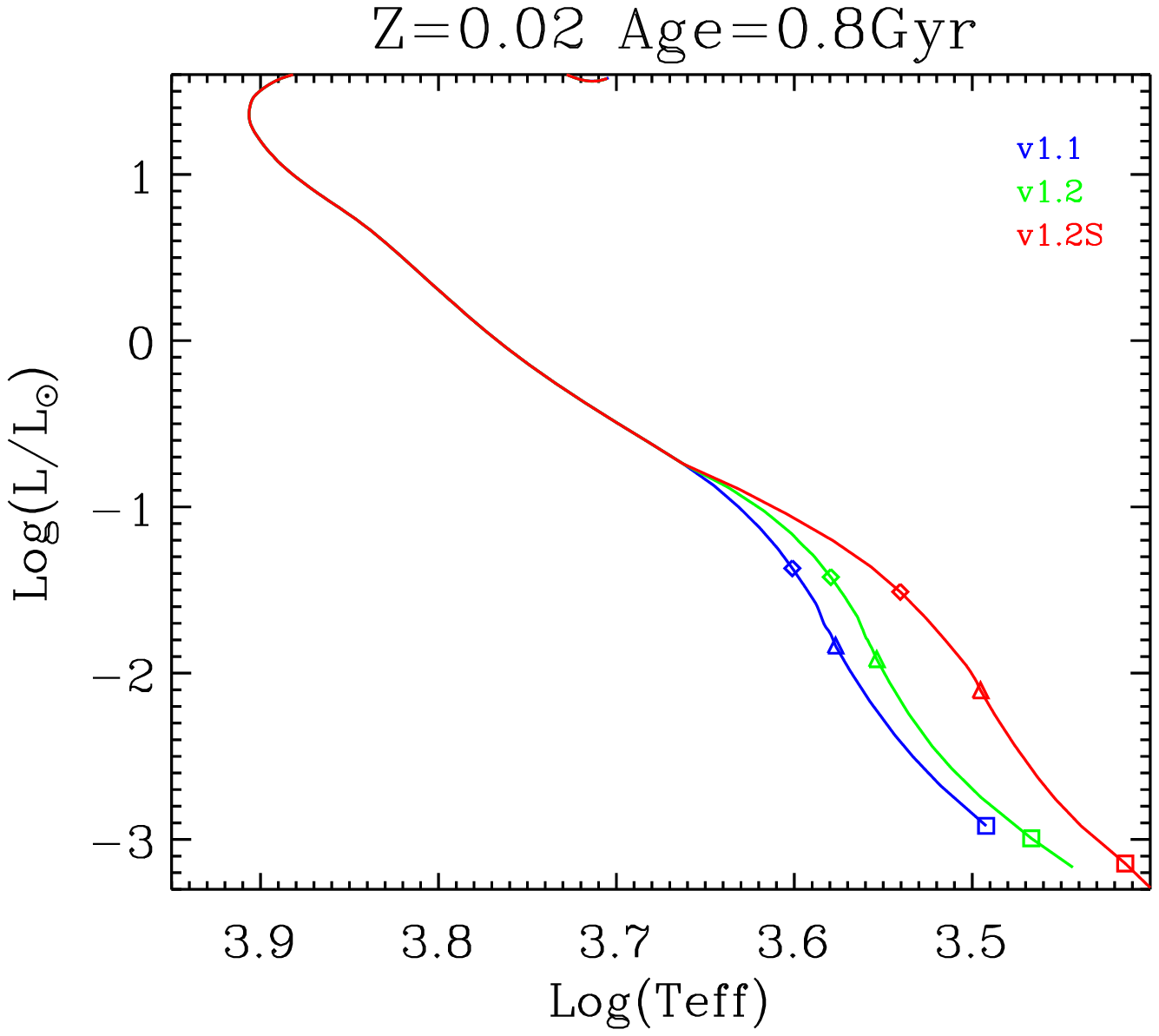}~     %L~Teff
\includegraphics[width=0.5\textwidth]{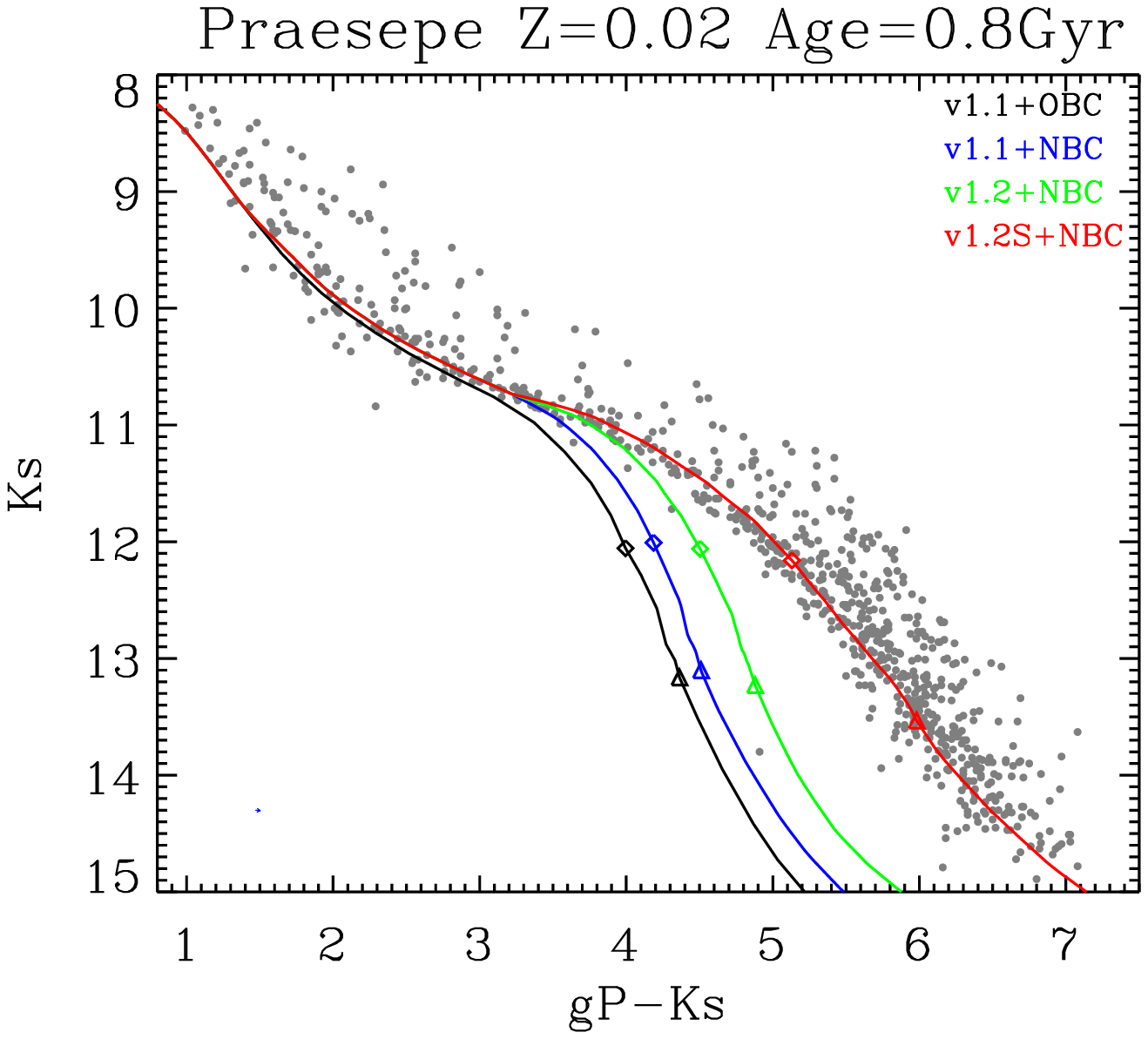}\\  %g - (g-Ks)
\includegraphics[width=0.5\textwidth]{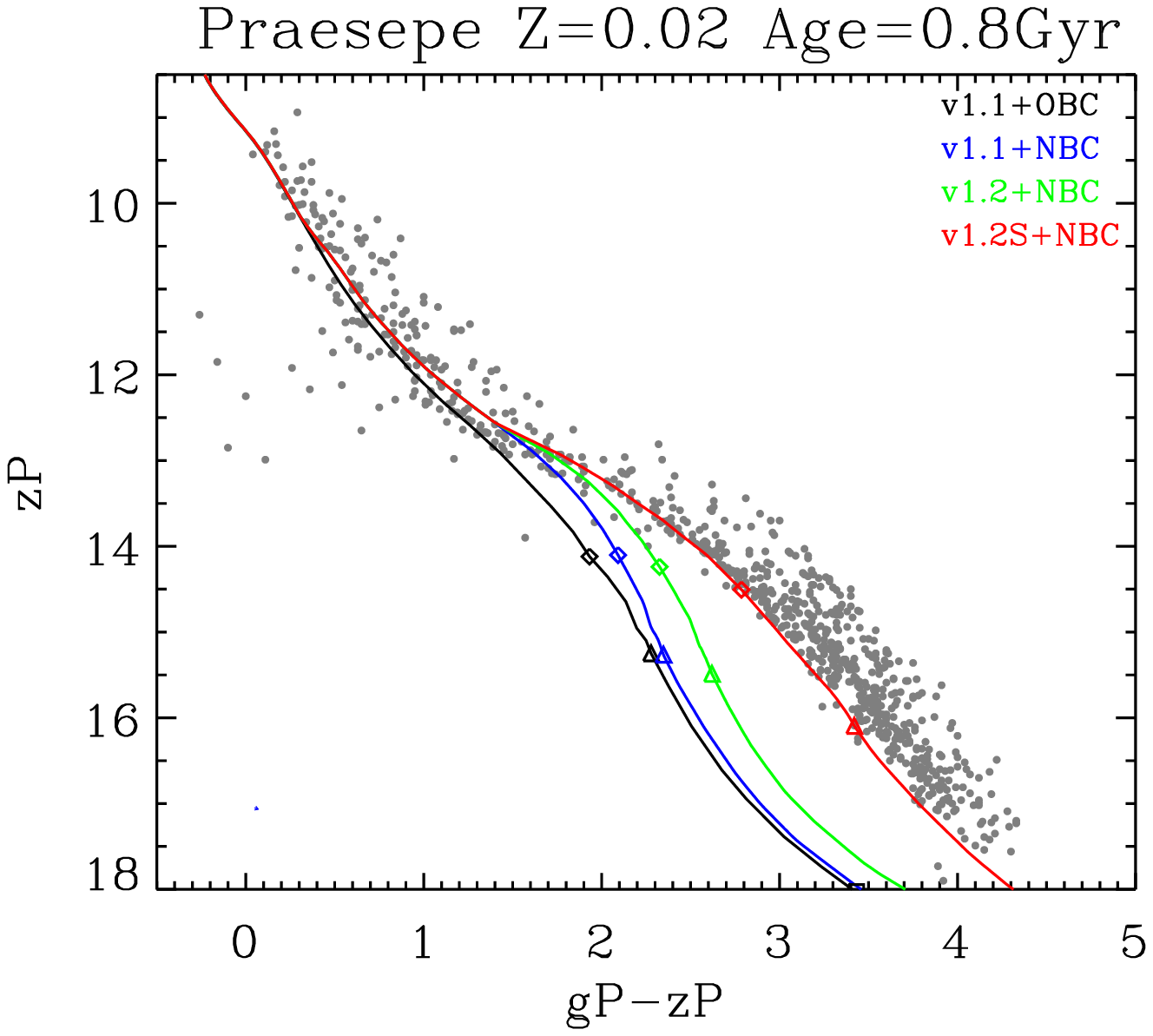}~  %z - (g-z)
\includegraphics[width=0.5\textwidth]{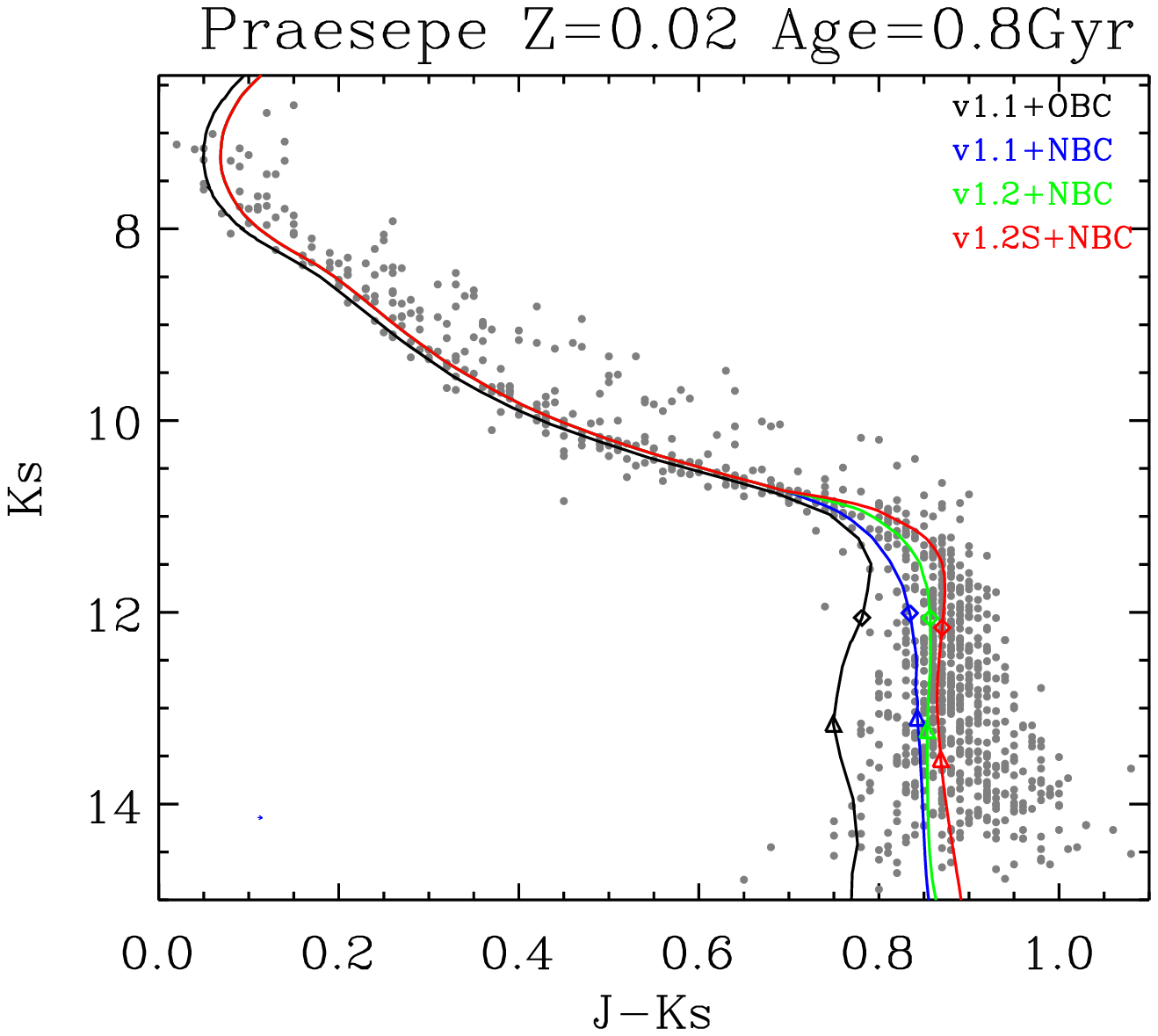}  %J - (J-Ks)
\caption{CMDs for Praesepe. The data points come from \citet{WangPF2014} with Pan-STARRS and 2MASS data. Black curves: PARSEC v1.1 isochrones 
with our previous BC tables; Blue curves: PARSEC v1.1 isochrones with new BC tables; Green curves: new isochrones (v1.2) with $T-\tau$ relation 
derived from PHOENIX (BT-Settl) models and interpolated with new BC tables; Red curves: new isochrones (v1.2S) with calibrated $T-\tau$ relation 
upon those derived from PHOENIX (BT-Settl) models and interpolated with new BC tables. 
The galactic reddening we use for Praesepe is $E(B-V)=0.01$ and the distance modulus is $\dmo=6.30$~mag 
\citep[from Hipparcos parallaxes, ][]{van_Leeuwen2009}. %Van Leeuwen 2009: log(age)=8.9, EBV=0.01, Modulus=6.30
The isochrones are for $Z=0.02$ and $\rm age=0.8$~Gyr. We also indicate initial the masses of 0.5, 0.3, 0.1~\Msun\ along the isochrones 
with open diamonds, triangles and squares respectively.
%van Leeuwen 2009: E(B-V)=0.01 and modulus=6.3.
}
\label{fig_Praesepecmds}
\end{figure*}

Fig.~\ref{fig_Praesepecmds} presents the \citet{WangPF2014} data in several diagrams
involving Pan-STARRS and 2MASS photometry, and as compared to a few sets of models. The models are initially shown in the theoretical H-R 
and mass-luminosity plots (top panels), for an age of $\log({\rm age/yr})=8.90$ \citep{van_Leeuwen2009} and a metallicity of $Z=0.02$ 
[\citet{Carrera2011} presents ${\rm [Fe/H]=0.16}$, which corresponds to $Z=0.0244$ with their adopted solar abundance]. Then, the other 
panels show the $g-z$ versus $z$, and the $g-\ks$ and $\jks$ versus $\ks$ CMDs.
It is evident that the previous PARSEC v1.1 models (black and blue lines, with OBC and NBC, respectively) fail to describe the
lower main sequence in CMDs involving the optical filters $g$ and $z$. In the case of the near-infrared $\jks$ versus $\ks$ CMD, the colour 
offset of the PARSEC v1.1 models almost completely disappear when we adopt the new bolometric corrections NBC (blue lines). In all the other 
CMDs, using the NBC just slightly moves the VLMS models towards the observed sequence.

The use of the $\Ttau$ relation, as illustrated by the green lines, causes the optical colours to move towards the observed sequences in a 
slightly more decisive way, but anyway, it is evident that no good agreement with the data is reached. The only exception seems to be the 
near-infrared $\jks$ versus $\ks$ diagram, in which all the model sequences in which the NBC tables are used appear with a satisfactory 
agreement with the data, being able to produce the vertical sequence observed at $\jks\simeq0.9$. Possible discrepancies with the 
data are at a level of just a few hundredths of magnitude in $\jks$.

Anyway, the important point that comes out of this comparison is the incapacity of models using the $\Ttau$ relation, and the latest 
tables of bolometric corrections, to reproduce the optical colours of VLMS in Praesepe, with discrepancies being as large as $\sim\!1$~mag in 
colours as $g-z$ and $g-\ks$. The models turn out to be far too blue, which suggests that some improvement could be reached by further 
decreasing the $\Teff$ -- hence increasing the stellar radius -- of the models, as we will see later.

%%%%%%%%%%%%%%%%%%%%%%%%%%%%%%%%%%%%%%%%%%%
\subsection{The lower main sequence in \texorpdfstring{M\,67}{M 67}}
\label{sec_m67}

The open cluster M\,67 constitutes another excellent testing ground for our
models, since it has extensive photometric and membership
data, added to well-determined global ages and metallicities \citep[e.g.][]{VandenbergStetson04, Randich_etal06}, and a small foreground reddening.

\begin{figure*}
\includegraphics[width=0.5\textwidth]{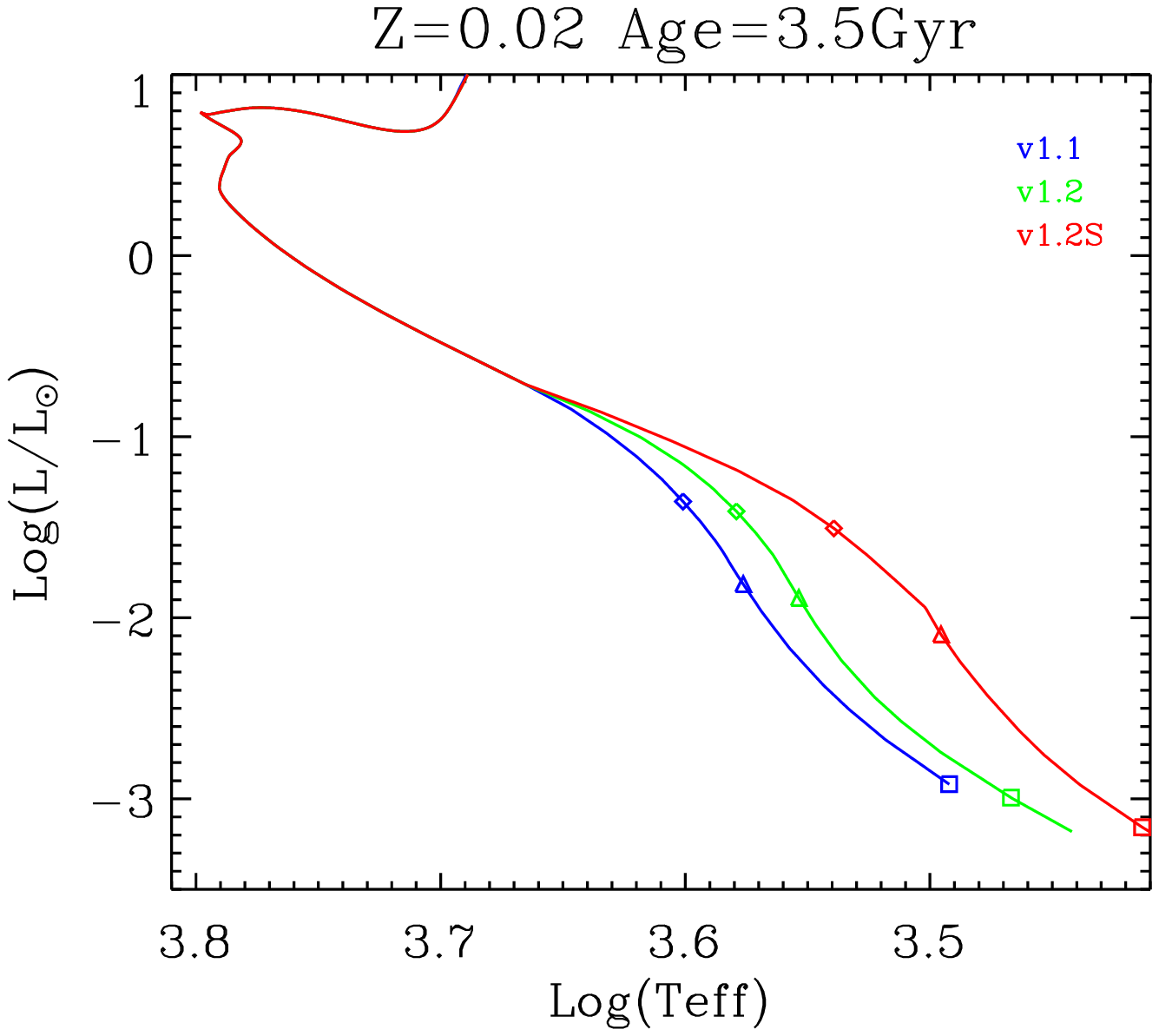}~    %L Teff
\includegraphics[width=0.5\textwidth]{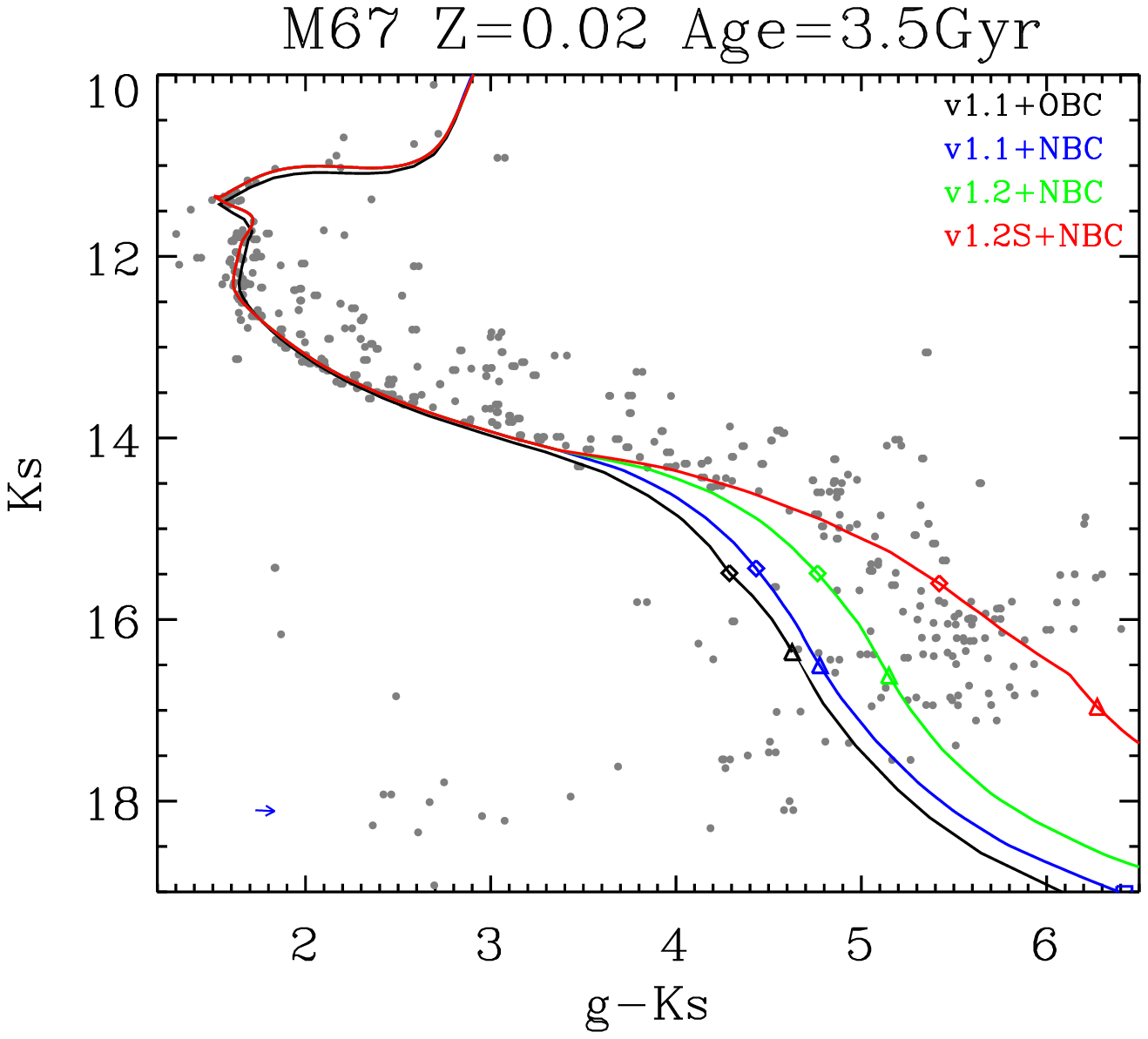}\\  %Ks (g-ks)
\includegraphics[width=0.5\textwidth]{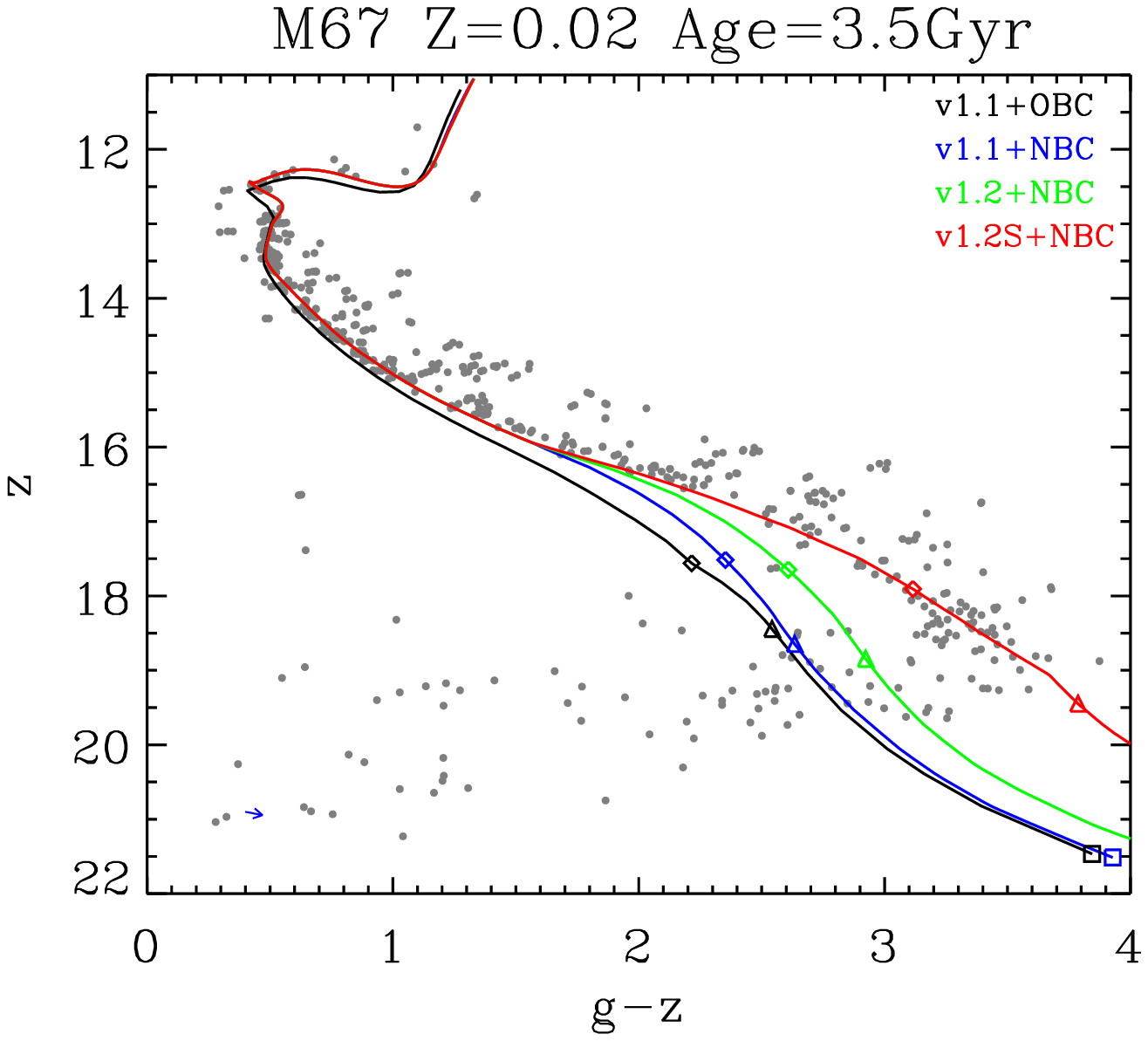}~ %g (g-z)
\includegraphics[width=0.5\textwidth]{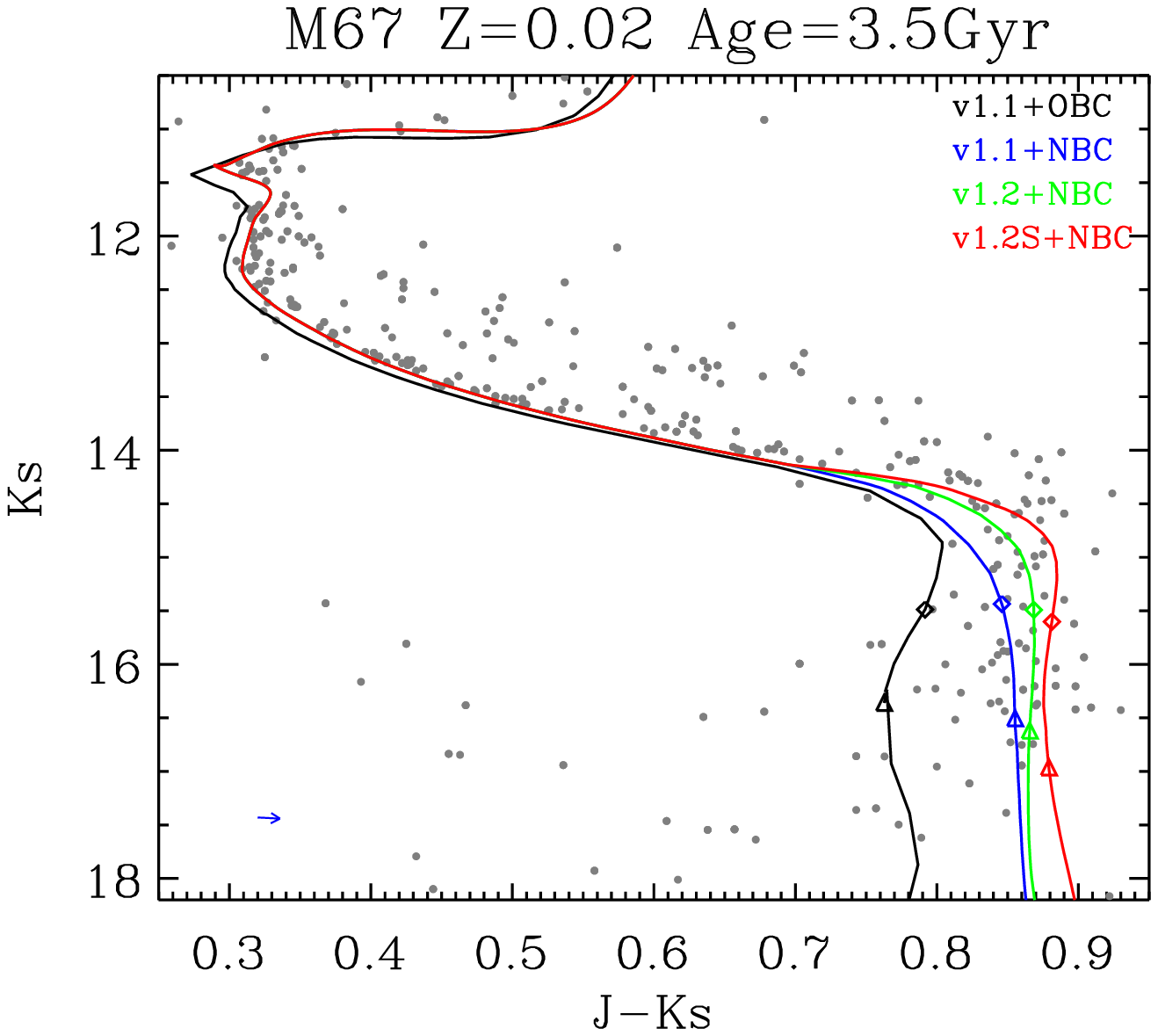}   %Ks (J-Ks)
\caption{M\,67 in several CMDs. The data points come from matching \citet{yadav_etal08} catalog with SDSS and 2MASS catalogs (see text). 
The labels are the same as in Fig.~\ref{fig_Praesepecmds}. The blue arrow in the left-lower corner of each panel is the reddening vector.
The galactic reddening we use for M\,67 is $E(B-V)=0.03$ and the distance modulus is $\dmo=9.75$~mag. 
The isochrones are for $Z=0.02$ and  $\rm age=3.5$~Gyr.}
\label{fig_m67cmds}
\end{figure*}

We have combined the following data sources for M\,67:
\begin{itemize}
\item the astrometry, $BVI$ photometry, and membership probability
  $P_{\rm mb}$ from \citet{yadav_etal08};
\item the 2MASS very deep photometry from the ``Combined 2MASS
  Calibration Scan'' (\citealt{2mass};
%\footnote{http://www.ipac.caltech.edu/2mass/releases/allsky/doc/seca7_4.html},
  see also section 2 of \citealt{Sarajedini2009});
\item the SDSS point spread function photometry as performed by
  \citet{An_etal08}.
\end{itemize}
The different catalogs were cross-matched with the Panoramic Survey Telescope \& Rapid Response System (Pan-STARRS) \citep{stilts}
revealing positional offsets typically smaller than
0.5\arcsec. Fig.~\ref{fig_m67cmds} shows a few of the resulting CMDs for
stars with $P_{\rm mb}>20~\%$. These diagrams are typically very clean
for all magnitudes brighter than $g=20$, whereas a significant
number of outliers appears at fainter magnitudes -- either due to the
more uncertain memberships or to the worse photometric quality in this
range of brightness. The important point for us is that the lower MS is very well
delineated. A parallel sequence of nearly-equal-mass binaries is also
evident, and located 0.7~mag above the MS.

In Fig.~\ref{fig_m67cmds} we present the fit of M\,67 for which we assume
a distance modulus $(m\!-\!M)_0=9.75$~mag and a reddening  $E(B\!-\!V)=0.03$~mag.
For the sake of simplicity we use models with $Z=0.02$ and,  
since this value is slightly higher than the observed metallicity \citep{Sarajedini2009},
we obtain an age of 3.5~Gyr which is a lower limit to the ages quoted in literature \citep{Sarajedini2009}.
As we will see in the following, this will not affect the results of our investigation.
In the H-R diagram (top left panel of Fig.~\ref{fig_m67cmds}), these isochrones appear nearly identical to those shown for Praesepe, 
for all luminosities below $\sim\!1~L_\odot$. The comparison with the models reveals essentially the same situation as for Praesepe: whereas the use of the NBC and $\Ttau$ relations 
both contribute to redden the model VLMS sequences, and reduce the disagreement with the data, the revised PARSEC v1.2+NBC models remain 
too blue at optical colours. The near-infrared colour \jks\ instead is little affected by the changes in 
the $\Ttau$ relation and reasonably well reproduced by all the NBC models.

The comparison with the blue band colours, such as the $B-V$, is more problematic and it is discussed in detail in Appendix~\ref{sec_m67BV}.

%%%%%%%%%%%%%%%%%%%%%%%%%%%%%%%%%%%%%%%%%%%
\subsection{Ultra-deep HST/ACS data for \texorpdfstring{47\,Tuc}{47 Tuc} and \texorpdfstring{NGC\,6397}{NGC 6397}}
\label{sec:47Tuc}
\begin{figure*}
\includegraphics[width=0.5\textwidth]{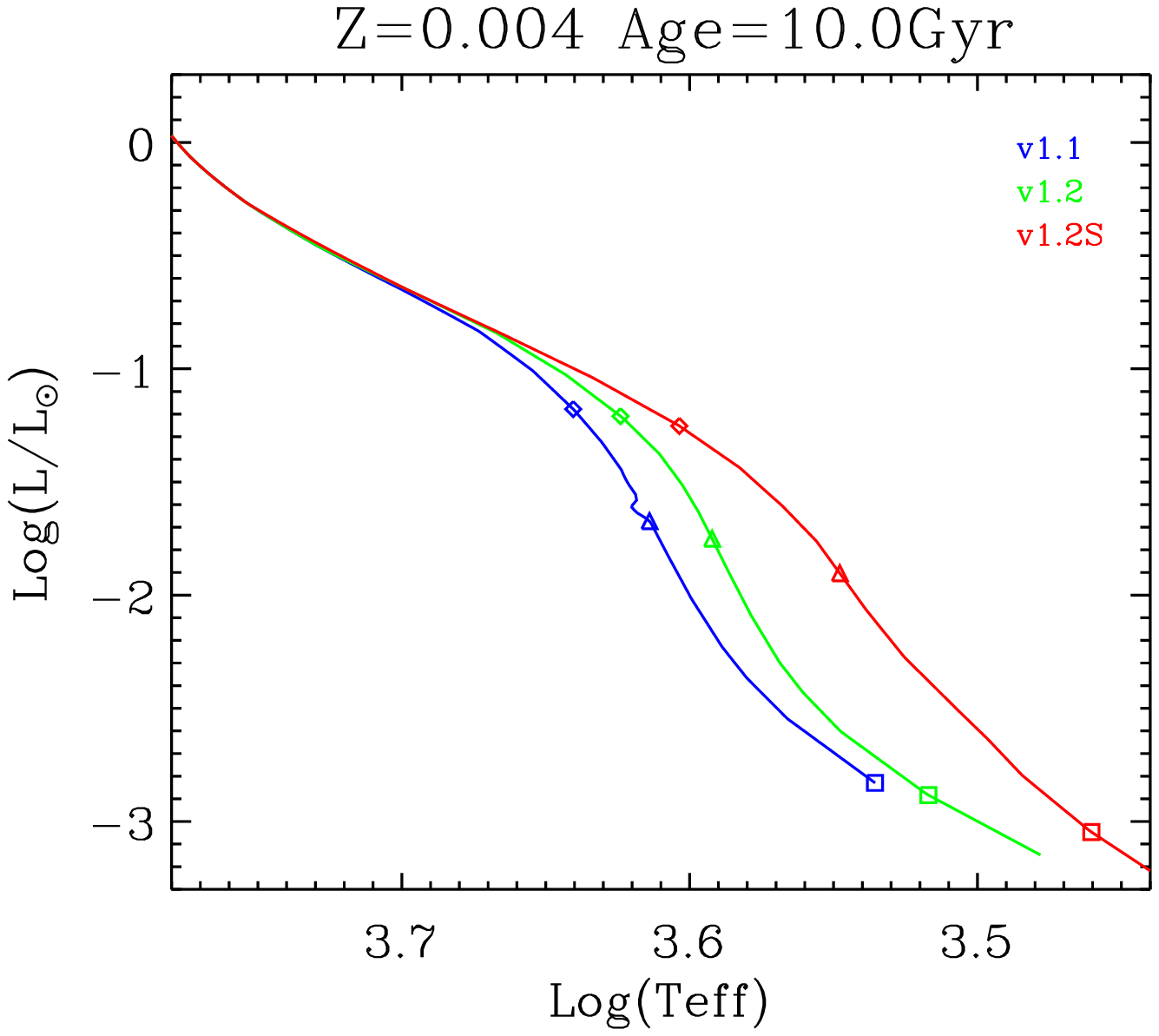}~   %L~Teff
\includegraphics[width=0.5\textwidth]{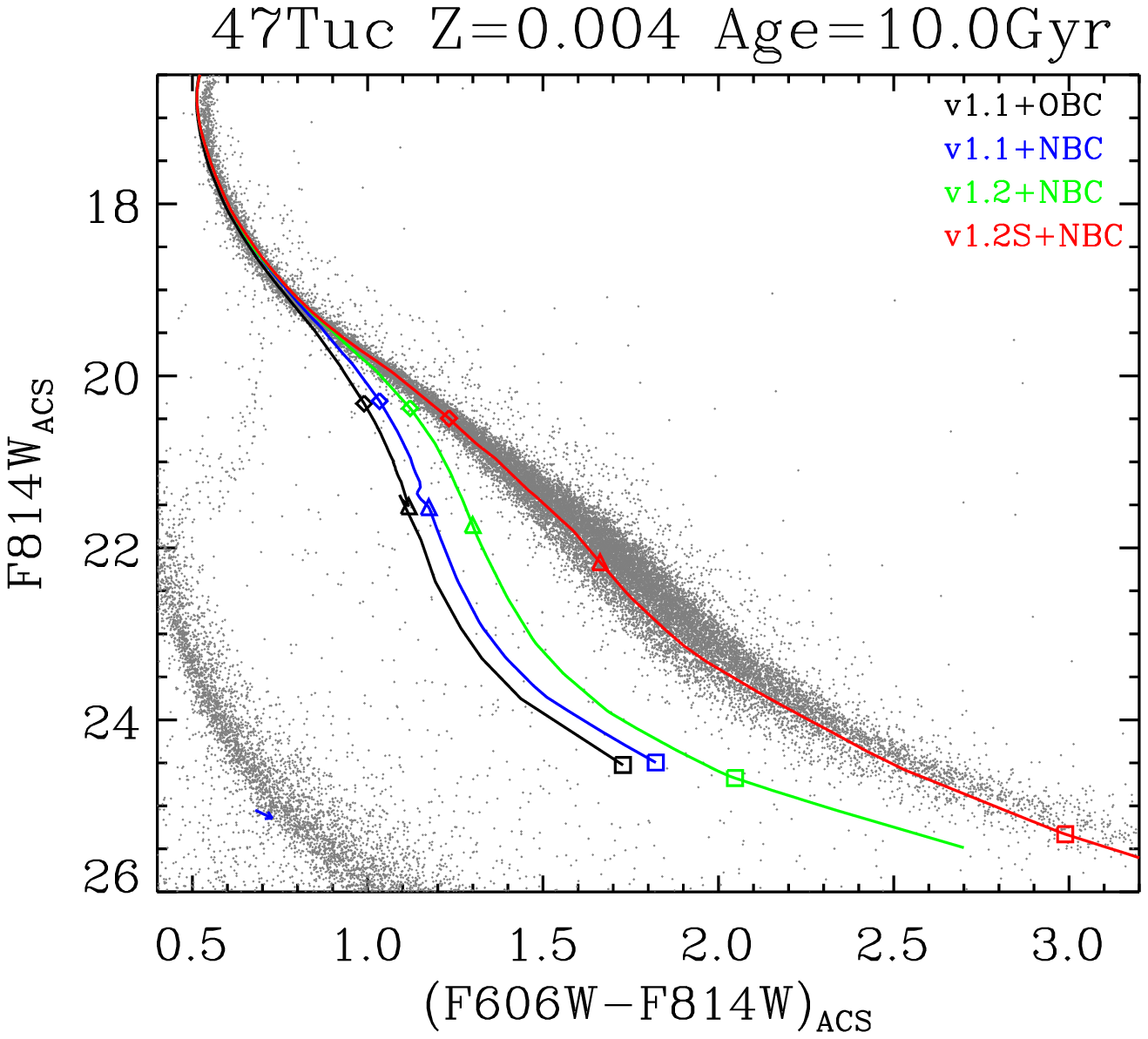}   %I - (V-I)
\caption{CMDs for 47\,Tuc. The data points are from \citet{Kalirai_etal12}. The sequences at the bottom-left corner of the CMD 
correspond to the background SMC population, and should be ignored. 
Model isochrones are presented with the same labels as in Fig.~\ref{fig_Praesepecmds}, but for a reddening of $E(B-V)=0.05$~mag, 
and a distance modulus of $\dmo=13.20$~mag. The isochrones are for $Z=0.004$ and an age of $10$~Gyr.}
\label{fig_47Tuc}
\end{figure*}

%\citet{Carretta2010} (Carretta 2010 A\&A 516 A55):
%modulus age  EBV  [Fe/H]  [a/Fe]  Z(derived)\\
%NGC6397   12.31   13.36  0.18  -1.988   0.36   ~0.000376\\
%47 TUC    13.32   12.83   0.04  -0.760   0.42   ~0.007195\\
47\,Tuc (NGC 104) is a relatively metal-rich globular cluster. The most recent abundance determination gives
$\feh=-0.79$ and a median value of  $\rm [O/Fe]\sim0.2$ \citep{Cordero2014}. \citet{Carretta2010} give a distance modulus of
$(m\!-\!M)_V=13.32$, an age of 12.83\,Gyr, and a reddening of $E(B\!-\!V)=0.04$~mag, while \cite{Hansen2013} determines an age of 9.7\,Gyr from 
the white dwarf cooling sequence.
\citet{Kalirai_etal12} obtained extremely deep {\sl Hubble Space Telescope (HST)}/Advanced Camera for Surveys (ACS) F606W and F814W data for this cluster (with $50\%$ completeness limits at $\sim$29.75 
and 28.75~mag for F606W and F814W, respectively), which makes it an excellent test bed for the lower main sequence. We compare our models 
with these observational data in Fig.~\ref{fig_47Tuc}. To fit the data with our models, we assume a metallicity of $Z=0.004$,
a distance modulus of $\dmo=13.20$, an age of 10\,Gyr and $E(B\!-\!V)=0.05$~mag.

It is evident that PARSEC v1.1 and v1.2 models fail to reproduce the lower main sequence in 47~Tuc, in a way similar to what was already noticed 
for Praesepe and M~67. It is also evident that the NBC and the use of \Ttau\ relations cause models to move on the right direction, but do not 
suffice for them to reach the observed sequences. In 47~Tuc, the discrepancy between models and data start at about $F814W=20$, which corresponds 
approximately to stellar models of mass $M=0.6$~\Msun. Finally we note that with an isochrone of 12.6\,Gyr we can also fit the 
turn-off very well, while with the isochrone of 10\,Gyr, we need to enhance the helium content. But since the goal of this paper is to improve the models for the lower main-sequence, we leave the detailed modelling of the turn-off and red giant sequences to a future work.

\begin{figure*}
\includegraphics[width=0.5\textwidth]{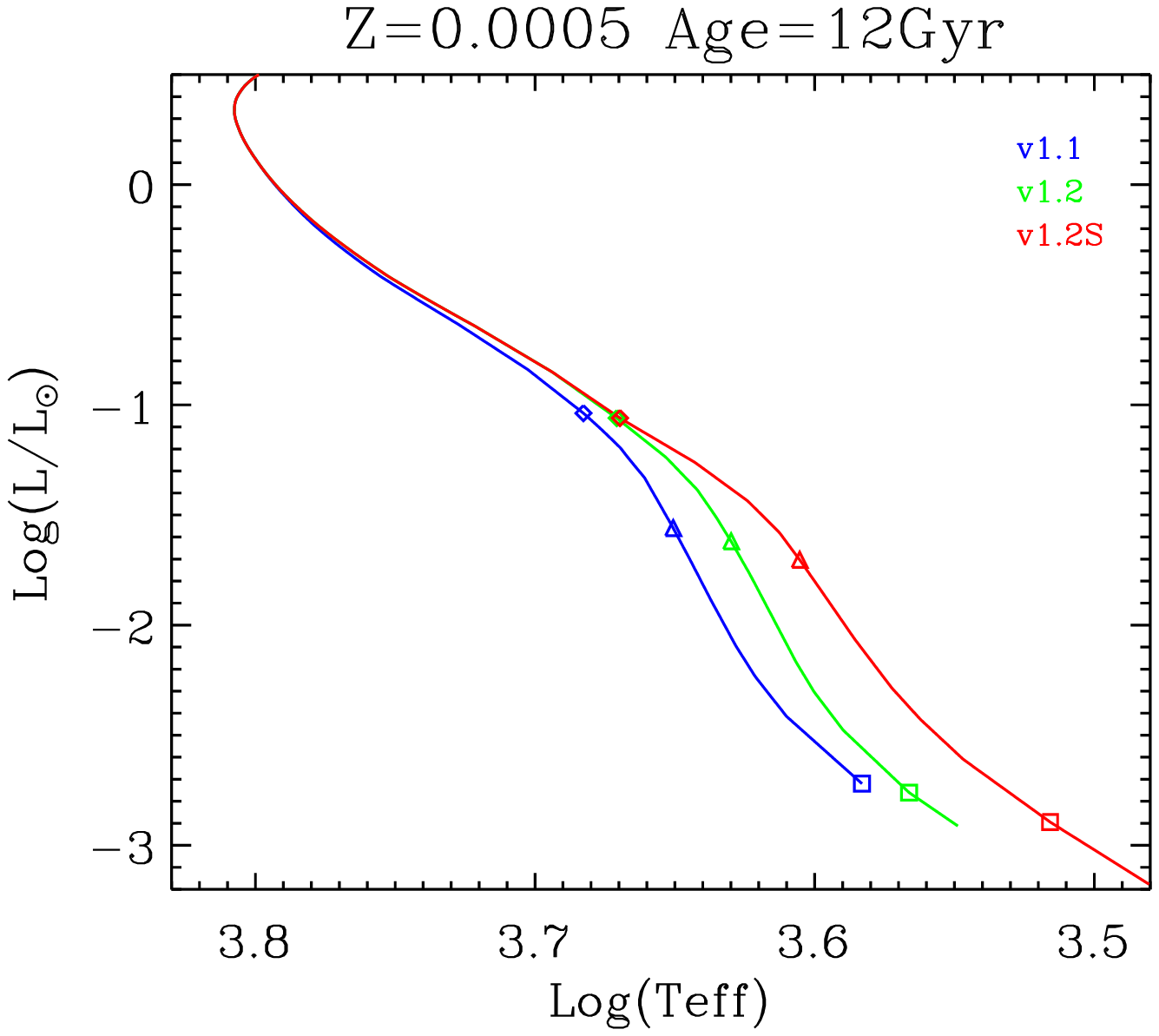}~   %L~Teff
\includegraphics[width=0.5\textwidth]{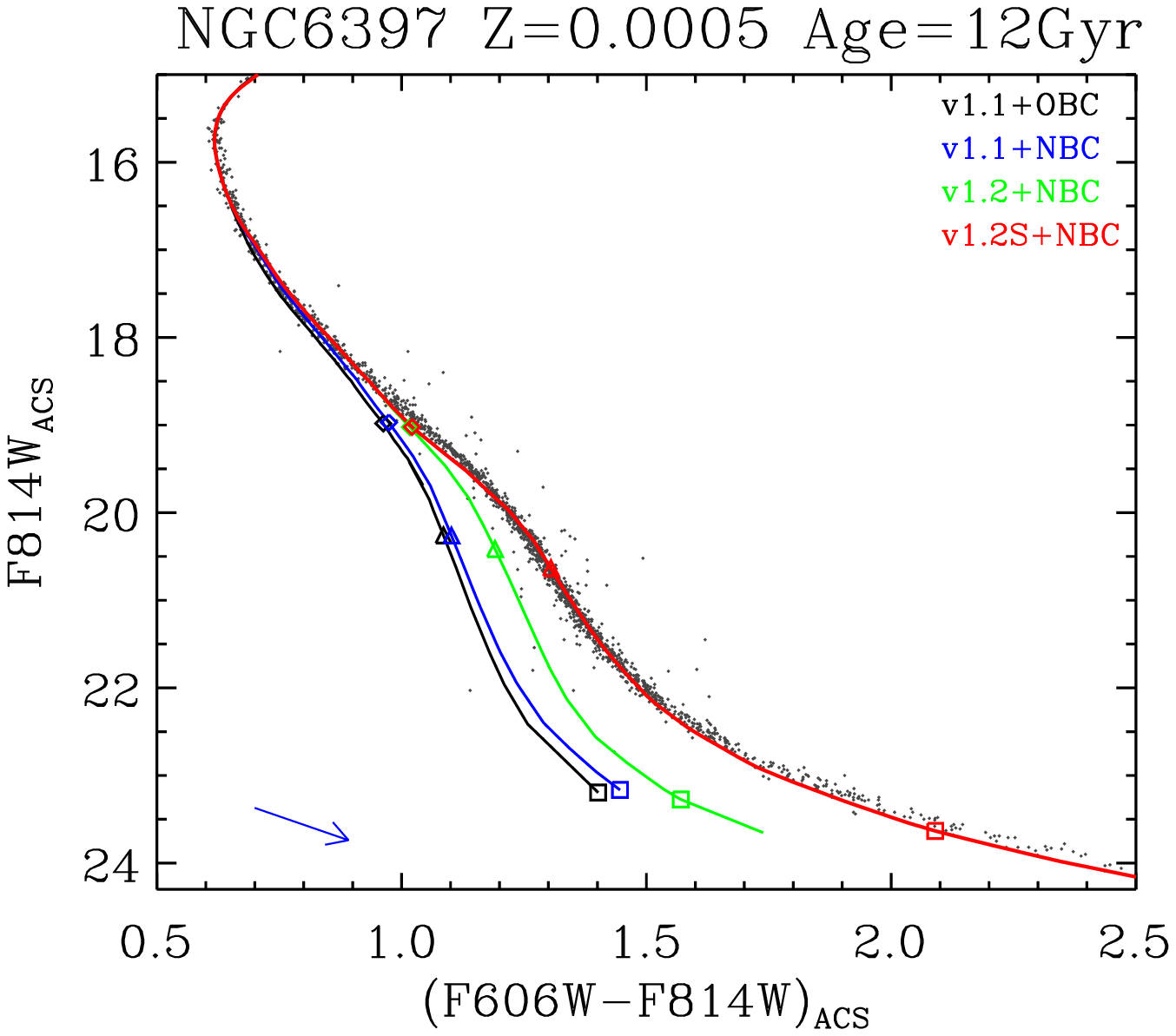}   %I - (V-I)
\caption{CMDs for NGC~6397. The data points come from \citet{Richer2008}. Model isochrones are presented with the same labels as in Fig.~\ref{fig_Praesepecmds}, but for a reddening of $E(B-V)=0.2$~mag, and a distance modulus of $\dmo=11.95$~mag. 
The isochrones are for $Z=0.0005$ and an age of $12$~Gyr.}
\label{fig_NGC6397}
\end{figure*}

Going to even smaller metallicities, we have the case of NGC~6397
with a measured metallicity of $\feh=-1.988$ (and $\rm [\alpha/Fe]=0.36$, cf. \citet{Carretta2010}, corresponding to $Z=0.000376$ with their 
adopted solar abundance). \citet{Carretta2010} also give a distance modulus of $(m\!-\!M)_V=12.31$, an age of 13.36 Gyr, and a reddening of 
$E(B\!-\!V)=0.18$~mag. \citet{Richer2008} observed this cluster with HST/ACS and, after proper motion cleaning, obtained a very narrow 
main-sequence down to F814W$\sim$26~mag, as shown in Fig.~\ref{fig_NGC6397}. To fit the data with our models we assume a 
metallicity of $Z=0.0005$ (the nearest metallicity in PARSEC v1.1),
a distance modulus of $\dmo=11.95$, an age of 12~Gyr and $E(B\!-\!V)=0.2$~mag. 
Comparison with PARSEC v1.1 and v1.2 models reveals about the same discrepancies as for 47~Tuc,
but now starting at $F814W\simeq19$~mag.

We remind the reader that the models used here adopt a solar abundance partition, without enhancement of $\alpha$ elements.
While the effect of $\alpha$-enhancement is the goal of a forthcoming detailed analysis
on globular clusters properties, we notice that its effects on the $\Ttau$ relation are of secondary importance,
after it is accounted for in deriving the global metallicity.\footnote{As first noted by \citet{Kalirai_etal12}, a dispersion in 
the abundance ratios might be at the origin of the colour dispersion observed at the bottom of the main sequence in 47~Tuc.}

%%%%%%%%%%%%%%%%%%%%%%%%%%%%%%%%%%%%%%%%%%%%%%%%%%
\subsection{Comparing the CMDs with other models}
\label{other_cmd}

\begin{figure*}
\includegraphics[width=0.5\textwidth]{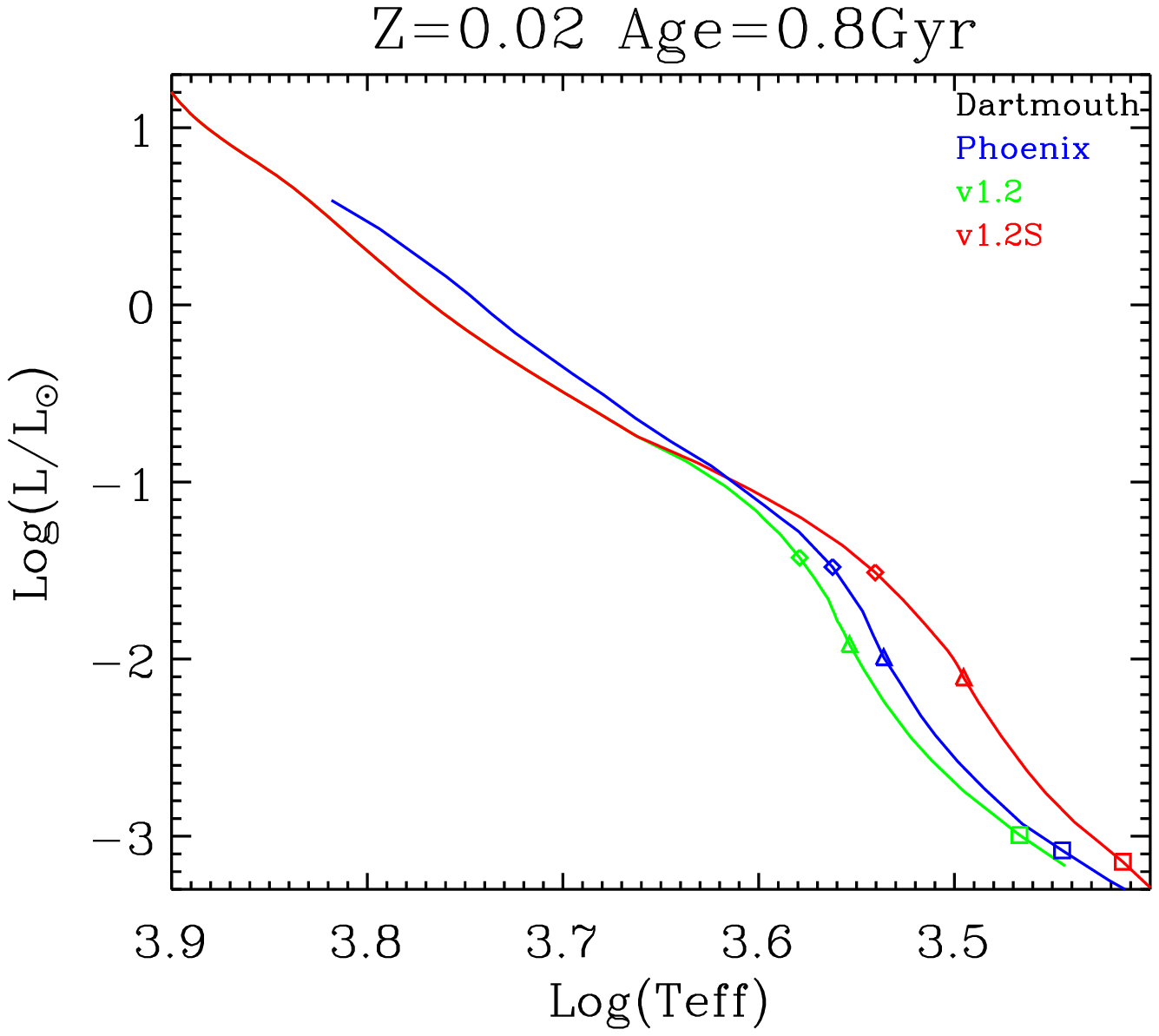}~     %L~Teff
\includegraphics[width=0.5\textwidth]{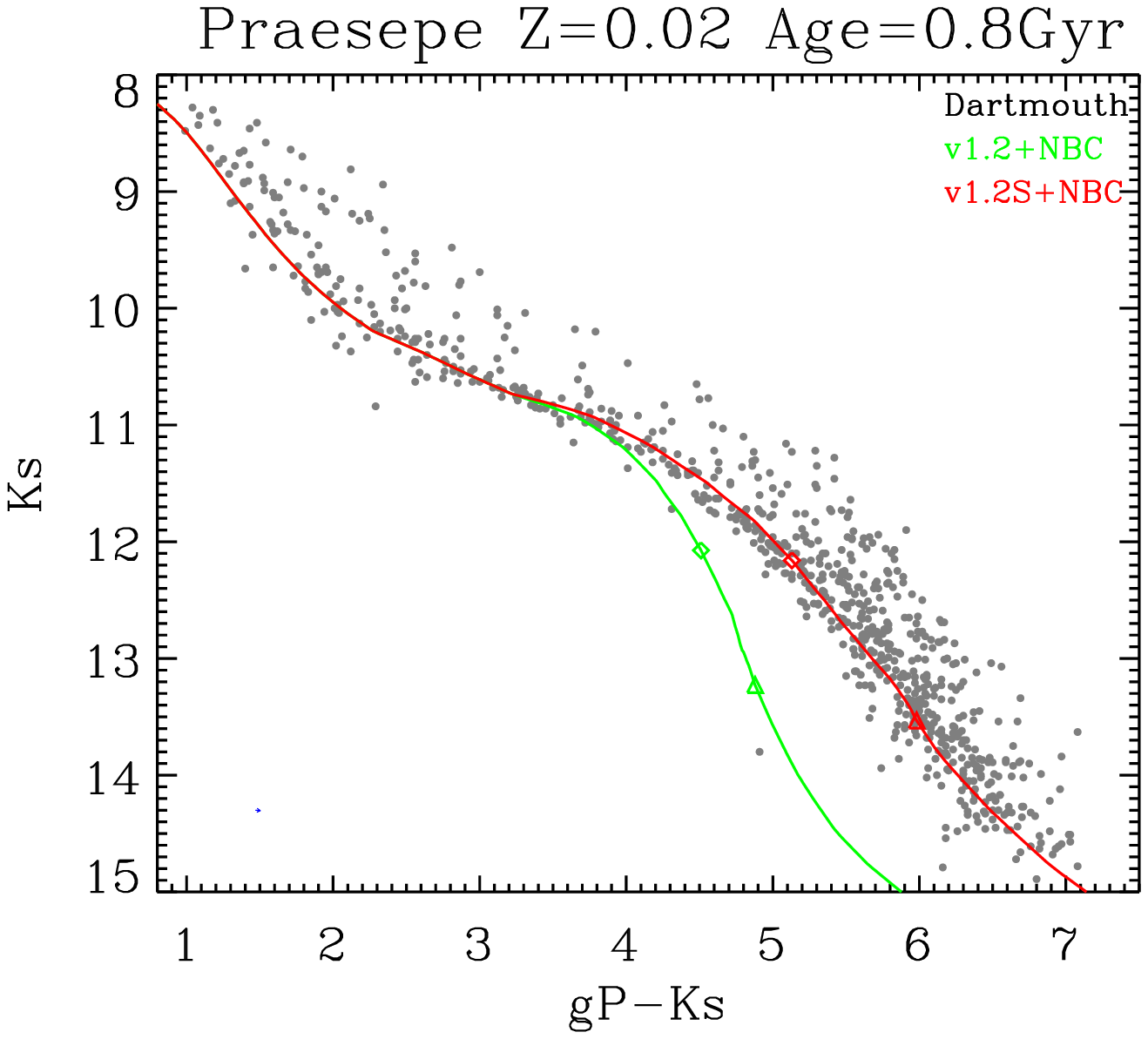}\\  %Ks - (g-Ks)
\includegraphics[width=0.5\textwidth]{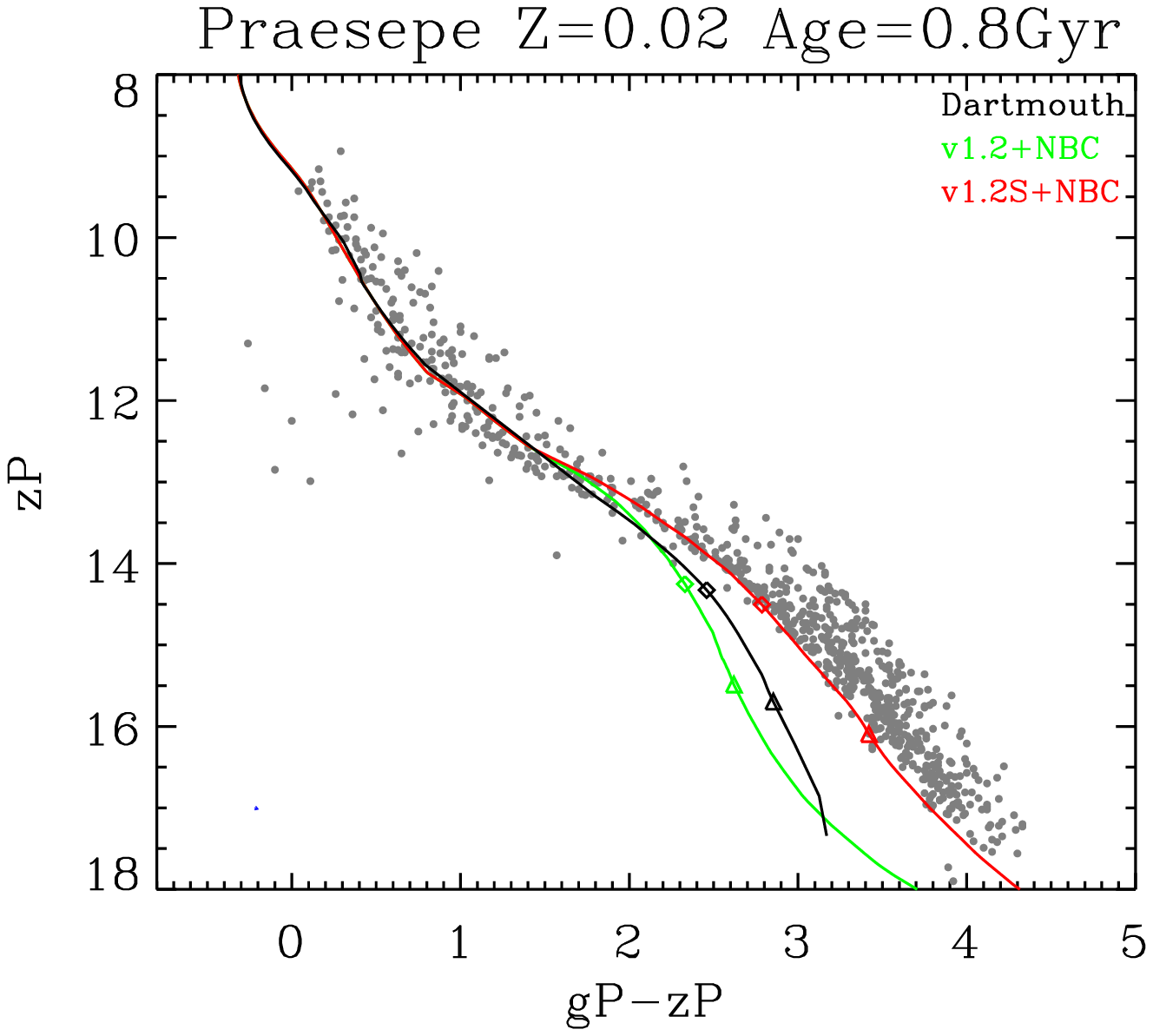}~   %z - (g-z)
\includegraphics[width=0.5\textwidth]{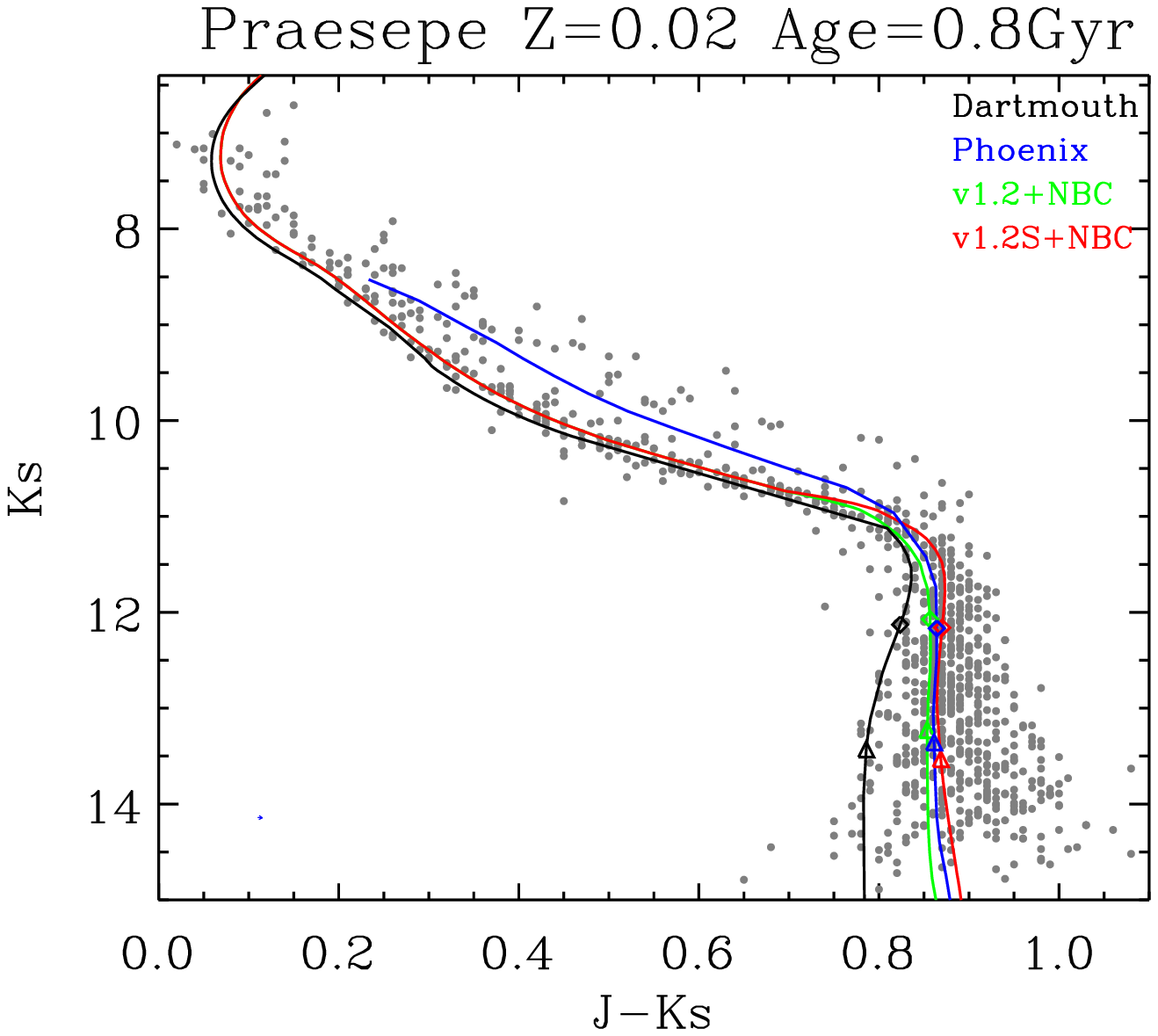}  %Ks - (J-Ks)
\caption{CMDs for Praesepe, same as in Fig.~\ref{fig_Praesepecmds}, except that the black lines are for Dartmouth with $Z=0.01885$ 
and $\rm age=0.8$~Gyr, and the blue ones are for PHOENIX  with $Z=0.02$ and $\rm age=0.8$~Gyr.}
\label{fig_Praesepe:other}
\end{figure*}

\begin{figure*}
\includegraphics[width=0.5\textwidth]{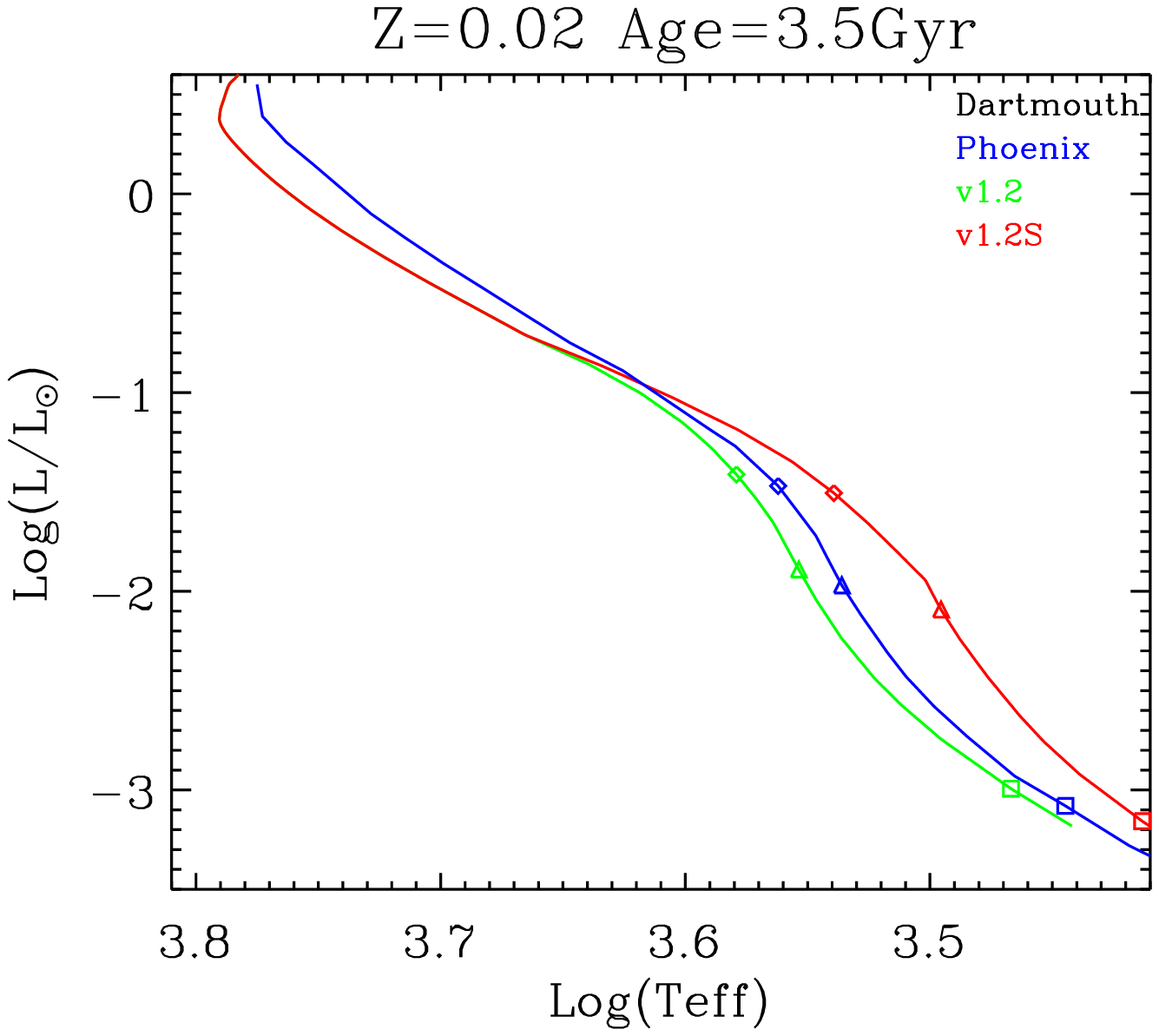}~   %L Teff
\includegraphics[width=0.5\textwidth]{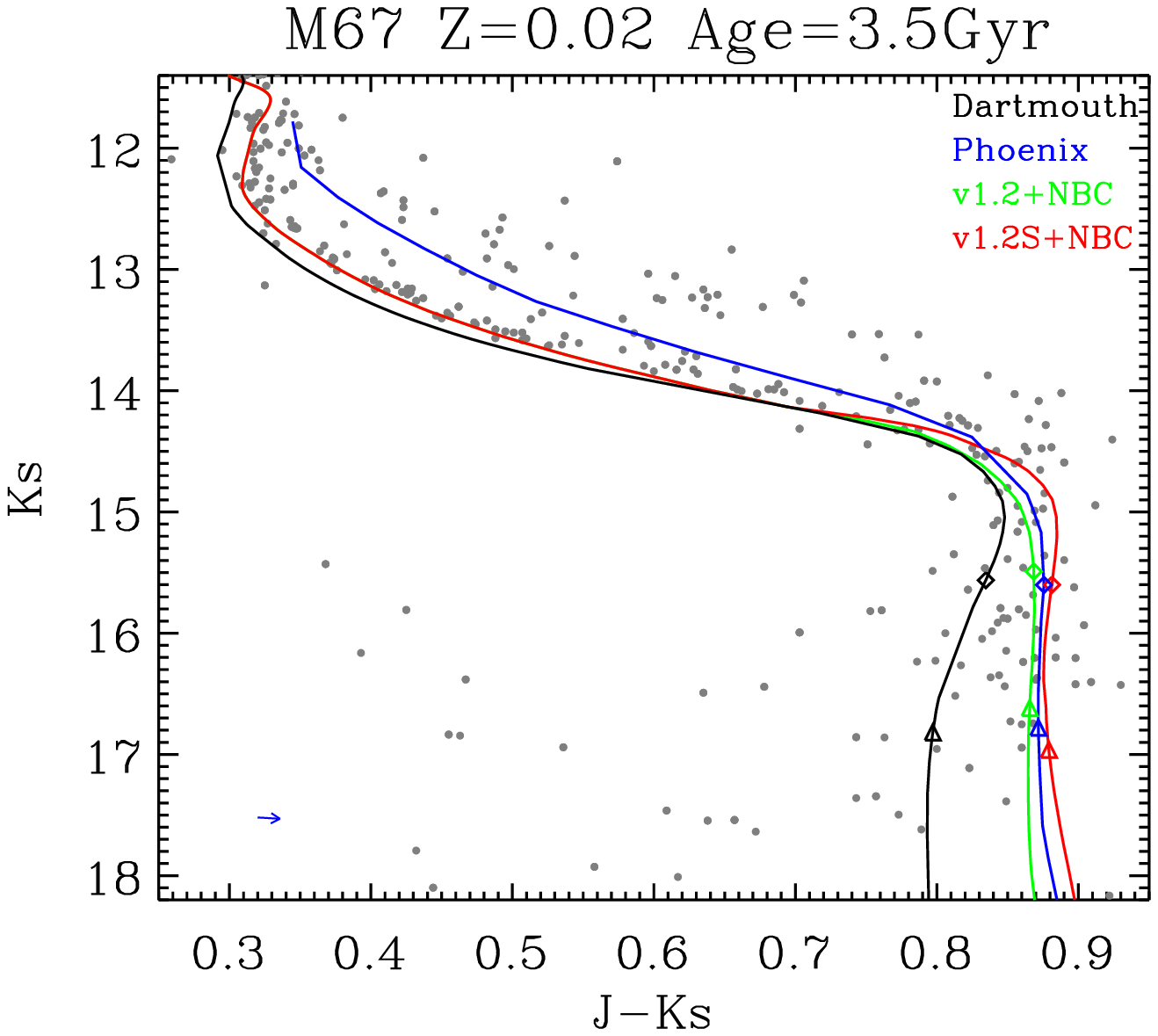}\\  %Ks (g-ks)
\includegraphics[width=0.5\textwidth]{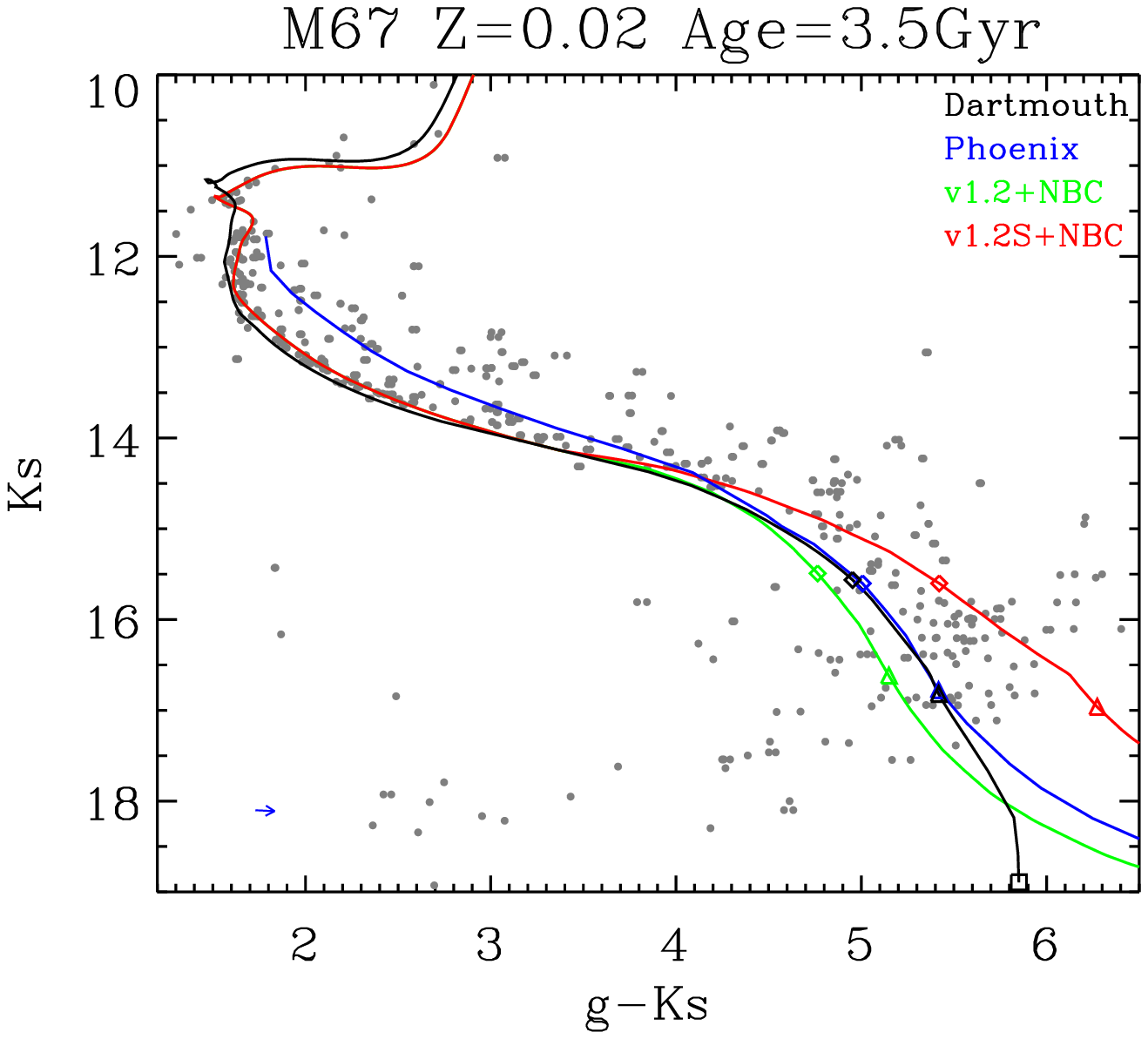}~  %g (g-z)
\includegraphics[width=0.5\textwidth]{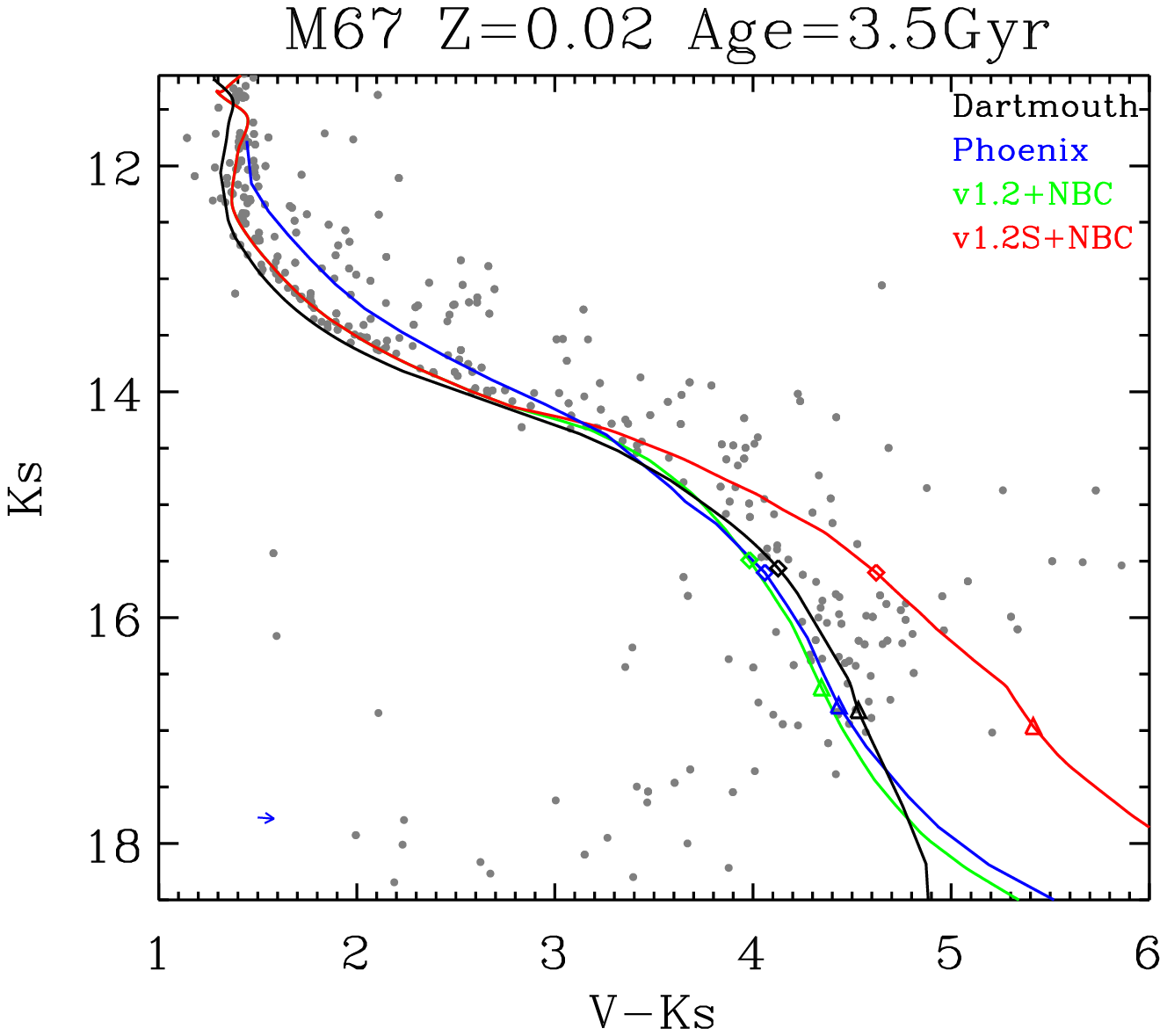}    %Ks (J-Ks)
\caption{M\,67 in several CMDs, same as in Fig.~\ref{fig_m67cmds}, except that the black lines are for Dartmouth with $Z=0.01885$ 
and $\rm age=3.5$~Gyr, and the blue ones is for PHOENIX with $Z=0.02$ and $\rm age=3.5$~Gyr.}
\label{fig_m67:other}
\end{figure*}

\begin{figure*}
\includegraphics[width=0.5\textwidth]{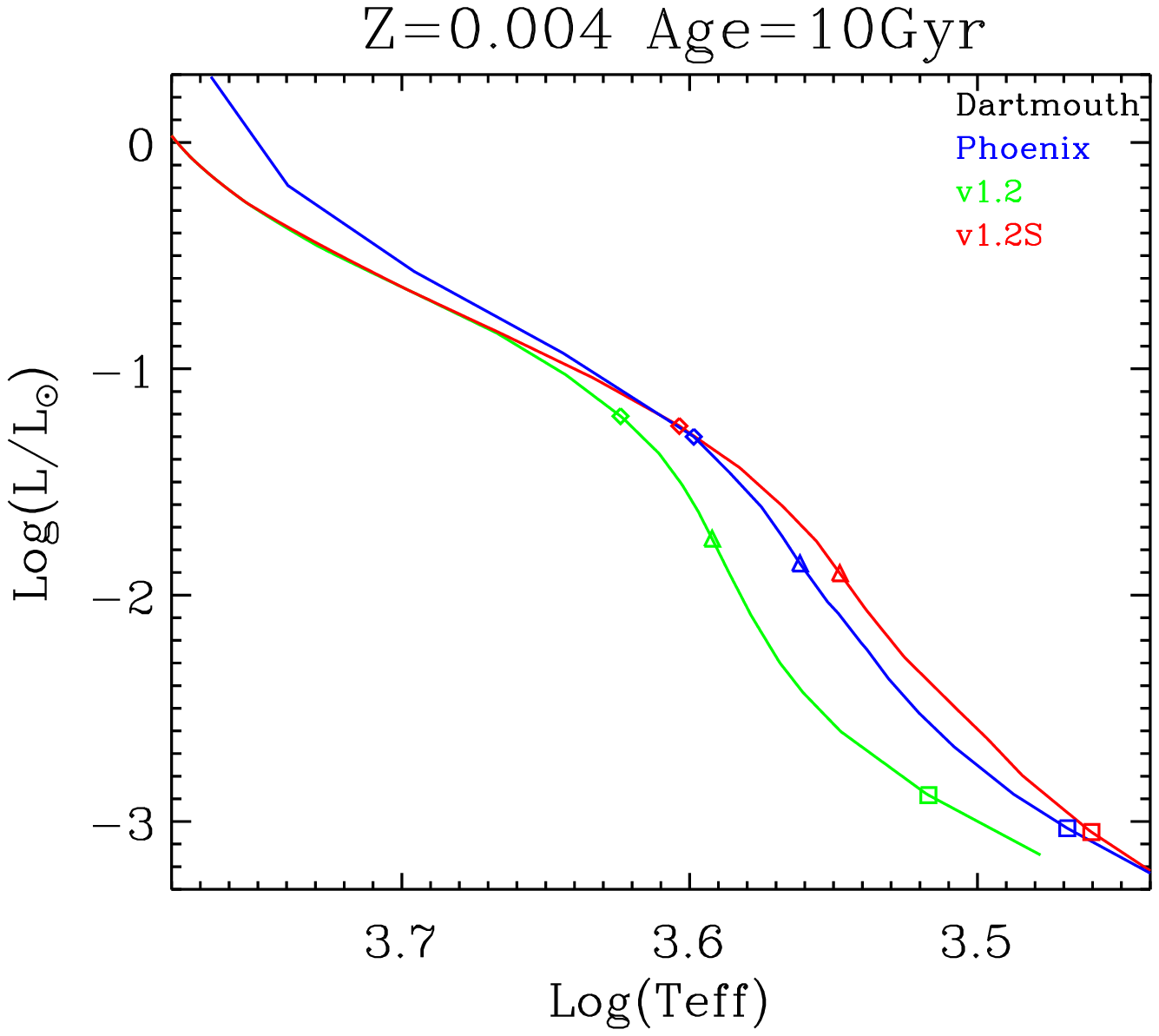}~  %L~Teff
\includegraphics[width=0.5\textwidth]{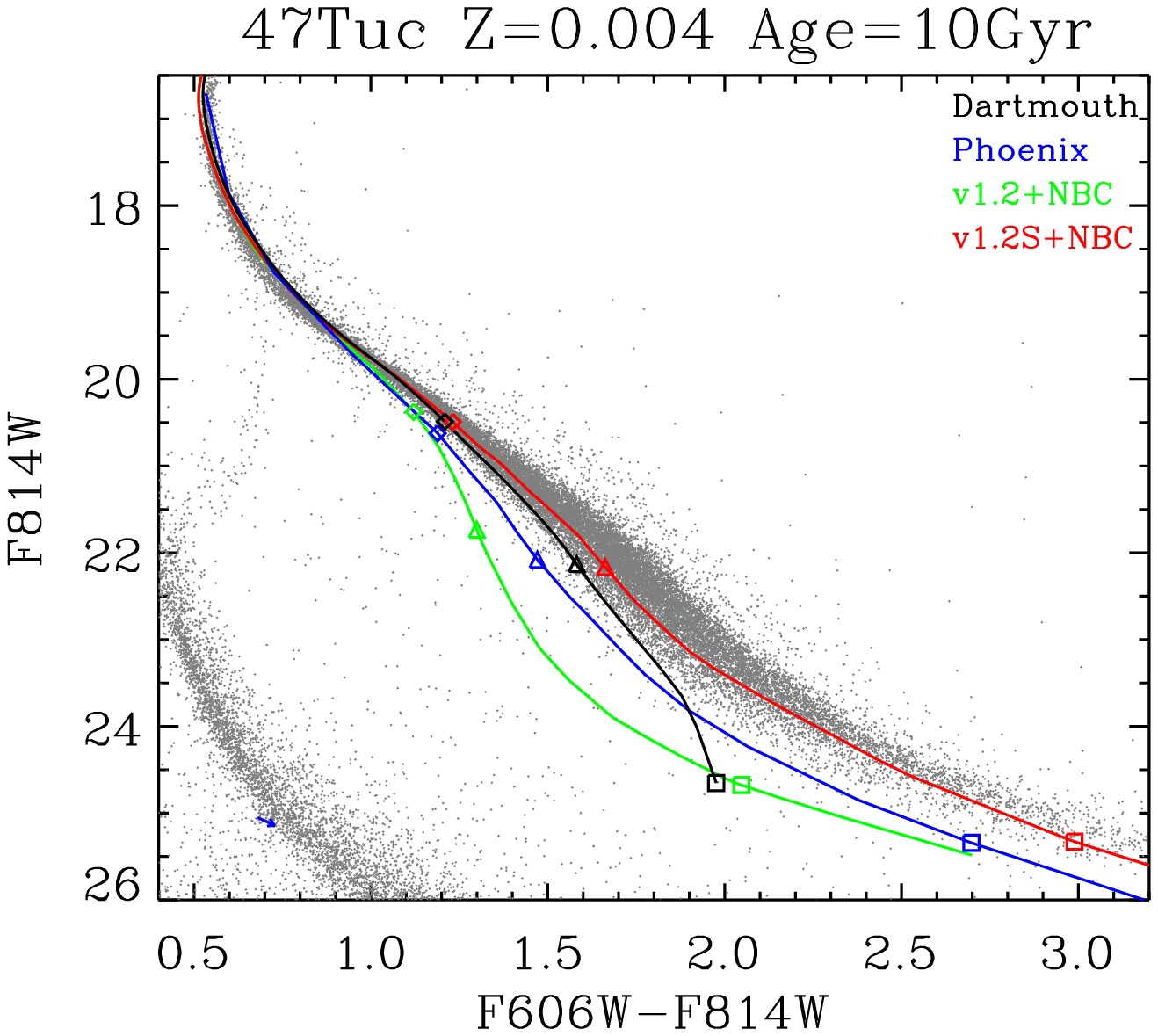}   %I - (V-I)
\caption{CMDs for 47\,Tuc, same as in Fig.~\ref{fig_47Tuc}, except that black lines are for Dartmouth ($Z=0.0053740$ and $\rm age=10$~Gyr) 
and blue ones are for PHOENIX ($Z=0.006340$ and $\rm age=10$~Gyr).}

\label{fig_47Tuc:other}
\end{figure*}

\begin{figure*}
\includegraphics[width=0.5\textwidth]{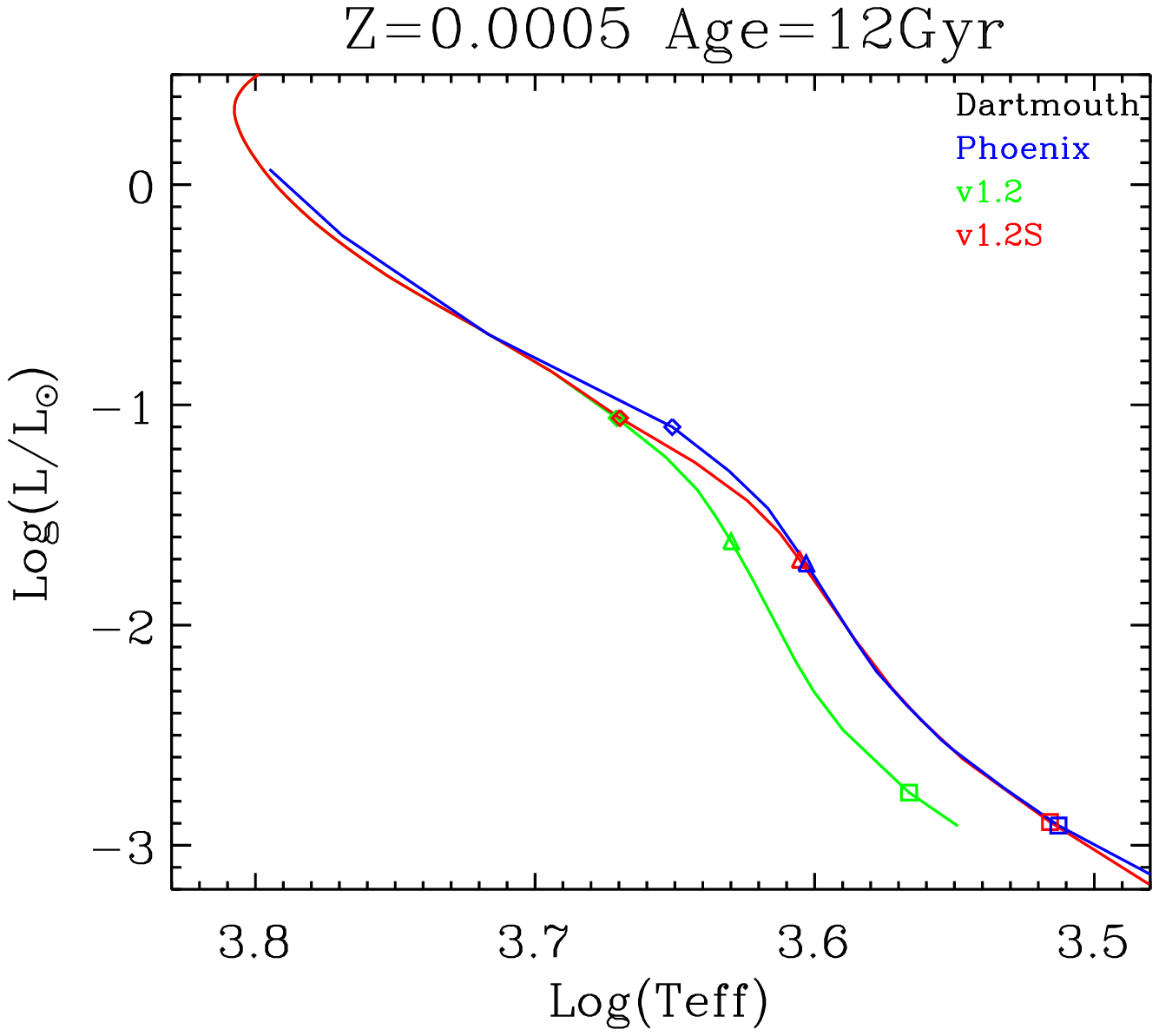}~  %L~Teff
\includegraphics[width=0.5\textwidth]{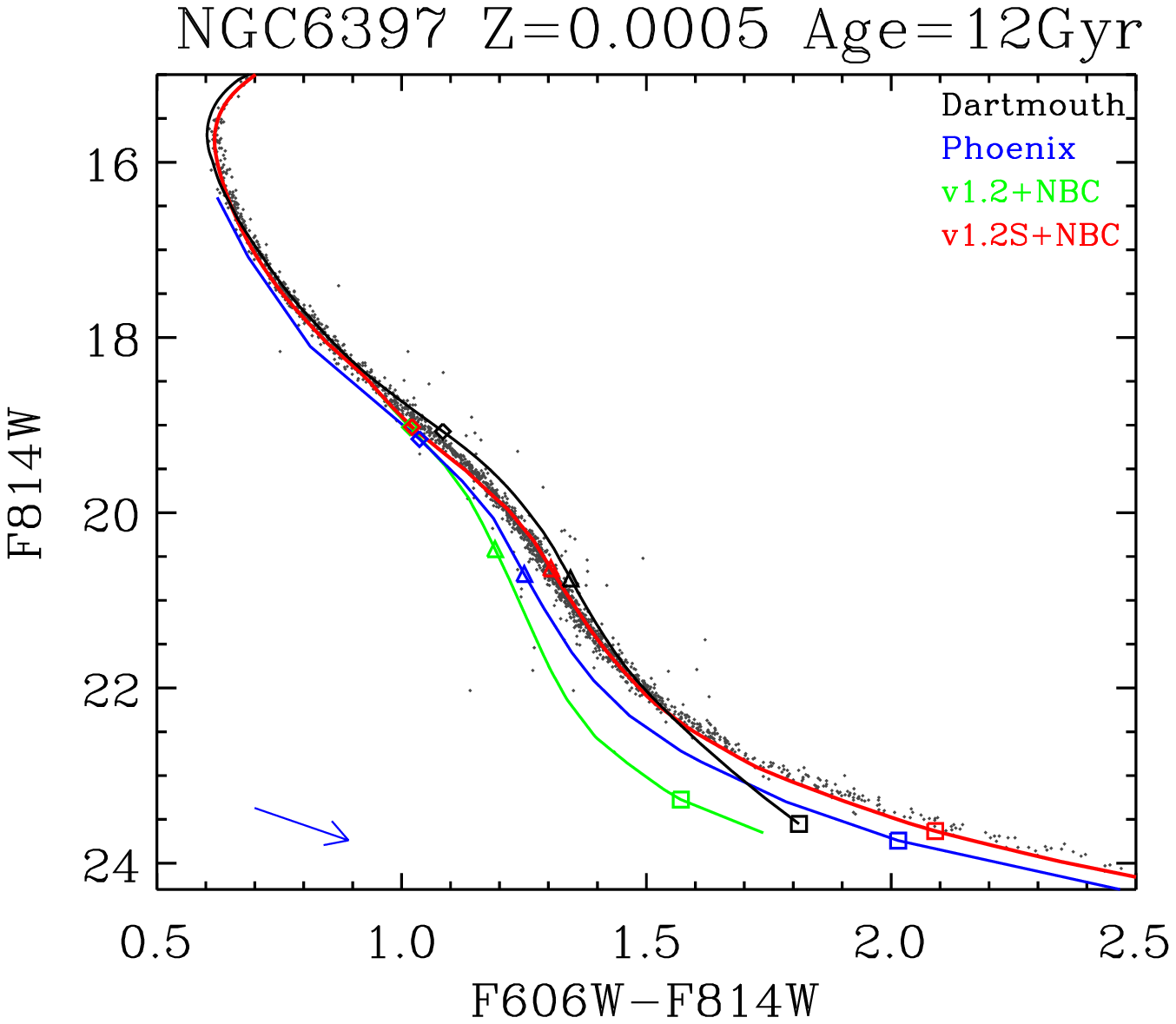}   %I - (V-I)
\caption{CMDs for NGC6397, same as in Fig.~\ref{fig_NGC6397}, except that black lines are for Dartmouth ($Z=0.00054651$ and $\rm age=12$~Gyr), 
and blue ones are for PHOENIX ($Z=0.000307$ and $\rm age=12$~Gyr).}
\label{fig_NGC6397:other}
\end{figure*}

The sequence of Figs.~\ref{fig_Praesepe:other} to \ref{fig_NGC6397:other} compares our best-performing isochrone set so 
far -- namely the PARSEC v1.2+NBC -- with those from other groups  already introduced in Sect.~\ref{sec_massradius}, and 
with the data of the four clusters we have just discussed. In doing so, we are by no means trying to find the best-fitting 
isochrone for each model set and cluster, we are just overplotting them for the same assumed distance and reddening, for a 
quick comparison of the several sets. The comparisons are first made in the H--R diagram, and later in the 
CMDs for which we have isochrones available in the same filter sets.

As can be seen in the figures, none of the sets being compared agree perfectly in the H--R diagram, even if the Y-Y and 
Dartmouth models implement similar $\Ttau$ relations as in our PARSEC v1.2 models. Our v1.2 models are slightly hotter 
than both Dartmouth and PHOENIX models in the lower main-sequence at near-solar metallicity (Figs.~\ref{fig_Praesepe:other} 
and \ref{fig_m67:other}), but still significantly hotter than those at low metallicities (Figs.~\ref{fig_47Tuc:other} 
and \ref{fig_NGC6397:other}). Tracing back the origin of these differences is difficult at this stage, and is beyond the scope of this paper.

When we look at the CMDs in Figs.~\ref{fig_Praesepe:other} to \ref{fig_NGC6397:other}, in addition to the intrinsic 
difference in the evolutionary tracks, we also see the effect of the different tables of bolometric corrections adopted 
by the different groups. What is more remarkable in the Preasepe and M\,67 plots is that all models seem to reproduce 
satisfactorily the kink that occurs at the bottom of the main sequence in the near-infrared colour $\jks$. However, in all 
cases the fit is far from satisfactory when we look at the colours which involve optical filters. The same 
applies to the two old globular clusters as shown in Figs.~\ref{fig_47Tuc:other} and \ref{fig_NGC6397:other}.

%%%%%%%%%%%%%%%%%%%%%%%%%%%%%%%%%%%%%%%%%%%%%%%%%%%%%%%%
\section{Discussion and conclusions}
\label{sec_final}

As is evident from the previous discussion, adopting better BC tables and \Ttau\ relations is not enough to bring 
models and data into agreement, and the problem seems to extend to other sets of models in the literature as well. 
Interestingly, we note that the changes requested in the mass-radius relation -- namely larger radii at a given 
mass -- go in the same sense of the changes required to improve the agreement with the CMDs -- namely lower \Teff\ (larger radii) 
for a given luminosity. Moreover, the discrepancies start to appear more or less at the same masses down the main sequence, namely 
at $\sim\!0.5$~\Msun. Therefore, it is natural to seek for changes that increase the stellar radii of the models, and check whether 
this causes better agreement with the CMD data. This is essentially the approach we will pursue in the following.

\subsection{A recalibration of the \texorpdfstring{$\Ttau$}{T-tau} relation\label{sec-rec}}

As can be seen  in Fig.~\ref{T_tau_nocorr},  the BT-Settl \Ttau\ relations are distributed in a relatively narrow 
region of the $T/\Teff$ versus $\tau$ plane and, at large values of $\tau$, they converge towards and even exceed 
the  gray \Ttau\ relation. This effect becomes more prominent at lower effective temperatures, 
where the \Ttau\ relations near the photosphere become significantly hotter than the gray atmosphere one.
The excess reaches $\Delta\log(T/\Teff)\sim0.04$~dex and is likely caused by the formation of molecules at low 
temperatures, which trap the radiation in the atmosphere. 
It is this shift of the \Ttau\ relation that causes some improvement in the mass--radius of the main sequence models
and, consequently, on the corresponding colour-magnitude relations.
However the agreement with observations of lower main sequence stars is far from being satisfactory. Thus we  wonder if  
(1) the mismatch can be due to an underestimate of the photospheric temperature by the \Ttau\ relations at smaller \Teff,
and if (2) we could use the observed mass-radius relation shown in Fig.~\ref{fig_massradius1}  to calibrate the  
\Ttau\ relations at low effective temperatures.

Concerning the first point  we can only say that there are many such relations in the literature and that
the empirically checked \citet{KS1966} relation predicts a significant shift already at $\sim$5000~K (dwarf stars), 
comparable to that obtained  by the BT-Settl \Ttau\ relation of the 2600~K model.

\begin{figure*}
\includegraphics[scale=0.8]{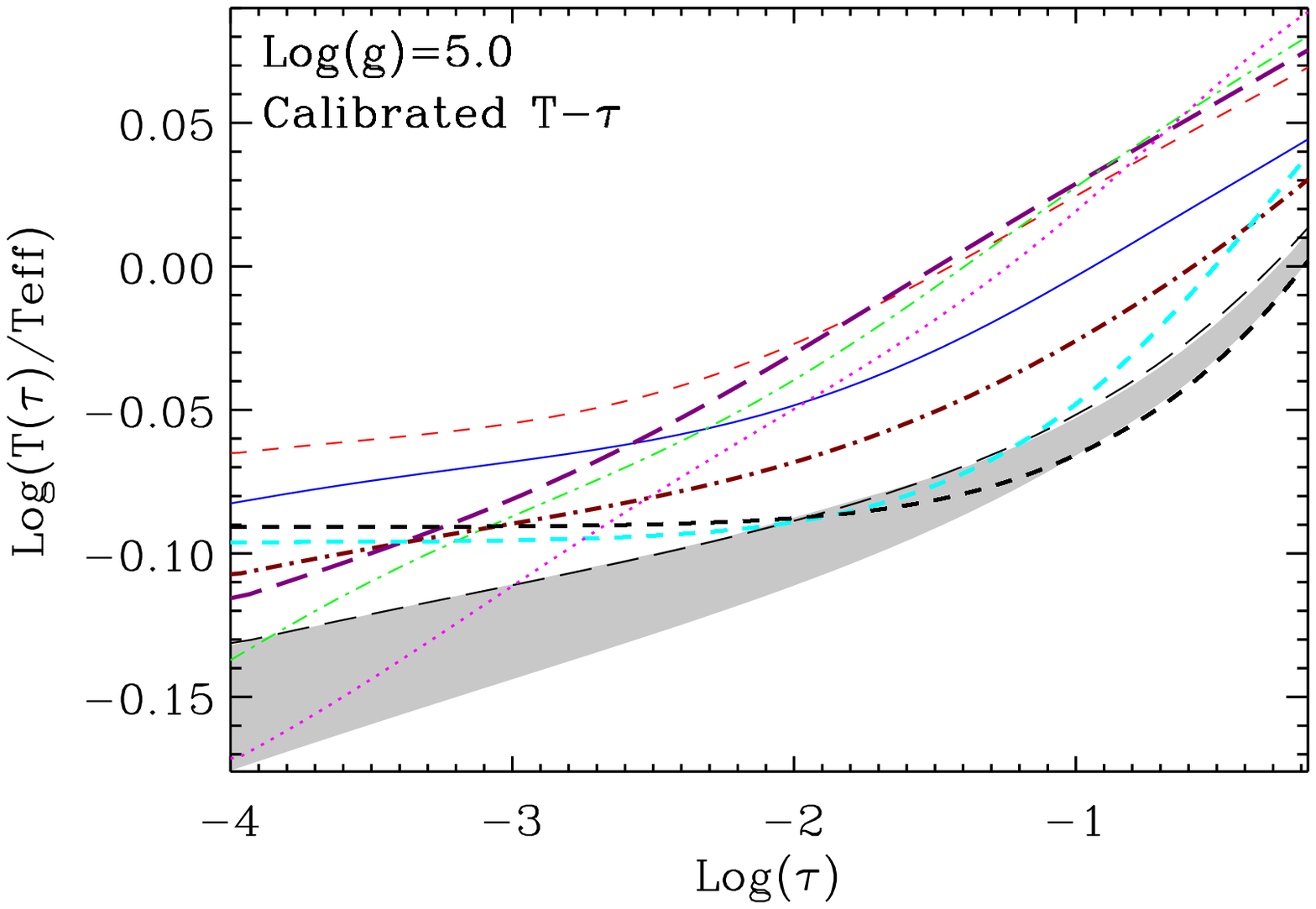}
\caption{As~Fig.~\ref{T_tau_nocorr}, but for calibrated $T-\tau$ relations, as described in section~\ref{sec-rec}. %\textbf{\boldmath (maximum $\rm \Delta Log(T(\tau)/T_{eff})=0.06$ for $T_{eff}/K \leq 3160$ )}
}
\label{T_tau}
\end{figure*}

\begin{figure*}
\includegraphics[width=0.7\textwidth]{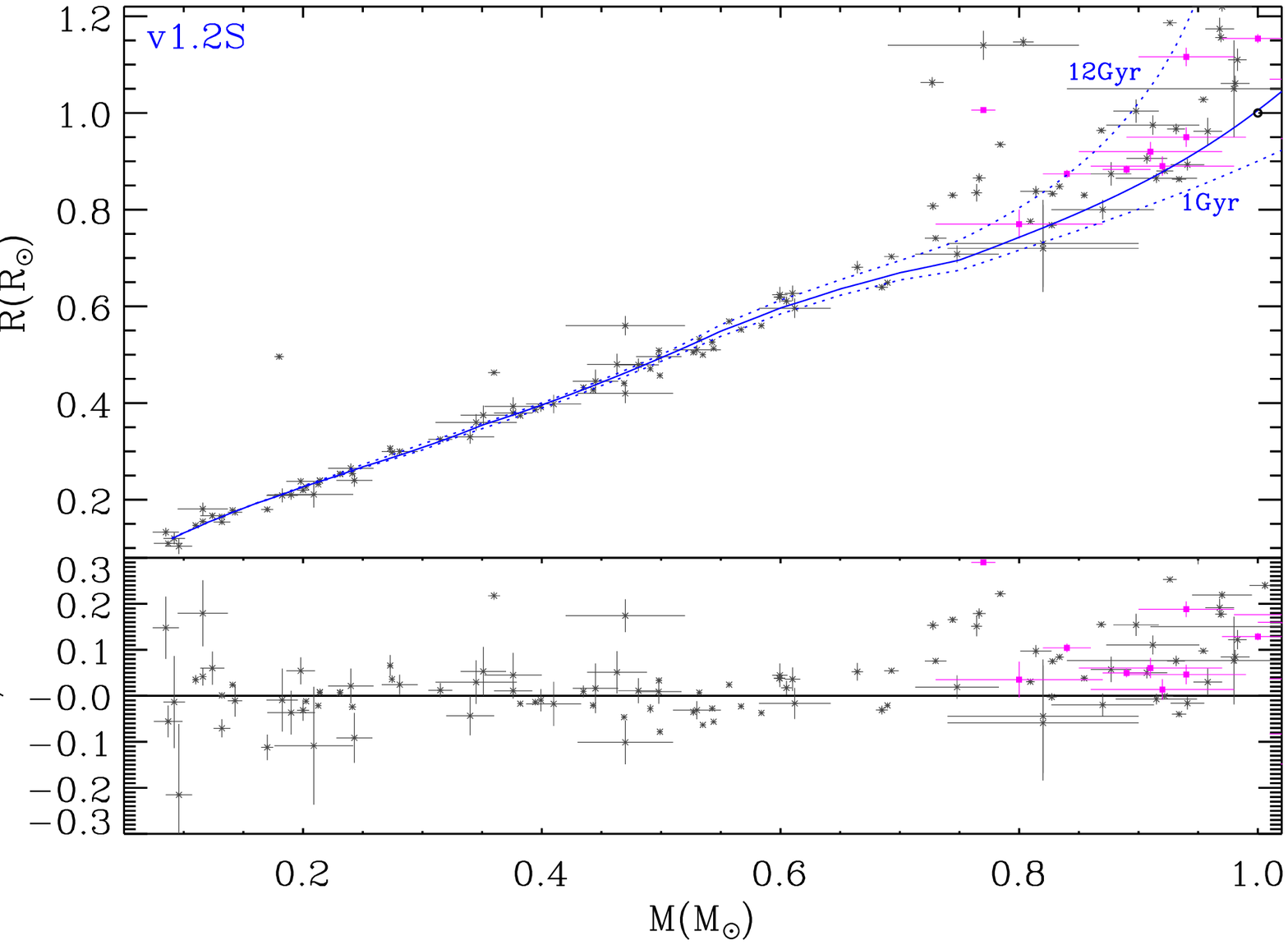}
\caption[]{As~Fig.~\ref{fig_massradius1}, but for calibrated $T-\tau$ relation. We also plot isochrones with ages of 12\,Gyr and 1\,Gyr.}
\label{fig_massradius3}
\end{figure*}

Concerning the second point, we have calculated a series of models for low mass stars
where we have applied a shift to the low temperature $\Ttau$ relations,
to reproduce the observed mass-radius relation. The correction factor depends on the effective temperature.
It is $\Delta\log(T/\Teff)=0$ at $\log(\Teff/{\rm K})=3.675$, and it increases linearly to
$\Delta\log(T/\Teff)\sim0.06$~dex ($\sim$14\%), at $\log(\Teff/{\rm K})=3.5$. The resulting $\Ttau$ relations are shown in Fig.~\ref{T_tau}.
Note that the correction is applied  {\em only} to the $\Ttau$ relation and we use our own EOS 
and opacity to get the pressure structure in the atmosphere.
The mass--radius calibration is shown in Fig.~\ref{fig_massradius3}, where it is indicated as PARSEC v1.2S models.
We checked that no shift is necessary at  \Teff\ higher than $\log(\Teff/{\rm K})=3.675$, since there is no need 
to alter the radii of stars with masses larger than $M=0.7$~\Msun\, at solar metallicities.

We now look at the effects of this calibration on the colour--magnitude relations of VLMS in clusters. 
The  results are illustrated by means of the PARSEC v1.2S+NBC models (red lines) overplotted in 
Figs.~\ref{fig_Praesepecmds},~\ref{fig_m67cmds},~\ref{fig_47Tuc} and~\ref{fig_NGC6397}, 
for Praesepe, M\,67, NGC~6398 and 47~Tuc, respectively. 
Notice that the same \Teff-dependent shift obtained from the calibration with the mass--radius relation,
is applied for all metallicities. Careful inspection of these plots reveals that the \Ttau-calibrated models provide an excellent fit 
to the lower MS in all these clusters, spanning a range of $\sim\!2$~dex in metallicity.

This result is remarkable. It is probably indicating that the problem at the origin of the too small radii in present VLMS models might be the same as that at the origin of their bad reproduction of the observed lower MS in cluster CMDs. 
Whether the present recipe of calibrating the available \Ttau\ relations in the way we described is an acceptable solution, 
is another question, which we open for discussion. Of course, we are well aware that, at this stage, this is not more than 
``a recipe that works'', rather than a recommendation we can do to stellar modellers. Work is  necessary to 
clarify whether more realistic descriptions of stellar atmospheres -- like for instance full 3D hydrodynamical 
models -- may lead to any indication of this kind, or to alternative recipes.

Another aspect revealed by Figs.~\ref{fig_Praesepecmds} and~\ref{fig_m67cmds}, is that 
the {\em near-infrared} CMDs of very low mass stars can be reproduced fairly well
with the  published \Ttau\ relations, provided that one uses the correct bolometric corrections. 
In the very low mass regime they are not as sensitive to \Teff\ as the optical CMDs.
Indeed we emphasize that  the optical CMDs should be considered as stronger diagnostic tools, together with the mass--radius 
relation which we used for the final calibration.
However we stress that, since the relation between mass and near-infrared luminosity is also affected by the adopted  \Ttau\ 
relation (see the top-left panel of Fig.~\ref{fig_Praesepecmds}), the use of different models, although reproducing equally well 
the near-infrared CMDs, may lead to different estimates of the present-day mass function in star clusters.

\subsection{Data release}

The VLMS models presented in this paper for $M< 0.75$~\Msun\ turn out to represent a significant improvement over the previous versions. 
The v1.2, by using a more realistic $\Ttau$ relation, clearly go in the direction of presenting larger radii and cooler \Teff, which is 
indicated by the data. Moreover, the ``calibrated'' v1.2S models fit very well the mass-radius and CMD data of VLMS in the Solar Neighbourhood 
and in star clusters over a wide range of metallicities, hence representing a good alternative to be applied in a series of astrophysical problems, 
going from the derivation of parameters of star clusters and transiting planets, to the interpretation of star counts in the Galaxy in terms of 
both their mass function and density variations across the galactic disk and halo.

Therefore, we are releasing these two new sets of evolutionary tracks through our web servers at \url{http://stev.oapd.inaf.it}. 
These VLMS models replace those in the previous database of PARSEC isochrones \citep[namely v1.1,][]{parsec}, producing the isochrones' 
version v1.2 and v1.2S. They become available using both the previous BC tables (OBC), and the revision based on BT-Settl models (NBC), 
for a wide variety of photometric systems, through our web interface \url{http://stev.oapd.inaf.it/cmd}. 
The models available are summarized in Table~\ref{tab_notation}.

Forthcoming papers will further extend these models towards higher masses, and other chemical mixtures, including the $\alpha$-enhanced 
ones. Moreover, work is ongoing to further improve the BC tables, especially for the coolest stars.

%%%%%%%%%%%%%%%%%%%%%%%%%%%%%%%%%%%%%%%%%%%%%%%%%%%%%

%%%%%%%%%%%%%%%%%%%%%%%
\section*{Acknowledgements}
We acknowledge financial support from contract ASI-INAF
I/009/10/0, and from the Progetto di Ateneo 2012, CPDA125588/12
funded by the University of Padova. YC and XK are supported by the National Natural Science Foundation of China 
(NSFC, Nos. 11225315, 11320101002), the Specialized Research Fund for the Doctoral Program of Higher Education 
(SRFDP, No. 20123402110037) and the Strategic Priority Research Program ``The Emergence of Cosmological Structures'' of the 
Chinese Academy of Sciences (No. XDB09000000). AB acknowledges financial
support from MIUR 2009. We thank Derek Homeier and France Allard for their help with PHOENIX BT-Settl models; Simone Zaggia, 
Guillermo Torres, Fab\'{i}ola Campos and S.O.\ Kepler for interesting discussions; Chen Wei-Ping and Jason Kalirai for 
sending their Praesepe and 47~Tuc data in computer-ready format. We are grateful to the referee for useful suggestions.

This publication makes use of data products from the Two Micron All
Sky Survey, which is a joint project of the University of
Massachusetts and the Infrared Processing and Analysis
Center/California Institute of Technology, funded by the National
Aeronautics and Space Administration and the National Science
Foundation.

%%%%%%%%%%%%%%%%%%%%%%%%%%%%%%%%%%%%%%%%%%%%%%
\input{mass_radius_tab.tex}

\clearpage
\bibliographystyle{mn2e/mn2e_new} %style mn.bst
\bibliography{mdwarfs} % your references file.bib

%\clearpage
\appendix
\section{Blue band colours}
\label{sec_m67BV}

As already discussed, we were able to obtain new stellar evolution models of low mass stars that can reproduce fairly well the observed MS of star clusters in a broad range of ages and metallicities. 
We show here that there remains some tension when we try to reproduce the optical colours, such as $B-V$, of selected nearby clusters, and discuss the possible origin of the remaining discrepancies.

\begin{figure*}
\includegraphics[width=0.6\textwidth]{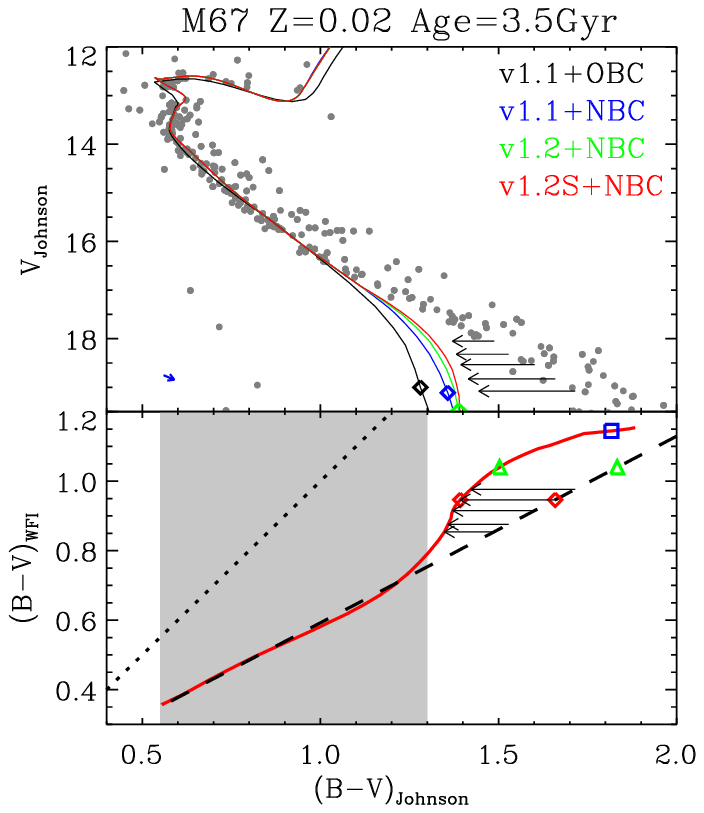}   %B-V
\caption{Top panel: The M\,67 data from \citet{yadav_etal08}, in the $V$ vs.\ $B-V$ diagram (gray dots). These data were originally collected in the ESO/Wide Field Imager (WFI) filters and then transformed to a Johnson system (see text). Our isochrones in the Johnson system are overlaid, using the same labels and parameters as in Fig.~\ref{fig_m67cmds}. 
Bottom panel: the difference between ESO/WFI $B-V$ and Johnson $B-V$, with our v1.2S isochrone shown as red solid line. The black dashed line represents the linear transformation between these two systems as defined in the colour range of $0.55<(B-V)_{\rm Johnson}<1.3$ (grey shaded region) and extrapolated to redder colour (or lower masses). The horizontal black arrows indicate the corrections that should have been applied to the linear transformation, in order to more correctly represent the colours of the redder stars observed with ESO/WFI filters. The dotted line is the identity line.}
\label{fig_m67BV}
\end{figure*}

In Fig.~\ref{fig_m67BV}, we compare our models with one of the deepest CMD of M\,67 in the $V$ versus $B-V$ diagram \citep{yadav_etal08}.
The models plotted are the same as used in Fig.~\ref{fig_m67cmds} and show the following characteristics.
First of all, they are able to reproduce the main sequence down to about three magnitudes below the turn-off, but then they run bluer than the observed data at fainter magnitudes.
It is also evident that the major difference between the models is due to the use of revised bolometric corrections rather than to the \Ttau\ relation.
The discrepancy with the data reaches $\delta(B-V) = 0.4$ at $V = 19$, even if our new models perform very well in other colours for the same cluster (as seen in Fig.~\ref{fig_m67cmds}). 

To find the origin of this large discrepancy we first remind the reader that 
\citet{yadav_etal08}'s data were obtained using a far-from-standard $B$ filter, namely the ESO\#842 filter available at the ESO/WFI camera in 2000. This $B$ filter has a transmission curve strongly skewed towards the red and with a sharp cut-off at $\lambda>5100$~\AA, instead of the more extended (and slightly skewed to the blue) curve expected from a Johnson filter\footnote{This difference can be appreciated in Fig.~3 of \citet{Girardi_etal02}, where ESO\#842 appears as the $B$ filter in the ESO Imaging Survey (ESO/EIS) photometric system, and is compared to the \citep{Bessell1990} representation of the Johnson $B$ filter.}. Indeed, the ESO\#842 mean wavelength is close to 4637~\AA, as compared to the 4460~\AA\ of the \citet{Bessell1990} $B$ filter. When coupled to a $V$ filter (with a mean wavelength of 5500~\AA), ESO\#842 provides a wavelength baseline of just 863~\AA, as compared to the $\sim\!$1040~\AA\ baseline expected from Johnson filters.\footnote{These and other mean wavelengths are provided at the web interface \url{http://stev.oapd.inaf.it/cmd}, together with the theoretical isochrones. }

Although collected in this very particular set of filters, \citet{yadav_etal08} observations were then ``calibrated'' using linear transformations between their instrumental magnitudes and the magnitudes of stars in common with the \citet{Sandquist2004} M\,67 catalogue, being the latter in a well-calibrated Johnson system. As \citealt{yadav_etal08} show in their paper, the stars used to derive the transformations are bluer than $(B-V)_{\rm Johnson} = 1.4$ so that for redder (and fainter) stars this linear transformation becomes an extrapolation. This step is critical for obtaining reliable magnitudes of the fainter stars and its validity must be carefully assessed -- especially in this case, where the $B$ filter is very different from a Johnson one.

To clarify this point we show the transformation between the ESO/WFI $B-V$ and Johnson's $B-V$ [hereafter $(B-V)_{\rm WFI}$ and $(B-V)_{\rm Johnson}$, respectively] by the red thick line in the lower panel of Fig.~\ref{fig_m67BV}. The colours in both systems were obtained using the ZAMS of our theoretical isochrones as a baseline, together with the PHOENIX BT-Settl spectral library and the corresponding filter transmission curves. Even if it is not exactly the same method used by \citet{yadav_etal08}, this procedure shows that the transformation between the two photometric systems is far from being linear just redward of $(B-V)_{\rm Johnson} = 1.3$. The analogue of the linear transformation used by \citet{yadav_etal08} is shown as a dashed line in the lower panel of the same figure. The dashed line is a linear fit of the actual transformation for $(B-V)_{\rm Johnson}\geq 1.3$. 
We evince from this exercise that, by extrapolating the linear transformation to stars redder than $(B-V)_{\rm Johnson}=1.3$, one gets $(B-V)_{\rm Johnson}$ colours that are significantly redder than those expected from the actual transformation. A measure of the possible error in the $(B-V)_{\rm Johnson}$ colours of VLMS is shown by the horizontal arrows in the lower panel, which are also reported in the upper panel.

%B-V temperature scale
\begin{figure*}
\includegraphics[width=0.5\textwidth]{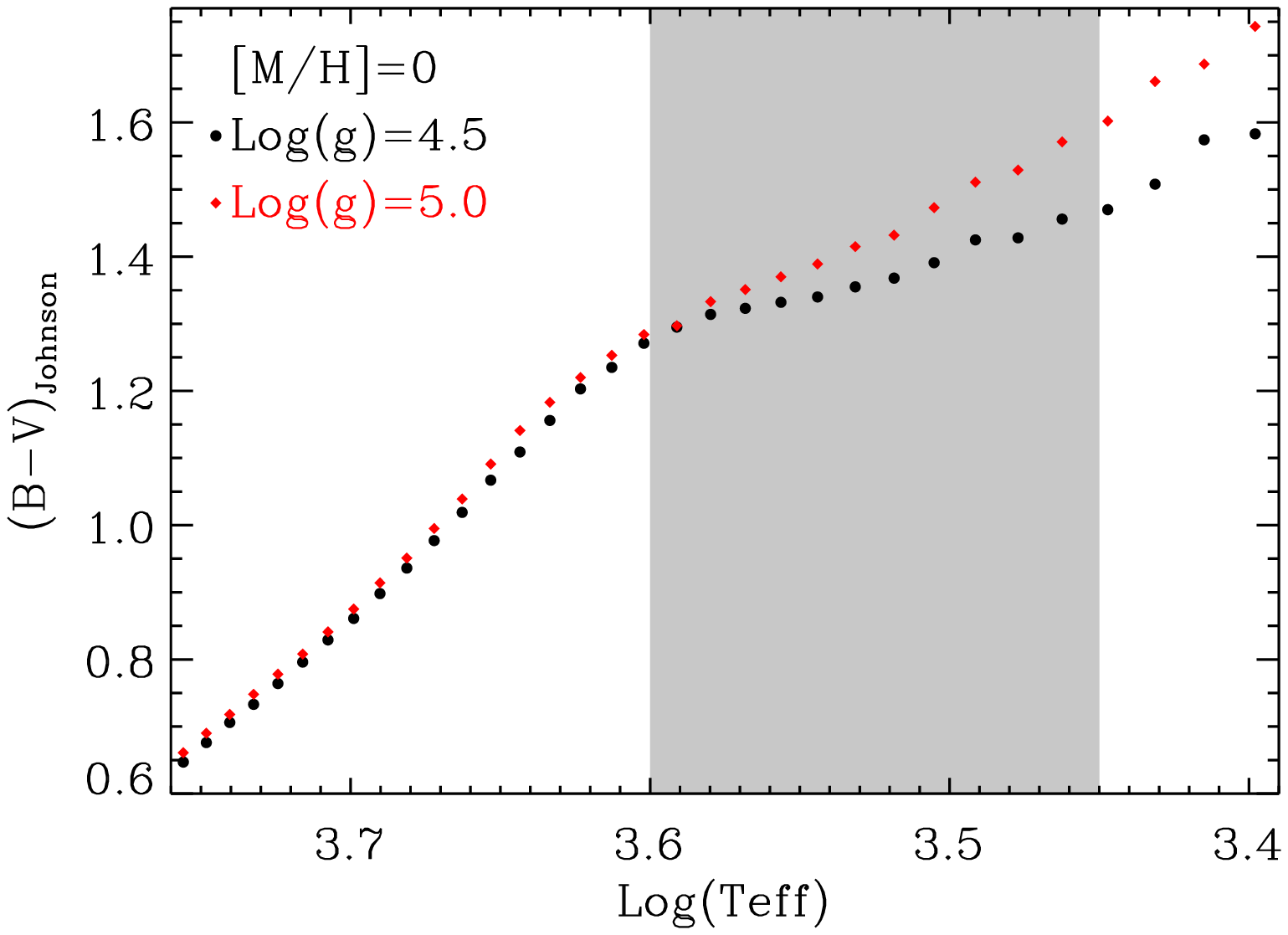}   %B-V
\caption{Temperature scale (\logT vs.\ $(B-V)_{\rm Johnson}$) for PHOENIX BT-Settl models of [M/H]=0. Red and black dots are for $\logg=5.0$ and 4.5 respectively. The grey area indicates the region where $B-V$ increases slowly as \logT\ decreases.}
\label{fig_BVscale}
\end{figure*}

These comparisons with \cite{yadav_etal08} data just emphasize the need of collecting data for VLMS in nearby open clusters using {\em standard} filters, together with a robust calibration of the photometry based on standard stars covering the widest-possible color range. Otherwise, any comparison with theoretical models \citep[as those presented in Fig.~10 of][]{yadav_etal08} may turn out to be misleading.

To reinforce this finding we plot in Fig.~\ref{fig_BVscale} the $(B-V)_{\rm Johnson}$ colours against $\logT$ of PHOENIX models (from their website).
Since below $\Teff\simeq4000$\,K the $B-V$ colours become quite flat, we expect a knee-like shape in the $V$ vs. $B-V$ diagram, as seen in our plotted isochrones, but not in the ``putative-Johnson'' CMD from \citet{yadav_etal08}.

%solar neighbourhood
\begin{figure*}
\includegraphics[width=0.5\textwidth]{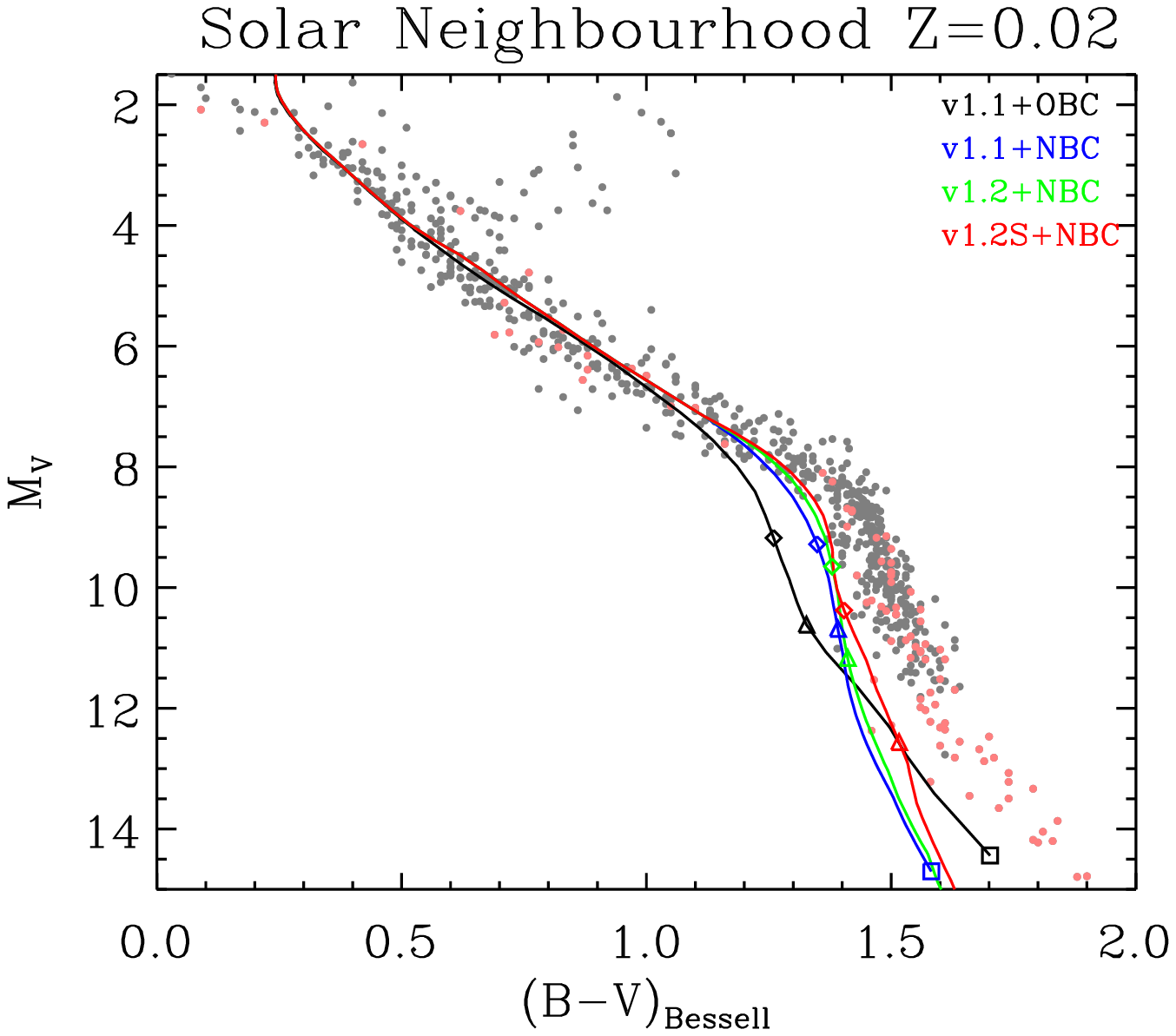}~   %B-V
\includegraphics[width=0.5\textwidth]{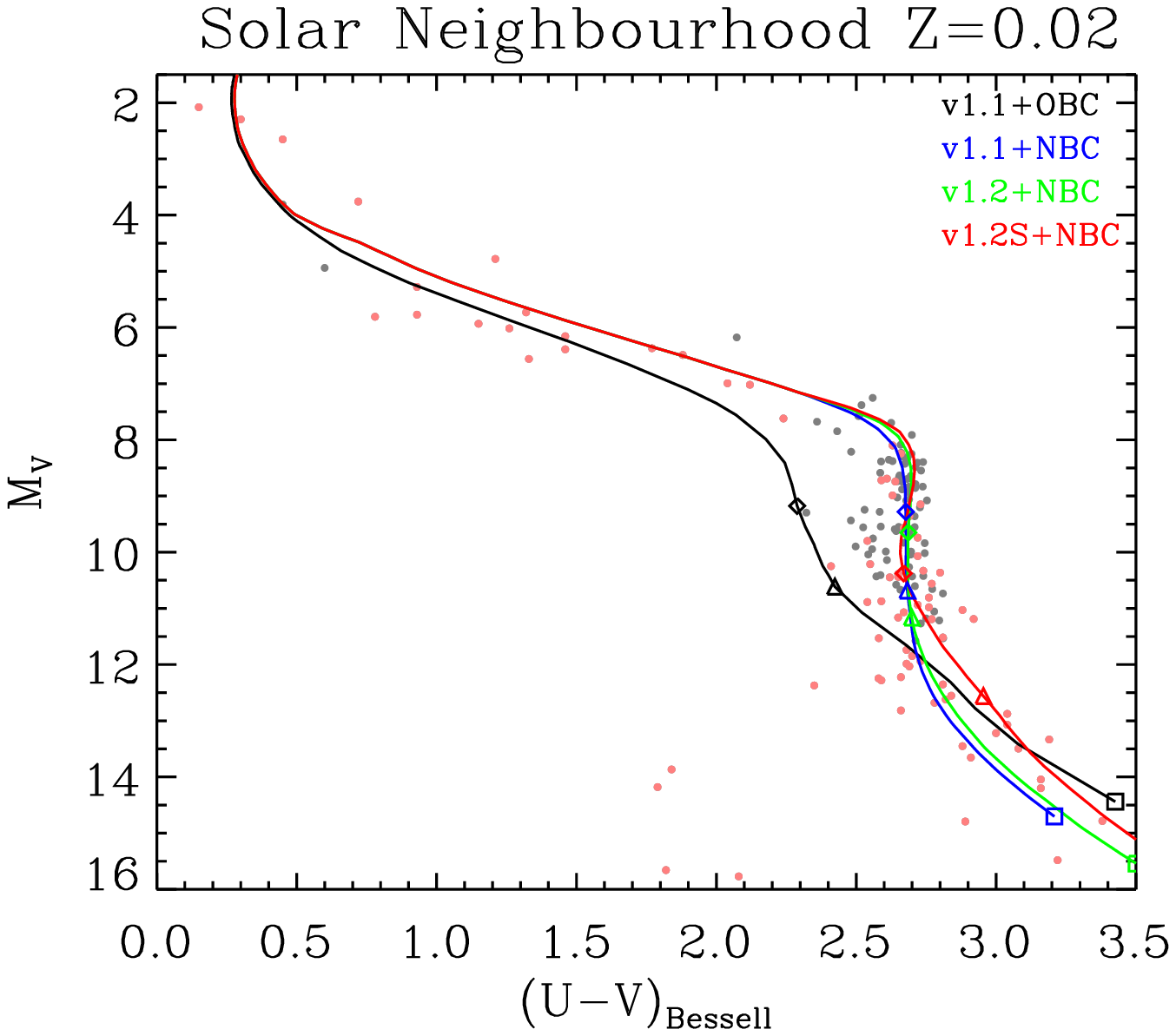}   %B-V
\caption{Solar neighbourhood stars in the $M_V$ vs.\ $B-V$ and $M_V$ vs.\ $U-V$ diagrams. The data points are from the CNS3 star sample; stars within 8\,pc of the Sun are marked with pink dots. The labels are the same as in Fig.~\ref{fig_Praesepecmds}. The absolute magnitudes are in the Bessell photometric system \citep{Bessell1990}. The galactic reddening is not considered in this plot. The isochrones are for $Z=0.02$. Although the stars have different ages, their evolution is quite slow, thereby we only show the isochrones of 1~Gyr.}
\label{fig_sn}
\end{figure*}

Another interesting -- and more conclusive -- comparison between models and data can be done using the Solar Neighbourhood data compiled by I.\ Neill Reid from \citet{Bessell91}, \citet{Leggett92} and \citet{Dahn_etal02} catalogues\footnote{\url{http://www.stsci.edu/~inr/cmd.html}}.
The $V$ vs.\ $B-V$ data in the Bessell system, are plotted in the left panel of Fig.~\ref{fig_sn}, together with our models. We first notice that these data show the expected knee-like shape in the very low mass range. But we also notice that our best model mismatch the data by $\sim 0.1$~mag at 0.3\,$\Msun$. 
We suspect that this discrepancy originates from some molecular absorption bands around 4400~\AA\ (AlH and NaH) being still unaccounted for in the models, as recently pointed out in \citet{Rajpurohit2013}. Indeed, from their Fig.~3
we notice that the observed spectra run slightly below the PHOENIX models at $\sim4400$~\AA\ in the temperature range $3000<\Teff/{\rm K}<3700$.
This shows that the real $B-V$ should be slightly redder than those obtained by the PHOENIX models, but more work is needed to check if the difference really amounts to $\sim0.04$~dex in flux, as expected from the observed mismatch.

Finally, the Solar Neighbourhood $V$ vs.\ $U-V$ data in the Bessell system, are plotted in the right panel of Fig.~\ref{fig_sn}, together with our models. In contrast to the $B-V$ colour, our new models can fit quite well the $U-V$ knee-like shape at $U-V\simeq2.7$, starting at $M_V\ga8$ and extending down to $M_V\simeq11$. There remain some possible mismatch at still fainter magnitudes, but the data are too scarce for drawing definitive conclusions.

\label{lastpage}

\end{document}

%% file: mass_radius_tab.tex
\onecolumn

%\addtolength{\tabcolsep}{-1.1pt}
%\renewcommand{\arraystretch}{1.5}
%\setlength\LTleft{0pt}            % default: \parindent
%\setlength\LTright{0pt}           % default: \fill
%\begin{tiny}
%\begin{longtable}[]{|l|l|l|l|l|l|l|l|l|}
\begin{longtable}[]{l|llllllll}
\caption[Mass-radius data]{Mass-radius data. Notations used are AS: asteroseismic; SYN: spin-orbit synchronisation; SB1: single-lined binary; EH: exoplanet host. 2MASS04463285: 2MASS04463285+1901432. LP133-373: mass ratio=1 assumed.  
\label{mass_radius_tab}}\\
%\begin{table*}
%\begin{landscape} %%%% does not work
\hline
Name& $M_*(\Msun) $&$\sigma(M_*(\Msun))$&$R_*(\Rsun)$& $\sigma(R_*(\Rsun))$& System type & Method & Ref.\footnotemark{} & Comment\\ 
\hline
KIC6521045  & 1.08 & $\pm$0.06 & 1.49 & $\pm$0.04 & single & AS & [1] & EH\\
KIC3544595  & 0.91 & $\pm$0.06 & 0.92 & $\pm$0.02 & single & AS & [1] & EH\\
KIC4914423  & 1.09 & $\pm$0.07 & 1.44 & $\pm$0.04 & single & AS & [1] & EH\\
KIC8349582  & 1.08 & $\pm$0.08 & 1.41 & $\pm$0.04 & single & AS & [1] & EH\\
KIC5094751  & 1.04 & $\pm$0.06 & 1.32 & $\pm$0.04 & single & AS & [1] & EH\\
KIC4349452  & 1.19 & $\pm$0.06 & 1.31 & $\pm$0.02 & single & AS & [1] & EH\\
KIC8478994  & 0.80 & $\pm$0.07 & 0.77 & $\pm$0.03 & single & AS & [1] & EH\\
KIC11295426 & 1.08 & $\pm$0.05 & 1.24 & $\pm$0.02 & single & AS & [1] & EH\\
KIC8753657  & 1.07 & $\pm$0.06 & 1.07 & $\pm$0.02 & single & AS & [1] & EH\\
KIC10963065 & 1.08 & $\pm$0.07 & 1.23 & $\pm$0.03 & single & AS & [1] & EH\\
KIC9955598  & 0.92 & $\pm$0.06 & 0.89 & $\pm$0.02 & single & AS & [1] & EH\\
%\hline
TrES-2 & 0.94 & $\pm$0.05 & 0.95 & $\pm$0.02 & single & RV & [2]  & EH\\
%\hline
HATS550-016\_P & 0.97  & $^{+0.05}_{-0.06}$   & 1.22  & $^{+0.02}_{-0.03}$   & binary & SYN & [3] & SB1 \\
HATS550-016\_S & 0.110 & $^{+0.005}_{-0.006}$ & 0.147 & $^{+0.003}_{-0.004}$ & binary & SYN & [3] & SB1 \\
HATS551-019\_P & 1.10  & $^{+0.05}_{-0.09}$   & 1.70  & $^{+0.09}_{-0.09}$   & binary & SYN & [3] & SB1 \\
HATS551-019\_S & 0.17  & $^{+0.01}_{-0.01}$   & 0.18  & $^{+0.01}_{-0.01}$   & binary & SYN & [3] & SB1 \\
HATS551-021\_P & 1.1   & $^{+0.1}_{-0.1}$     & 1.20  & $^{+0.08}_{-0.01}$   & binary & SYN & [3] & SB1 \\
HATS551-021\_S & 0.132 & $^{+0.014}_{-0.005}$ & 0.154 & $^{+0.006}_{-0.008}$ & binary & SYN & [3] & SB1 \\
HATS553-001\_P & 1.2   & $^{+0.1}_{-0.1}$     & 1.58  & $^{+0.08}_{-0.03}$   & binary & SYN & [3] & SB1 \\
HATS553-001\_S & 0.20  & $^{+0.01}_{-0.02}$   & 0.22  & $^{+0.01}_{-0.01}$   & binary & SYN & [3] & SB1 \\
%\hline
HP\_Aur\_P & 0.9543 & $\pm$0.0041 & 1.0278 & $\pm$0.0042 & binary & RV & [4] & \\
HP\_Aur\_S & 0.8094 & $\pm$0.0036 & 0.7758 & $\pm$0.0034 & binary & RV & [4] & \\
%\hline
V65\_P & 0.8035 & $\pm$0.0086 & 1.1470 & $\pm$0.0104 & binary & RV & [5] & \\
V65\_S & 0.6050 & $\pm$0.0044 & 0.6110 & $\pm$0.0092 & binary & RV & [5] & \\
V66\_P & 0.7842 & $\pm$0.0045 & 0.9347 & $\pm$0.0048 & binary & RV & [5] & \\
V66\_S & 0.7443 & $\pm$0.0042 & 0.8298 & $\pm$0.0053 & binary & RV & [5] & \\
V69\_P & 0.7665 & $\pm$0.0053 & 0.8655 & $\pm$0.0097 & binary & RV & [5] & \\
V69\_S & 0.7278 & $\pm$0.0048 & 0.8074 & $\pm$0.0080 & binary & RV & [5] & \\
%\hline
HD181068\_A  & 3.0   & $\pm$0.1   & 12.46 & $\pm$0.15  & triple & RV & [6] & \\
HD181068\_Ba & 0.915 & $\pm$0.034 & 0.865 & $\pm$0.010 & triple & RV & [6] & \\
HD181068\_Bb & 0.870 & $\pm$0.043 & 0.800 & $\pm$0.020 & triple & RV & [6] & \\
%\hline
C4780Bb & 0.096 & $\pm$0.011 & 0.104 & $\pm$0.0160 & binary & RV & [7] & primary:F-star\\ 
%\hline
NSVS07394765\_P & 0.360 & $\pm$0.005 & 0.463 & $\pm$0.004 & binary & RV & [8] & \\
NSVS07394765\_S & 0.180 & $\pm$0.004 & 0.496 & $\pm$0.005 & binary & RV & [8] & \\
%\hline
WTS19g-4-02069\_P & 0.53  & $\pm$0.02  & 0.51  & $\pm$0.01  & binary & RV & [9] & \\
WTS19g-4-02069\_S & 0.143 & $\pm$0.006 & 0.174 & $\pm$0.006 & binary & RV & [9] & \\
%\hline
KOI-126A & 1.3470 & $\pm$0.0320 & 2.0254 & $\pm$0.0098 & triple & RV & [10] & \\
KOI-126B & 0.2413 & $\pm$0.0030 & 0.2543 & $\pm$0.0014 & triple & RV & [10] & \\
KOI-126C & 0.2127 & $\pm$0.0026 & 0.2318 & $\pm$0.0013 & triple & RV & [10] & \\
%\hline
KIC6131659\_P & 0.922 & $\pm$0.007 & 0.8800 & $\pm$0.0028 & binary & RV & [11] & \\
KIC6131659\_S & 0.685 & $\pm$0.005 & 0.6395 & $\pm$0.0061 & binary & RV & [11] & \\
%\hline
MG1-78457\_P   & 0.527 & $\pm$0.002 & 0.505 & $^{+0.008}_{-0.007}$ & binary & RV & [12] & \\
MG1-78457\_S   & 0.491 & $\pm$0.001 & 0.471 & $^{+0.009}_{-0.007}$ & binary & RV & [12] & \\
MG1-116309\_P  & 0.567 & $\pm$0.002 & 0.552 & $^{+0.004}_{-0.013}$ & binary & RV & [12] & \\
MG1-116309\_S  & 0.532 & $\pm$0.002 & 0.532 & $^{+0.004}_{-0.008}$ & binary & RV & [12] & \\
MG1-506664\_P  & 0.584 & $\pm$0.002 & 0.560 & $^{+0.001}_{-0.004}$ & binary & RV & [12] & \\
MG1-506664\_S  & 0.544 & $\pm$0.002 & 0.513 & $^{+0.001}_{-0.008}$ & binary & RV & [12] & \\
MG1-646680\_P  & 0.499 & $\pm$0.002 & 0.457 & $^{+0.006}_{-0.004}$ & binary & RV & [12] & \\
MG1-646680\_S  & 0.443 & $\pm$0.002 & 0.427 & $^{+0.006}_{-0.002}$ & binary & RV & [12] & \\
MG1-1819499\_P & 0.557 & $\pm$0.001 & 0.569 & $^{+0.002}_{-0.023}$ & binary & RV & [12] & \\
MG1-1819499\_S & 0.535 & $\pm$0.001 & 0.500 & $^{+0.003}_{-0.014}$ & binary & RV & [12] & \\
MG1-2056316\_P & 0.469 & $\pm$0.002 & 0.441 & $^{+0.002}_{-0.002}$ & binary & RV & [12] & \\
MG1-2056316\_S & 0.382 & $\pm$0.001 & 0.374 & $^{+0.002}_{-0.002}$ & binary & RV & [12] & \\
%\hline
SDSSJ1212−0123 & 0.273 & $\pm$0.002 & 0.306 & $\pm$0.007 & binary & RV & [13] & primary:WD\\
GK-Vir & 0.116 & $\pm$0.003 & 0.155 & $\pm$0.003 & binary & RV & [13] & primary:WD\\
SDSSJ0857+0342 & 0.087 & $\pm$0.012 & 0.1096 & $\pm$0.0038 & binary & RV & [14] & primary:WD\\
SDSS0138−0016 & 0.132 & $\pm$0.003 & 0.165 & $\pm$0.001 & binary & RV & [15] & primary:WD\\
%\hline
Kepler-16\_P & 0.6897 & $^{+0.0035}_{-0.0034}$ & 0.6489 & $^{+0.0013}_{-0.0013}$ & binary & RV & [16] & \\
Kepler-16\_S & 0.20255 & $^{+0.00066}_{-0.000654}$ & 0.22623 & $^{+0.00059}_{-0.00053}$ & binary & RV & [16] & \\
%\hline
CM-Dra\_P & 0.2310 & $\pm$0.0009 & 0.2534 & $\pm$0.0019 & binary & RV & [17] & \\
CM-Dra\_S & 0.2141 & $\pm$0.0010 & 0.2396 & $\pm$0.0015 & binary & RV & [17] & \\
%\hline
T-Boo0-00080\_P & 1.49  & $\pm$0.07  & 1.83  & $\pm$0.03  & binary & SYN & [18] & \\
T-Boo0-00080\_S & 0.315 & $\pm$0.010 & 0.325 & $\pm$0.005 & binary & SYN & [18] & \\
T-Lyr1-01662\_P & 0.77  & $\pm$0.08  & 1.14  & $\pm$0.03  & binary & SYN & [18] & \\
T-Lyr1-01662\_S & 0.198 & $\pm$0.012 & 0.238 & $\pm$0.007 & binary & SYN & [18] & \\
T-Lyr0-08070\_P & 0.95  & $\pm$0.11  & 1.36  & $\pm$0.05  & binary & SYN & [18] & \\
T-Lyr0-08070\_S & 0.240 & $\pm$0.019 & 0.265 & $\pm$0.010 & binary & SYN & [18] & \\
T-Cyg1-01385\_P & 0.91  & $\pm$0.15  & 1.63  & $\pm$0.08  & binary & SYN & [18] & \\
T-Cyg1-01385\_S & 0.345 & $\pm$0.034 & 0.360 & $\pm$0.017 & binary & SYN & [18] & \\
%\hline
HAT-TR-205-013\_P & 1.04  & $\pm$0.13  & 1.28  & $\pm$0.04  & binary & RV & [19] & SB1 \\
HAT-TR-205-013\_S & 0.124 & $\pm$0.010 & 0.167 & $\pm$0.006 & binary & RV & [19] & SB1 \\
%\hline
ASAS-01A\_P & 0.612 & $\pm$0.030 & 0.596 & $\pm$0.020 & multiple & RV & [20] & \\
ASAS-01A\_P & 0.445 & $\pm$0.019 & 0.445 & $\pm$0.024 & multiple & RV & [20] & \\
%\hline
LSPM-J1112+7626\_P & 0.3946 & $\pm$0.0023 & 0.3860 & $^{+0.0055}_{-0.0028}$ & binary & RV & [21] & \\ 
LSPM-J1112+7626\_S & 0.2745 & $\pm$0.0012 & 0.2978 & $^{+0.0049}_{-0.0046}$ & binary & RV & [21] & \\ 
%\hline
WTS19b-2-01387\_P & 0.498 & $\pm$0.019 & 0.496 & $\pm$0.013 & binary & RV & [22] & \\ 
WTS19b-2-01387\_S & 0.481 & $\pm$0.017 & 0.479 & $\pm$0.013 & binary & RV & [22] & \\ 
WTS19c-3-01405\_P & 0.410 & $\pm$0.023 & 0.398 & $\pm$0.019 & binary & RV & [22] & \\ 
WTS19c-3-01405\_S & 0.376 & $\pm$0.024 & 0.393 & $\pm$0.019 & binary & RV & [22] & \\ 
WTS19e-3-08413\_P & 0.463 & $\pm$0.025 & 0.480 & $\pm$0.022 & binary & RV & [22] & \\ 
WTS19e-3-08413\_S & 0.351 & $\pm$0.019 & 0.375 & $\pm$0.020 & binary & RV & [22] & \\ 
%\hline
V1061-Cyg\_P  & 1.282   & $\pm$0.016   & 1.616   & $\pm$0.017  & binary & RV & [23] & \\
V1061-Cyg\_S  & 0.9315  & $\pm$0.0074  &  0.967  & $\pm$0.011  & binary & RV & [23] & \\
RT-And\_P     & 1.240   & $\pm$0.030   & 1.256   & $\pm$0.015  & binary & RV & [23] & \\
RT-And\_S     & 0.907   & $\pm$0.017   & 0.906   & $\pm$0.011  & binary & RV & [23] & \\
FL-Lyr\_P     & 1.218   & $\pm$0.016   & 1.283   & $\pm$0.028  & binary & RV & [23] & \\
FL-Lyr\_S     & 0.958   & $\pm$0.012   & 0.962   & $\pm$0.028  & binary & RV & [23] & \\
ZZ-UMa\_P     & 1.1386  & $\pm$0.0052  &  1.513  & $\pm$0.019  & binary & RV & [23] & \\
ZZ-UMa\_S     & 0.9691  & $\pm$0.0048  & 1.1562  & $\pm$0.0096 & binary & RV & [23] & \\
$\alpha$-Cen\_P & 1.105   & $\pm$0.007   & 1.224   & $\pm$0.003  & binary & RV & [23] & \\
$\alpha$-Cen\_S & 0.934   & $\pm$0.006   & 0.863   & $\pm$0.005  & binary & RV & [23] & \\
V568-Lyr\_P   & 1.0745  & $\pm$0.0077  &  1.400  & $\pm$0.016  & binary & RV & [23] & \\
V568-Lyr\_S   & 0.8273  & $\pm$0.0042  & 0.7679  & $\pm$0.0064 & binary & RV & [23] & \\
V636Cen\_P    & 1.0518  & $\pm$0.0048  & 1.0186  & $\pm$0.0043 & binary & RV & [23] & \\
V636Cen\_S    & 0.8545  & $\pm$0.0030  & 0.8300  & $\pm$0.0043 & binary & RV & [23] & \\
CV-Boo\_P     & 1.032   & $\pm$0.013   & 1.263   & $\pm$0.023  & binary & RV & [23] & \\
CV-Boo\_S     & 0.968   & $\pm$0.012   & 1.174   & $\pm$0.023  & binary & RV & [23] & \\
V1174-Ori\_P  & 1.006   & $\pm$0.013   & 1.338   & $\pm$0.011  & binary & RV & [23] & \\
V1174-Ori\_S  & 0.7271  & $\pm$0.0096  &  1.063  & $\pm$0.011  & binary & RV & [23] & \\
UV-Psc\_P     & 0.9829  & $\pm$0.0077  &  1.110  & $\pm$0.023  & binary & RV & [23] & \\
UV-Psc\_S     & 0.7644  & $\pm$0.0045  &  0.835  & $\pm$0.018  & binary & RV & [23] & \\
CG-Cyg\_P     & 0.941   & $\pm$0.014   & 0.893   & $\pm$0.012  & binary & RV & [23] & \\
CG-Cyg\_S     & 0.814   & $\pm$0.013   & 0.838   & $\pm$0.011  & binary & RV & [23] & \\
RW-Lac\_P     & 0.9263  & $\pm$0.0057  & 1.1864  & $\pm$0.0038 & binary & RV & [23] & \\
RW-Lac\_S     & 0.8688  & $\pm$0.0040  & 0.9638  & $\pm$0.0040 & binary & RV & [23] & \\
HS-Aur\_P     & 0.898   & $\pm$0.019   & 1.004   & $\pm$0.024  & binary & RV & [23] & \\
HS-Aur\_S     & 0.877   & $\pm$0.017   & 0.874   & $\pm$0.024  & binary & RV & [23] & \\
GU-Boo\_P     & 0.6101  & $\pm$0.0064  &  0.627  & $\pm$0.016  & binary & RV & [23] & \\
GU-Boo\_S     & 0.5995  & $\pm$0.0064  & 0.624   & $\pm$0.016  & binary & RV & [23] & \\
YY-Gem\_P     & 0.5992  & $\pm$0.0047  & 0.6194  & $\pm$0.0057 & binary & RV & [23] & \\
YY-Gem\_S     & 0.5992  & $\pm$0.0047  & 0.6194  & $\pm$0.0057 & binary & RV & [23] & \\
CU-Cnc\_P     & 0.4349  & $\pm$0.0012  & 0.4323  & $\pm$0.0055 & binary & RV & [23] & \\
CU-Cnc\_S     & 0.39922 & $\pm$0.00089 &  0.3916 & $\pm$0.0094 & binary & RV & [23] & \\
CM-Dra\_P     & 0.23102 & $\pm$0.00089 &  0.2534 & $\pm$0.0019 & binary & RV & [23] & \\
CM-Dra\_S     & 0.21409 & $\pm$0.00083 &  0.2398 & $\pm$0.0018 & binary & RV & [23] & \\ 
%\hline
LP133-373 & 0.34 & $\pm$0.02 & 0.330 & $\pm$0.014 & binary & RV & [24] & \\ 
%\hline
ASAS-04\_P & 0.8338 & $\pm$0.0036 & 0.848 & $\pm$0.005 & binary & RV & [25] & \\ 
ASAS-04\_S & 0.8280 & $\pm$0.0040 & 0.833 & $\pm$0.005 & binary & RV & [25] & \\ 
%\hline
GJ3236\_P & 0.376 & $\pm$0.016 & 0.3795 & $\pm$0.0064 & binary & RV & [26] & \\ 
GJ3236\_S & 0.281 & $\pm$0.015 & 0.2996 & $\pm$0.0064 & binary & RV & [26] & \\ 
%\hline
AP-And\_P   & 1.211 & $\pm$0.024 & 1.218 & $\pm$0.013 & binary & RV & [27] & \\
AP-And\_S   & 1.222 & $\pm$0.024 & 1.226 & $\pm$0.061 & binary & RV & [27] & \\
VZ-Cep\_P   & 1.376 & $\pm$0.027 & 1.622 & $\pm$0.019 & binary & RV & [27] & \\
VZ-Cep\_S   & 1.073 & $\pm$0.023 & 0.934 & $\pm$0.025 & binary & RV & [27] & \\
V881-Per\_P & 0.912 & $\pm$0.039 & 0.975 & $\pm$0.020 & binary & RV & [27] & \\
V881-Per\_S & 0.748 & $\pm$0.035 & 0.708 & $\pm$0.018 & binary & RV & [27] & \\
%\hline
IM-Vir\_P & 0.981  & $\pm$0.012  & 1.061 & $\pm$0.016 & binary & RV & [28] & \\
IM-Vir\_S & 0.6644 & $\pm$0.0048 & 0.681 & $\pm$0.013 & binary & RV & [28] & \\
%\hline
RXJ0239.1\_P & 0.730 & $\pm$0.009 & 0.741 & $\pm$0.004 & binary & RV & [29]\ & \\
RXJ0239.1\_S & 0.693 & $\pm$0.006 & 0.703 & $\pm$0.002 & binary & RV & [29] & \\
%\hline
NSVS0103\_P & 0.5428 & $\pm$0.0027 & 0.5260 & $\pm$0.0028 & binary & RV & [30]  & \\
NSVS0103\_S & 0.4982 & $\pm$0.0025 & 0.5088 & $\pm$0.0030 & binary & RV & [30]  & \\
%\hline
2MASS04463285\_P & 0.47 & $\pm$0.05 & 0.56 & $\pm$0.02 & binary & RV & [31] & \\
2MASS04463285\_S & 0.19 & $\pm$0.02 & 0.21 & $\pm$0.01 & binary & RV & [31] & \\
%\hline
KIC1571511\_P & 1.265 & $^{+0.036}_{-0.030}$ & 1.343  & $^{+0.012}_{-0.010}$   & binary & RV & [32] & SB1 \\
KIC1571511\_S & 0.141 & $^{+0.005}_{-0.004}$ & 0.1783 & $^{+0.0014}_{-0.0017}$ & binary & RV & [32] & SB1 \\
%\hline
RR-Cae & 0.1825 & $\pm$0.0131 & 0.2090 & $\pm$0.0143 & binary & RV & [33] & primary:WD\\
%\hline
OGLE-TR-123\_P & 1.29  & $\pm$0.26  & 1.55  & $\pm$0.10  & binary & RV & [34] & SB1\\
OGLE-TR-123\_S & 0.085 & $\pm$0.011 & 0.133 & $\pm$0.009 & binary & RV & [34] & SB1\\
%\hline
OGLE-TR-122\_P & 0.98  & $\pm$0.14  & 1.05  & $^{+0.20}_{-0.09}$  & binary & RV & [35] & SB1\\
OGLE-TR-122\_S & 0.092 & $\pm$0.009 & 0.120 & $^{+0.024}_{-0.013}$ & binary & RV & [35]  & SB1\\
%\hline
OGLE-TR-125\_S & 0.209 & $\pm$0.033 & 0.211 & $\pm$0.027 & binary & RV & [36] & SB1\\1
OGLE-TR-120\_S & 0.47 & $\pm$0.04 & 0.42 & $\pm$0.02 & binary & RV & [36] & SB1\\1
OGLE-TR-114\_P & 0.82 & $\pm$0.08 & 0.73 & $\pm$0.09 & triple & RV & [36] & \\
OGLE-TR-114\_S & 0.82 & $\pm$0.08 & 0.72 & $\pm$0.09 & triple & RV & [36] & \\
OGLE-TR-106\_S & 0.116 & $\pm$0.021 & 0.181 & $\pm$0.013 & binary & RV & [36] & SB1\\1
OGLE-TR-78\_S & 0.243 & $\pm$0.015 & 0.24 & $\pm$0.013 & binary & RV & [36] & SB1\\1
OGLE-TR-65\_P & 1.15 & $\pm$0.03 & 1.58 & $\pm$0.07 & triple & RV & [36] & \\
OGLE-TR-65\_S & 1.11 & $\pm$0.03 & 1.59 & $\pm$0.05 & triple & RV & [36]  & \\
%\hline
KIC7871531 & 0.84 & $\pm$0.02 & 0.874 & $\pm$0.008 & single & AS & [37] & \\
KIC8006161 & 1.04 & $\pm$0.02 & 0.947 & $\pm$0.007 & single & AS & [37] & \\
KIC8394589 & 0.94 & $\pm$0.04 & 1.116 & $\pm$0.019 & single & AS & [37] & \\
KIC8694723 & 0.96 & $\pm$0.03 & 1.436 & $\pm$0.024 & single & AS & [37] & \\
KIC8760414 & 0.77 & $\pm$0.01 & 1.006 & $\pm$0.004 & single & AS & [37] & \\
KIC9098294 & 1.00 & $\pm$0.03 & 1.154 & $\pm$0.009 & single & AS & [37] & \\
KIC9955598 & 0.89 & $\pm$0.02 & 0.883 & $\pm$0.008 & single & AS & [37] & \\
\hline
\footnotetext{References: [1] \citet{Marcy2014} table 1; [2] \citet{Barclay2012} table 1; [3] \citet{Zhou2014} table 4; [4] \citet{Lacy2014} table 7; [5] \citet{Kaluzny2013} table 12; [6] \citet{Borkovits2013} table 4; [7] \citet{Tal-Or2013} table 4; [8] \citet{Cakirli2013} table 5; [9] \citet{Nefs2013} table 5; [10] \citet{Carter2011} table 1; [11] \citet{Bass2012} table 6; [12] \citet{Kraus2011} table 8; [13] \citet{Parsons2012a} table 9; [14] \citet{Parsons2012b} table 5; [15] \citet{Parsons2012c} table 4; [16] \citet{Doyle2011} table 1; [17] \citet{Morales2009a} table 9; [18] \citet{Fernandez2009} table 13; [19] \citet{Beatty2007} table 8; [20] \citet{Helminiak2012} table 5; [21] \citet{Irwin2011} table 10; [22] \citet{Birkby2012} table 11; [23] \citet{Torres2010} table 1; [24] \citet{Vaccaro2007} table 1; [25] \citet{Helminiak2011} table3; [26] \citet{Irwin2009} table 9; [27] \citet{Zola2014} table 6; [28] \citet{Morales2009b} table 11; [29] \citet{Lopez-Morales2007} table 2; [30] \citet{Lopez-Morales2006} table 5; [31] \citet{Hebb2006} table 2; [32] \citet{Ofir2012} table 3; [33] \citet{Maxted2007} table 5; [34] \citet{Pont2006} table 2; [35] \citet{Pont2005b} table 2; [36] \citet{Pont2005a} table 7; [37] \citet{Metcalfe2014} table 1.
} 
\end{longtable}
%\end{tiny}
%%%%\end{landscape}
%\end{table*}
\twocolumn

%% file: draft0.bbl
\begin{thebibliography}{112}
\expandafter\ifx\csname natexlab\endcsname\relax\def\natexlab#1{#1}\fi

\bibitem[{{Allard} \& {Hauschildt}(1995)}]{AllardHauschildt95}
{Allard} F., {Hauschildt} P.~H., 1995, \apj, 445, 433

\bibitem[{{Allard} {et~al}\mbox{.}(2000){Allard}, {Hauschildt}, {Alexander},
  {Ferguson}, \& {Tamanai}}]{Allard_etal00}
{Allard} F., {Hauschildt} P.~H., {Alexander} D.~R., {Ferguson} J.~W., {Tamanai}
  A., 2000, in Astronomical Society of the Pacific Conference Series, Vol. 212,
  From Giant Planets to Cool Stars, {Griffith} C.~A., {Marley} M.~S., eds., p.
  127

\bibitem[{{Allard} {et~al}\mbox{.}(1997){Allard}, {Hauschildt}, {Alexander}, \&
  {Starrfield}}]{Allard_etal97}
{Allard} F., {Hauschildt} P.~H., {Alexander} D.~R., {Starrfield} S., 1997,
  \araa, 35, 137

\bibitem[{{Allard}, {Homeier} \& {Freytag}(2012){Allard}, {Homeier}, \&
  {Freytag}}]{Allard2012}
{Allard} F., {Homeier} D., {Freytag} B., 2012, Royal Society of London
  Philosophical Transactions Series A, 370, 2765

\bibitem[{{An} {et~al}\mbox{.}(2008){An}, {Johnson}, {Clem}, {Yanny},
  {Rockosi}, {Morrison}, {Harding}, {Gunn}, {Allende Prieto}, {Beers},
  {Cudworth}, {Ivans}, {Ivezi{\'c}}, {Lee}, {Lupton}, {Bizyaev}, {Brewington},
  {Malanushenko}, {Malanushenko}, {Oravetz}, {Pan}, {Simmons}, {Snedden},
  {Watters}, \& {York}}]{An_etal08}
{An} D. {et~al.}, 2008, \apjs, 179, 326

\bibitem[{{Arnaboldi} {et~al}\mbox{.}(2012){Arnaboldi}, {Rejkuba}, {Retzlaff},
  {Delmotte}, {Hanuschik}, {Hilker}, {H{\"u}mmel}, {Hussain}, {Ivanov},
  {Micol}, {Neeser}, {Petr-Gotzens}, {Szeifert}, {Comeron}, {Primas}, \&
  {Romaniello}}]{vistapublicsurveys}
{Arnaboldi} M. {et~al.}, 2012, The Messenger, 149, 7

\bibitem[{{Asplund} {et~al}\mbox{.}(2009){Asplund}, {Grevesse}, {Sauval}, \&
  {Scott}}]{AGSS2009}
{Asplund} M., {Grevesse} N., {Sauval} A.~J., {Scott} P., 2009, \araa, 47, 481

\bibitem[{{Baraffe} {et~al}\mbox{.}(1997){Baraffe}, {Chabrier}, {Allard}, \&
  {Hauschildt}}]{Baraffe1997}
{Baraffe} I., {Chabrier} G., {Allard} F., {Hauschildt} P.~H., 1997, \aap, 327,
  1054

\bibitem[{{Baraffe} {et~al}\mbox{.}(1998){Baraffe}, {Chabrier}, {Allard}, \&
  {Hauschildt}}]{Baraffe_etal98}
---, 1998, \aap, 337, 403

\bibitem[{{Barclay} {et~al}\mbox{.}(2012){Barclay}, {Huber}, {Rowe}, {Fortney},
  {Morley}, {Quintana}, {Fabrycky}, {Barentsen}, {Bloemen}, {Christiansen},
  {Demory}, {Fulton}, {Jenkins}, {Mullally}, {Ragozzine}, {Seader}, {Shporer},
  {Tenenbaum}, \& {Thompson}}]{Barclay2012}
{Barclay} T. {et~al.}, 2012, \apj, 761, 53

\bibitem[{{Bass} {et~al}\mbox{.}(2012){Bass}, {Orosz}, {Welsh}, {Windmiller},
  {Ames Gregg}, {Fetherolf}, {Wade}, \& {Quinn}}]{Bass2012}
{Bass} G., {Orosz} J.~A., {Welsh} W.~F., {Windmiller} G., {Ames Gregg} T.,
  {Fetherolf} T., {Wade} R.~A., {Quinn} S.~N., 2012, \apj, 761, 157

\bibitem[{{Basu} {et~al}\mbox{.}(2009){Basu}, {Chaplin}, {Elsworth}, {New}, \&
  {Serenelli}}]{Basuetal2009}
{Basu} S., {Chaplin} W.~J., {Elsworth} Y., {New} R., {Serenelli} A.~M., 2009,
  \apj, 699, 1403

\bibitem[{{Beatty} {et~al}\mbox{.}(2007){Beatty}, {Fern{\'a}ndez}, {Latham},
  {Bakos}, {Kov{\'a}cs}, {Noyes}, {Stefanik}, {Torres}, {Everett}, \&
  {Hergenrother}}]{Beatty2007}
{Beatty} T.~G. {et~al.}, 2007, \apj, 663, 573

\bibitem[{{Bertelli} {et~al}\mbox{.}(2008){Bertelli}, {Girardi}, {Marigo}, \&
  {Nasi}}]{Bertelli_etal08}
{Bertelli} G., {Girardi} L., {Marigo} P., {Nasi} E., 2008, \aap, 484, 815

\bibitem[{{Bessel}(1990{\natexlab{a}})}]{Bessell91}
{Bessel} M.~S., 1990{\natexlab{a}}, \aaps, 83, 357

\bibitem[{{Bessel}(1990{\natexlab{b}})}]{Bessell1990}
---, 1990{\natexlab{b}}, \pasp, 102, 1181

\bibitem[{{Birkby} {et~al}\mbox{.}(2012){Birkby}, {Nefs}, {Hodgkin},
  {Kov{\'a}cs}, {Sip{\H o}cz}, {Pinfield}, {Snellen}, {Mislis}, {Murgas},
  {Lodieu}, {de Mooij}, {Goulding}, {Cruz}, {Stoev}, {Cappetta}, {Palle},
  {Barrado}, {Saglia}, {Martin}, \& {Pavlenko}}]{Birkby2012}
{Birkby} J. {et~al.}, 2012, \mnras, 426, 1507

\bibitem[{{B{\"o}hm-Vitense}(1958)}]{BV1958}
{B{\"o}hm-Vitense} E., 1958, \zap, 46, 108

\bibitem[{{Borkovits} {et~al}\mbox{.}(2013){Borkovits}, {Derekas}, {Kiss},
  {Kir{\'a}ly}, {Forg{\'a}cs-Dajka}, {B{\'{\i}}r{\'o}}, {Bedding}, {Bryson},
  {Huber}, \& {Szab{\'o}}}]{Borkovits2013}
{Borkovits} T. {et~al.}, 2013, \mnras, 428, 1656

\bibitem[{{Borucki} {et~al}\mbox{.}(2011){Borucki}, {Koch}, {Basri}, {Batalha},
  {Brown}, {Bryson}, {Caldwell}, {Christensen-Dalsgaard}, {Cochran}, {DeVore},
  {Dunham}, {Gautier}, {Geary}, {Gilliland}, {Gould}, {Howell}, {Jenkins},
  {Latham}, {Lissauer}, {Marcy}, {Rowe}, {Sasselov}, {Boss}, {Charbonneau},
  {Ciardi}, {Doyle}, {Dupree}, {Ford}, {Fortney}, {Holman}, {Seager},
  {Steffen}, {Tarter}, {Welsh}, {Allen}, {Buchhave}, {Christiansen}, {Clarke},
  {Das}, {D{\'e}sert}, {Endl}, {Fabrycky}, {Fressin}, {Haas}, {Horch},
  {Howard}, {Isaacson}, {Kjeldsen}, {Kolodziejczak}, {Kulesa}, {Li}, {Lucas},
  {Machalek}, {McCarthy}, {MacQueen}, {Meibom}, {Miquel}, {Prsa}, {Quinn},
  {Quintana}, {Ragozzine}, {Sherry}, {Shporer}, {Tenenbaum}, {Torres},
  {Twicken}, {Van Cleve}, {Walkowicz}, {Witteborn}, \& {Still}}]{Borucki2011}
{Borucki} W.~J. {et~al.}, 2011, \apj, 736, 19

\bibitem[{{Boyajian} {et~al}\mbox{.}(2012){Boyajian}, {von Braun}, {van Belle},
  {McAlister}, {ten Brummelaar}, {Kane}, {Muirhead}, {Jones}, {White},
  {Schaefer}, {Ciardi}, {Henry}, {L{\'o}pez-Morales}, {Ridgway}, {Gies}, {Jao},
  {Rojas-Ayala}, {Parks}, {Sturmann}, {Sturmann}, {Turner}, {Farrington},
  {Goldfinger}, \& {Berger}}]{Boyajian_etal12}
{Boyajian} T.~S. {et~al.}, 2012, \apj, 757, 112

\bibitem[{{Bressan} {et~al}\mbox{.}(2012){Bressan}, {Marigo}, {Girardi},
  {Salasnich}, {Dal Cero}, {Rubele}, \& {Nanni}}]{parsec}
{Bressan} A., {Marigo} P., {Girardi} L., {Salasnich} B., {Dal Cero} C.,
  {Rubele} S., {Nanni} A., 2012, \mnras, 427, 127

\bibitem[{{Bressan}, {Chiosi} \& {Bertelli}(1981){Bressan}, {Chiosi}, \&
  {Bertelli}}]{Bressan_etal81}
{Bressan} A.~G., {Chiosi} C., {Bertelli} G., 1981, \aap, 102, 25

\bibitem[{{{\c C}ak{\i}rl{\i}}(2013)}]{Cakirli2013}
{{\c C}ak{\i}rl{\i}} {\"O}., 2013, \na, 22, 15

\bibitem[{{Caffau} {et~al}\mbox{.}(2008){Caffau}, {Ludwig}, {Steffen}, {Ayres},
  {Bonifacio}, {Cayrel}, {Freytag}, \& {Plez}}]{Caffau_etal08}
{Caffau} E., {Ludwig} H., {Steffen} M., {Ayres} T.~R., {Bonifacio} P., {Cayrel}
  R., {Freytag} B., {Plez} B., 2008, \aap, 488, 1031

\bibitem[{{Caffau} {et~al}\mbox{.}(2009){Caffau}, {Maiorca}, {Bonifacio},
  {Faraggiana}, {Steffen}, {Ludwig}, {Kamp}, \& {Busso}}]{Caffau_etal09}
{Caffau} E., {Maiorca} E., {Bonifacio} P., {Faraggiana} R., {Steffen} M.,
  {Ludwig} H., {Kamp} I., {Busso} M., 2009, \aap, 498, 877

\bibitem[{{Campos} {et~al}\mbox{.}(2013){Campos}, {Kepler}, {Bonatto}, \&
  {Ducati}}]{Campos_etal13}
{Campos} F., {Kepler} S.~O., {Bonatto} C., {Ducati} J.~R., 2013, \mnras, 433,
  243

\bibitem[{{Carrera} \& {Pancino}(2011)}]{Carrera2011}
{Carrera} R., {Pancino} E., 2011, \aap, 535, A30

\bibitem[{{Carretta} {et~al}\mbox{.}(2010){Carretta}, {Bragaglia}, {Gratton},
  {Recio-Blanco}, {Lucatello}, {D'Orazi}, \& {Cassisi}}]{Carretta2010}
{Carretta} E., {Bragaglia} A., {Gratton} R.~G., {Recio-Blanco} A., {Lucatello}
  S., {D'Orazi} V., {Cassisi} S., 2010, \aap, 516, A55

\bibitem[{{Carter} {et~al}\mbox{.}(2011){Carter}, {Fabrycky}, {Ragozzine},
  {Holman}, {Quinn}, {Latham}, {Buchhave}, {Van Cleve}, {Cochran}, {Cote},
  {Endl}, {Ford}, {Haas}, {Jenkins}, {Koch}, {Li}, {Lissauer}, {MacQueen},
  {Middour}, {Orosz}, {Rowe}, {Steffen}, \& {Welsh}}]{Carter2011}
{Carter} J.~A. {et~al.}, 2011, Science, 331, 562

\bibitem[{{Casagrande}, {Flynn} \& {Bessell}(2008){Casagrande}, {Flynn}, \&
  {Bessell}}]{Casagrande_etal08}
{Casagrande} L., {Flynn} C., {Bessell} M., 2008, \mnras, 389, 585

\bibitem[{{Castelli} \& {Kurucz}(2003)}]{CastelliKurucz03}
{Castelli} F., {Kurucz} R.~L., 2003, in IAU Symposium, Vol. 210, Modelling of
  Stellar Atmospheres, {Piskunov} N., {Weiss} W.~W., {Gray} D.~F., eds., p. 20P

\bibitem[{{Chabrier}(2001)}]{Chabrier01}
{Chabrier} G., 2001, \apj, 554, 1274

\bibitem[{{Copeland}, {Jensen} \& {Jorgensen}(1970){Copeland}, {Jensen}, \&
  {Jorgensen}}]{Copeland_etal70}
{Copeland} H., {Jensen} J.~O., {Jorgensen} H.~E., 1970, \aap, 5, 12

\bibitem[{{Cordero} {et~al}\mbox{.}(2014){Cordero}, {Pilachowski}, {Johnson},
  {McDonald}, {Zijlstra}, \& {Simmerer}}]{Cordero2014}
{Cordero} M.~J., {Pilachowski} C.~A., {Johnson} C.~I., {McDonald} I.,
  {Zijlstra} A.~A., {Simmerer} J., 2014, \apj, 780, 94

\bibitem[{{Cutri} {et~al}\mbox{.}(2003){Cutri}, {Skrutskie}, {van Dyk},
  {Beichman}, {Carpenter}, {Chester}, {Cambresy}, {Evans}, {Fowler}, {Gizis},
  {Howard}, {Huchra}, {Jarrett}, {Kopan}, {Kirkpatrick}, {Light}, {Marsh},
  {McCallon}, {Schneider}, {Stiening}, {Sykes}, {Weinberg}, {Wheaton},
  {Wheelock}, \& {Zacarias}}]{2mass}
{Cutri} R.~M. {et~al.}, 2003, {2MASS All Sky Catalog of point sources.}

\bibitem[{{Cyburt} {et~al}\mbox{.}(2010){Cyburt}, {Amthor}, {Ferguson},
  {Meisel}, {Smith}, {Warren}, {Heger}, {Hoffman}, {Rauscher}, {Sakharuk},
  {Schatz}, {Thielemann}, \& {Wiescher}}]{Cyburt_etal10}
{Cyburt} R.~H. {et~al.}, 2010, \apjs, 189, 240

\bibitem[{{Dahn} {et~al}\mbox{.}(2002){Dahn}, {Harris}, {Vrba}, {Guetter},
  {Canzian}, {Henden}, {Levine}, {Luginbuhl}, {Monet}, {Monet}, {Pier},
  {Stone}, {Walker}, {Burgasser}, {Gizis}, {Kirkpatrick}, {Liebert}, \&
  {Reid}}]{Dahn_etal02}
{Dahn} C.~C. {et~al.}, 2002, \aj, 124, 1170

\bibitem[{{Dotter} {et~al}\mbox{.}(2008){Dotter}, {Chaboyer}, {Jevremovi{\'c}},
  {Kostov}, {Baron}, \& {Ferguson}}]{Dotter2008}
{Dotter} A., {Chaboyer} B., {Jevremovi{\'c}} D., {Kostov} V., {Baron} E.,
  {Ferguson} J.~W., 2008, \apjs, 178, 89

\bibitem[{{Doyle} {et~al}\mbox{.}(2011){Doyle}, {Carter}, {Fabrycky},
  {Slawson}, {Howell}, {Winn}, {Orosz}, {Pr{caron}sa}, {Welsh}, {Quinn},
  {Latham}, {Torres}, {Buchhave}, {Marcy}, {Fortney}, {Shporer}, {Ford},
  {Lissauer}, {Ragozzine}, {Rucker}, {Batalha}, {Jenkins}, {Borucki}, {Koch},
  {Middour}, {Hall}, {McCauliff}, {Fanelli}, {Quintana}, {Holman}, {Caldwell},
  {Still}, {Stefanik}, {Brown}, {Esquerdo}, {Tang}, {Furesz}, {Geary},
  {Berlind}, {Calkins}, {Short}, {Steffen}, {Sasselov}, {Dunham}, {Cochran},
  {Boss}, {Haas}, {Buzasi}, \& {Fischer}}]{Doyle2011}
{Doyle} L.~R. {et~al.}, 2011, Science, 333, 1602

\bibitem[{{Eker} {et~al}\mbox{.}(2014){Eker}, {Bilir}, {Soydugan}, {G{\"o}k{\c
  c}e}, {Soydugan}, {T{\"u}ys{\"u}z}, {{\c S}eny{\"u}z}, \&
  {Demircan}}]{Eker2014}
{Eker} Z., {Bilir} S., {Soydugan} F., {G{\"o}k{\c c}e} E.~Y., {Soydugan} E.,
  {T{\"u}ys{\"u}z} M., {{\c S}eny{\"u}z} T., {Demircan} O., 2014, \pasa, 31, 24

\bibitem[{{Feiden} \& {Chaboyer}(2012)}]{Feiden2012}
{Feiden} G.~A., {Chaboyer} B., 2012, \apj, 757, 42

\bibitem[{{Feiden} \& {Chaboyer}(2013)}]{Feiden2013}
---, 2013, \apj, 779, 183

\bibitem[{{Fernandez} {et~al}\mbox{.}(2009){Fernandez}, {Latham}, {Torres},
  {Everett}, {Mandushev}, {Charbonneau}, {O'Donovan}, {Alonso}, {Esquerdo},
  {Hergenrother}, \& {Stefanik}}]{Fernandez2009}
{Fernandez} J.~M. {et~al.}, 2009, \apj, 701, 764

\bibitem[{{Girardi} {et~al}\mbox{.}(2002){Girardi}, {Bertelli}, {Bressan},
  {Chiosi}, {Groenewegen}, {Marigo}, {Salasnich}, \& {Weiss}}]{Girardi_etal02}
{Girardi} L., {Bertelli} G., {Bressan} A., {Chiosi} C., {Groenewegen} M.~A.~T.,
  {Marigo} P., {Salasnich} B., {Weiss} A., 2002, \aap, 391, 195

\bibitem[{{Girardi} {et~al}\mbox{.}(2000){Girardi}, {Bressan}, {Bertelli}, \&
  {Chiosi}}]{Girardi_etal00}
{Girardi} L., {Bressan} A., {Bertelli} G., {Chiosi} C., 2000, \aaps, 141, 371

\bibitem[{{Green} {et~al}\mbox{.}(2014){Green}, {Schlafly}, {Finkbeiner},
  {Juri{\'c}}, {Rix}, {Burgett}, {Chambers}, {Draper}, {Flewelling},
  {Kudritzki}, {Magnier}, {Martin}, {Metcalfe}, {Tonry}, {Wainscoat}, \&
  {Waters}}]{Green_etal14}
{Green} G.~M. {et~al.}, 2014, \apj, 783, 114

\bibitem[{{Hansen} {et~al}\mbox{.}(2013){Hansen}, {Kalirai}, {Anderson},
  {Dotter}, {Richer}, {Rich}, {Shara}, {Fahlman}, {Hurley}, {King}, {Reitzel},
  \& {Stetson}}]{Hansen2013}
{Hansen} B.~M.~S. {et~al.}, 2013, \nat, 500, 51

\bibitem[{{Hebb} {et~al}\mbox{.}(2006){Hebb}, {Wyse}, {Gilmore}, \&
  {Holtzman}}]{Hebb2006}
{Hebb} L., {Wyse} R.~F.~G., {Gilmore} G., {Holtzman} J., 2006, \aj, 131, 555

\bibitem[{{He{\l}miniak} \& {Konacki}(2011)}]{Helminiak2011}
{He{\l}miniak} K.~G., {Konacki} M., 2011, \aap, 526, A29

\bibitem[{{He{\l}miniak} {et~al}\mbox{.}(2012){He{\l}miniak}, {Konacki},
  {R{\'o}{\.Z}yczka}, {Ka{\l}u{\.Z}ny}, {Ratajczak}, {Borkowski}, {Sybilski},
  {Muterspaugh}, {Reichart}, {Ivarsen}, {Haislip}, {Crain}, {Foster},
  {Nysewander}, \& {LaCluyze}}]{Helminiak2012}
{He{\l}miniak} K.~G. {et~al.}, 2012, \mnras, 425, 1245

\bibitem[{{Henyey}, {Forbes} \& {Gould}(1964){Henyey}, {Forbes}, \&
  {Gould}}]{Henyey}
{Henyey} L.~G., {Forbes} J.~E., {Gould} N.~L., 1964, \apj, 139, 306

\bibitem[{{Hofmeister}, {Kippenhahn} \& {Weigert}(1964){Hofmeister},
  {Kippenhahn}, \& {Weigert}}]{Hofmeister_etal64}
{Hofmeister} E., {Kippenhahn} R., {Weigert} A., 1964, \zap, 59, 215

\bibitem[{{Iglesias} \& {Rogers}(1996)}]{IglesiasRogers96}
{Iglesias} C.~A., {Rogers} F.~J., 1996, \apj, 464, 943

\bibitem[{{Irwin} {et~al}\mbox{.}(2009){Irwin}, {Charbonneau}, {Berta},
  {Quinn}, {Latham}, {Torres}, {Blake}, {Burke}, {Esquerdo}, {F{\"u}r{\'e}sz},
  {Mink}, {Nutzman}, {Szentgyorgyi}, {Calkins}, {Falco}, {Bloom}, \&
  {Starr}}]{Irwin2009}
{Irwin} J. {et~al.}, 2009, \apj, 701, 1436

\bibitem[{{Irwin} {et~al}\mbox{.}(2011){Irwin}, {Quinn}, {Berta}, {Latham},
  {Torres}, {Burke}, {Charbonneau}, {Dittmann}, {Esquerdo}, {Stefanik},
  {Oksanen}, {Buchhave}, {Nutzman}, {Berlind}, {Calkins}, \&
  {Falco}}]{Irwin2011}
{Irwin} J.~M. {et~al.}, 2011, \apj, 742, 123

\bibitem[{{Ivezi{\'c}}, {Beers} \& {Juri{\'c}}(2012){Ivezi{\'c}}, {Beers}, \&
  {Juri{\'c}}}]{Ivezic_etal12}
{Ivezi{\'c}} {\v Z}., {Beers} T.~C., {Juri{\'c}} M., 2012, \araa, 50, 251

\bibitem[{{Jackson} \& {Jeffries}(2014)}]{Jackson2014}
{Jackson} R.~J., {Jeffries} R.~D., 2014, \mnras, 441, 2111

\bibitem[{{Juri{\'c}} {et~al}\mbox{.}(2008){Juri{\'c}}, {Ivezi{\'c}}, {Brooks},
  {Lupton}, {Schlegel}, {Finkbeiner}, {Padmanabhan}, {Bond}, {Sesar},
  {Rockosi}, {Knapp}, {Gunn}, {Sumi}, {Schneider}, {Barentine}, {Brewington},
  {Brinkmann}, {Fukugita}, {Harvanek}, {Kleinman}, {Krzesinski}, {Long},
  {Neilsen}, {Nitta}, {Snedden}, \& {York}}]{Juric_etal08}
{Juri{\'c}} M. {et~al.}, 2008, \apj, 673, 864

\bibitem[{{Kalirai} {et~al}\mbox{.}(2012){Kalirai}, {Richer}, {Anderson},
  {Dotter}, {Fahlman}, {Hansen}, {Hurley}, {King}, {Reitzel}, {Rich}, {Shara},
  {Stetson}, \& {Woodley}}]{Kalirai_etal12}
{Kalirai} J.~S. {et~al.}, 2012, \aj, 143, 11

\bibitem[{{Kaluzny} {et~al}\mbox{.}(2013){Kaluzny}, {Thompson}, {Rozyczka},
  {Dotter}, {Krzeminski}, {Pych}, {Rucinski}, {Burley}, \&
  {Shectman}}]{Kaluzny2013}
{Kaluzny} J. {et~al.}, 2013, \aj, 145, 43

\bibitem[{{Kippenhahn}, {Weigert} \& {Weiss}(2013){Kippenhahn}, {Weigert}, \&
  {Weiss}}]{Kippenhahn_book}
{Kippenhahn} R., {Weigert} A., {Weiss} A., 2013, {Stellar Structure and
  Evolution}

\bibitem[{{Kraus} {et~al}\mbox{.}(2011){Kraus}, {Tucker}, {Thompson}, {Craine},
  \& {Hillenbrand}}]{Kraus2011}
{Kraus} A.~L., {Tucker} R.~A., {Thompson} M.~I., {Craine} E.~R., {Hillenbrand}
  L.~A., 2011, \apj, 728, 48

\bibitem[{{Krishna Swamy}(1966)}]{KS1966}
{Krishna Swamy} K.~S., 1966, \apj, 145, 174

\bibitem[{{Kroupa}(2001)}]{Kroupa01}
{Kroupa} P., 2001, \mnras, 322, 231

\bibitem[{{Lawrence} {et~al}\mbox{.}(2007){Lawrence}, {Warren}, {Almaini},
  {Edge}, {Hambly}, {Jameson}, {Lucas}, {Casali}, {Adamson}, {Dye}, {Emerson},
  {Foucaud}, {Hewett}, {Hirst}, {Hodgkin}, {Irwin}, {Lodieu}, {McMahon},
  {Simpson}, {Smail}, {Mortlock}, \& {Folger}}]{Lawrence_etal07}
{Lawrence} A. {et~al.}, 2007, \mnras, 379, 1599

\bibitem[{{Leggett}(1992)}]{Leggett92}
{Leggett} S.~K., 1992, \apjs, 82, 351

\bibitem[{{Lopez-Morales} {et~al}\mbox{.}(2006){Lopez-Morales}, {Orosz},
  {Shaw}, {Havelka}, {Arevalo}, {McIntyre}, \& {Lazaro}}]{Lopez-Morales2006}
{Lopez-Morales} M., {Orosz} J.~A., {Shaw} J.~S., {Havelka} L., {Arevalo} M.~J.,
  {McIntyre} T., {Lazaro} C., 2006, ArXiv

\bibitem[{{L{\'o}pez-Morales} \& {Shaw}(2007)}]{Lopez-Morales2007}
{L{\'o}pez-Morales} M., {Shaw} J.~S., 2007, in Astronomical Society of the
  Pacific Conference Series, Vol. 362, The Seventh Pacific Rim Conference on
  Stellar Astrophysics, {Kang} Y.~W., {Lee} H.-W., {Leung} K.-C., {Cheng}
  K.-S., eds., p.~26

\bibitem[{{MacDonald} \& {Mullan}(2013)}]{MacDonald2013}
{MacDonald} J., {Mullan} D.~J., 2013, \apj, 765, 126

\bibitem[{{Marcy} {et~al}\mbox{.}(2014){Marcy}, {Isaacson}, {Howard}, {Rowe},
  {Jenkins}, {Bryson}, {Latham}, {Howell}, {Gautier}, {Batalha}, {Rogers},
  {Ciardi}, {Fischer}, {Gilliland}, {Kjeldsen}, {Christensen-Dalsgaard},
  {Huber}, {Chaplin}, {Basu}, {Buchhave}, {Quinn}, {Borucki}, {Koch}, {Hunter},
  {Caldwell}, {Van Cleve}, {Kolbl}, {Weiss}, {Petigura}, {Seager}, {Morton},
  {Johnson}, {Ballard}, {Burke}, {Cochran}, {Endl}, {MacQueen}, {Everett},
  {Lissauer}, {Ford}, {Torres}, {Fressin}, {Brown}, {Steffen}, {Charbonneau},
  {Basri}, {Sasselov}, {Winn}, {Sanchis-Ojeda}, {Christiansen}, {Adams},
  {Henze}, {Dupree}, {Fabrycky}, {Fortney}, {Tarter}, {Holman}, {Tenenbaum},
  {Shporer}, {Lucas}, {Welsh}, {Orosz}, {Bedding}, {Campante}, {Davies},
  {Elsworth}, {Handberg}, {Hekker}, {Karoff}, {Kawaler}, {Lund}, {Lundkvist},
  {Metcalfe}, {Miglio}, {Silva Aguirre}, {Stello}, {White}, {Boss}, {Devore},
  {Gould}, {Prsa}, {Agol}, {Barclay}, {Coughlin}, {Brugamyer}, {Mullally},
  {Quintana}, {Still}, {Thompson}, {Morrison}, {Twicken}, {D{\'e}sert},
  {Carter}, {Crepp}, {H{\'e}brard}, {Santerne}, {Moutou}, {Sobeck}, {Hudgins},
  {Haas}, {Robertson}, {Lillo-Box}, \& {Barrado}}]{Marcy2014}
{Marcy} G.~W. {et~al.}, 2014, \apjs, 210, 20

\bibitem[{{Marigo} \& {Aringer}(2009)}]{MarigoAringer09}
{Marigo} P., {Aringer} B., 2009, \aap, 508, 1539

\bibitem[{{Maxted} {et~al}\mbox{.}(2007){Maxted}, {O'Donoghue},
  {Morales-Rueda}, {Napiwotzki}, \& {Smalley}}]{Maxted2007}
{Maxted} P.~F.~L., {O'Donoghue} D., {Morales-Rueda} L., {Napiwotzki} R.,
  {Smalley} B., 2007, \mnras, 376, 919

\bibitem[{{Metcalfe} {et~al}\mbox{.}(2014){Metcalfe}, {Creevey}, {Dogan},
  {Mathur}, {Xu}, {Bedding}, {Chaplin}, {Christensen-Dalsgaard}, {Karoff},
  {Trampedach}, {Benomar}, {Brown}, {Buzasi}, {Campante}, {Celik}, {Cunha},
  {Davies}, {Deheuvels}, {Derekas}, {Di Mauro}, {Garcia}, {Guzik}, {Howe},
  {MacGregor}, {Mazumdar}, {Montalban}, {Monteiro}, {Salabert}, {Serenelli},
  {Stello}, {Steslicki}, {Suran}, {Yildiz}, {Aksoy}, {Elsworth}, {Gruberbauer},
  {Guenther}, {Lebreton}, {Molaverdikhani}, {Pricopi}, {Simoniello}, \&
  {White}}]{Metcalfe2014}
{Metcalfe} T.~S. {et~al.}, 2014, ArXiv e-prints

\bibitem[{{Mihalas}(1978)}]{Mihalas1978}
{Mihalas} D., 1978, {Stellar atmospheres /2nd edition/}

\bibitem[{{Mihalas} {et~al}\mbox{.}(1990){Mihalas}, {Hummer}, {Mihalas}, \&
  {Daeppen}}]{Mihalas1990}
{Mihalas} D., {Hummer} D.~G., {Mihalas} B.~W., {Daeppen} W., 1990, \apj, 350,
  300

\bibitem[{{Morales} {et~al}\mbox{.}(2009{\natexlab{a}}){Morales}, {Ribas},
  {Jordi}, {Torres}, {Gallardo}, {Guinan}, {Charbonneau}, {Wolf}, {Latham},
  {Anglada-Escud{\'e}}, {Bradstreet}, {Everett}, {O'Donovan}, {Mandushev}, \&
  {Mathieu}}]{Morales2009a}
{Morales} J.~C. {et~al.}, 2009{\natexlab{a}}, \apj, 691, 1400

\bibitem[{{Morales} {et~al}\mbox{.}(2009{\natexlab{b}}){Morales}, {Torres},
  {Marschall}, \& {Brehm}}]{Morales2009b}
{Morales} J.~C., {Torres} G., {Marschall} L.~A., {Brehm} W.,
  2009{\natexlab{b}}, \apj, 707, 671

\bibitem[{{Nefs} {et~al}\mbox{.}(2013){Nefs}, {Birkby}, {Snellen}, {Hodgkin},
  {Sip{\H o}cz}, {Kov{\'a}cs}, {Mislis}, {Pinfield}, \& {Martin}}]{Nefs2013}
{Nefs} S.~V. {et~al.}, 2013, \mnras, 431, 3240

\bibitem[{{Nikolaev} \& {Weinberg}(2000)}]{NikolaevWeinberg00}
{Nikolaev} S., {Weinberg} M.~D., 2000, \apj, 542, 804

\bibitem[{{Ofir} {et~al}\mbox{.}(2012){Ofir}, {Gandolfi}, {Buchhave}, {Lacy},
  {Hatzes}, \& {Fridlund}}]{Ofir2012}
{Ofir} A., {Gandolfi} D., {Buchhave} L., {Lacy} C.~H.~S., {Hatzes} A.~P.,
  {Fridlund} M., 2012, \mnras, 423, L1

\bibitem[{{Parsons} {et~al}\mbox{.}(2012{\natexlab{a}}){Parsons},
  {G{\"a}nsicke}, {Marsh}, {Bergeron}, {Copperwheat}, {Dhillon}, {Bento},
  {Littlefair}, \& {Schreiber}}]{Parsons2012c}
{Parsons} S.~G. {et~al.}, 2012{\natexlab{a}}, \mnras, 426, 1950

\bibitem[{{Parsons} {et~al}\mbox{.}(2012{\natexlab{b}}){Parsons}, {Marsh},
  {G{\"a}nsicke}, {Dhillon}, {Copperwheat}, {Littlefair}, {Pyrzas}, {Drake},
  {Koester}, {Schreiber}, \& {Rebassa-Mansergas}}]{Parsons2012b}
---, 2012{\natexlab{b}}, \mnras, 419, 304

\bibitem[{{Parsons} {et~al}\mbox{.}(2012{\natexlab{c}}){Parsons}, {Marsh},
  {G{\"a}nsicke}, {Rebassa-Mansergas}, {Dhillon}, {Littlefair}, {Copperwheat},
  {Hickman}, {Burleigh}, {Kerry}, {Koester}, {Nebot G{\'o}mez-Mor{\'a}n},
  {Pyrzas}, {Savoury}, {Schreiber}, {Schmidtobreick}, {Schwope}, {Steele}, \&
  {Tappert}}]{Parsons2012a}
---, 2012{\natexlab{c}}, \mnras, 420, 3281

\bibitem[{{Pont} {et~al}\mbox{.}(2005{\natexlab{a}}){Pont}, {Bouchy}, {Melo},
  {Santos}, {Mayor}, {Queloz}, \& {Udry}}]{Pont2005a}
{Pont} F., {Bouchy} F., {Melo} C., {Santos} N.~C., {Mayor} M., {Queloz} D.,
  {Udry} S., 2005{\natexlab{a}}, \aap, 438, 1123

\bibitem[{{Pont} {et~al}\mbox{.}(2005{\natexlab{b}}){Pont}, {Melo}, {Bouchy},
  {Udry}, {Queloz}, {Mayor}, \& {Santos}}]{Pont2005b}
{Pont} F., {Melo} C.~H.~F., {Bouchy} F., {Udry} S., {Queloz} D., {Mayor} M.,
  {Santos} N.~C., 2005{\natexlab{b}}, \aap, 433, L21

\bibitem[{{Pont} {et~al}\mbox{.}(2006){Pont}, {Moutou}, {Bouchy}, {Behrend},
  {Mayor}, {Udry}, {Queloz}, {Santos}, \& {Melo}}]{Pont2006}
{Pont} F. {et~al.}, 2006, \aap, 447, 1035

\bibitem[{{Quintana} {et~al}\mbox{.}(2014){Quintana}, {Barclay}, {Raymond},
  {Rowe}, {Bolmont}, {Caldwell}, {Howell}, {Kane}, {Huber}, {Crepp},
  {Lissauer}, {Ciardi}, {Coughlin}, {Everett}, {Henze}, {Horch}, {Isaacson},
  {Ford}, {Adams}, {Still}, {Hunter}, {Quarles}, \& {Selsis}}]{Quintana2014}
{Quintana} E.~V. {et~al.}, 2014, Science, 344, 277

\bibitem[{{Rajpurohit} {et~al}\mbox{.}(2013){Rajpurohit}, {Reyl{\'e}},
  {Allard}, {Homeier}, {Schultheis}, {Bessell}, \& {Robin}}]{Rajpurohit2013}
{Rajpurohit} A.~S., {Reyl{\'e}} C., {Allard} F., {Homeier} D., {Schultheis} M.,
  {Bessell} M.~S., {Robin} A.~C., 2013, \aap, 556, A15

\bibitem[{{Randich} {et~al}\mbox{.}(2006){Randich}, {Sestito}, {Primas},
  {Pallavicini}, \& {Pasquini}}]{Randich_etal06}
{Randich} S., {Sestito} P., {Primas} F., {Pallavicini} R., {Pasquini} L., 2006,
  \aap, 450, 557

\bibitem[{{Richer} {et~al}\mbox{.}(2008){Richer}, {Dotter}, {Hurley},
  {Anderson}, {King}, {Davis}, {Fahlman}, {Hansen}, {Kalirai}, {Paust}, {Rich},
  \& {Shara}}]{Richer2008}
{Richer} H.~B. {et~al.}, 2008, \aj, 135, 2141

\bibitem[{{Rogers}, {Swenson} \& {Iglesias}(1996){Rogers}, {Swenson}, \&
  {Iglesias}}]{opaleos}
{Rogers} F.~J., {Swenson} F.~J., {Iglesias} C.~A., 1996, \apj, 456, 902

\bibitem[{{Sandberg Lacy} {et~al}\mbox{.}(2014){Sandberg Lacy}, {Torres},
  {Wolf}, \& {Burks}}]{Lacy2014}
{Sandberg Lacy} C.~H., {Torres} G., {Wolf} M., {Burks} C.~L., 2014, \aj, 147, 1

\bibitem[{{Sandquist}(2004)}]{Sandquist2004}
{Sandquist} E.~L., 2004, \mnras, 347, 101

\bibitem[{{Sarajedini}, {Dotter} \& {Kirkpatrick}(2009){Sarajedini}, {Dotter},
  \& {Kirkpatrick}}]{Sarajedini2009}
{Sarajedini} A., {Dotter} A., {Kirkpatrick} A., 2009, \apj, 698, 1872

\bibitem[{{Siegel} {et~al}\mbox{.}(2002){Siegel}, {Majewski}, {Reid}, \&
  {Thompson}}]{Siegel_etal02}
{Siegel} M.~H., {Majewski} S.~R., {Reid} I.~N., {Thompson} I.~B., 2002, \apj,
  578, 151

\bibitem[{{Skrutskie} {et~al}\mbox{.}(2006){Skrutskie}, {Cutri}, {Stiening},
  {Weinberg}, {Schneider}, {Carpenter}, {Beichman}, {Capps}, {Chester},
  {Elias}, {Huchra}, {Liebert}, {Lonsdale}, {Monet}, {Price}, {Seitzer},
  {Jarrett}, {Kirkpatrick}, {Gizis}, {Howard}, {Evans}, {Fowler}, {Fullmer},
  {Hurt}, {Light}, {Kopan}, {Marsh}, {McCallon}, {Tam}, {Van Dyk}, \&
  {Wheelock}}]{2masspaper}
{Skrutskie} M.~F. {et~al.}, 2006, \aj, 131, 1163

\bibitem[{{Spada} {et~al}\mbox{.}(2013){Spada}, {Demarque}, {Kim}, \&
  {Sills}}]{Spada_etal13}
{Spada} F., {Demarque} P., {Kim} Y.-C., {Sills} A., 2013, \apj, 776, 87

\bibitem[{{Spruit} \& {Weiss}(1986)}]{Spruit1986}
{Spruit} H.~C., {Weiss} A., 1986, \aap, 166, 167

\bibitem[{{Tal-Or} {et~al}\mbox{.}(2013){Tal-Or}, {Mazeh}, {Alonso}, {Bouchy},
  {Cabrera}, {Deeg}, {Deleuil}, {Faigler}, {Fridlund}, {H{\'e}brard}, {Moutou},
  {Santerne}, \& {Tingley}}]{Tal-Or2013}
{Tal-Or} L. {et~al.}, 2013, \aap, 553, A30

\bibitem[{{Taylor}(2006)}]{stilts}
{Taylor} M.~B., 2006, in Astronomical Society of the Pacific Conference Series,
  Vol. 351, Astronomical Data Analysis Software and Systems XV, {Gabriel} C.,
  {Arviset} C., {Ponz} D., {Enrique} S., eds., p. 666

\bibitem[{{Torres}, {Andersen} \& {Gim{\'e}nez}(2010){Torres}, {Andersen}, \&
  {Gim{\'e}nez}}]{Torres2010}
{Torres} G., {Andersen} J., {Gim{\'e}nez} A., 2010, \aapr, 18, 67

\bibitem[{{Vaccaro} {et~al}\mbox{.}(2007){Vaccaro}, {Rudkin}, {Kawka},
  {Vennes}, {Oswalt}, {Silver}, {Wood}, \& {Smith}}]{Vaccaro2007}
{Vaccaro} T.~R., {Rudkin} M., {Kawka} A., {Vennes} S., {Oswalt} T.~D., {Silver}
  I., {Wood} M., {Smith} J.~A., 2007, \apj, 661, 1112

\bibitem[{{van Leeuwen}(2009)}]{van_Leeuwen2009}
{van Leeuwen} F., 2009, \aap, 497, 209

\bibitem[{{VandenBerg} {et~al}\mbox{.}(2008){VandenBerg}, {Edvardsson},
  {Eriksson}, \& {Gustafsson}}]{VandenBerg_etal08}
{VandenBerg} D.~A., {Edvardsson} B., {Eriksson} K., {Gustafsson} B., 2008,
  \apj, 675, 746

\bibitem[{{VandenBerg} \& {Stetson}(2004)}]{VandenbergStetson04}
{VandenBerg} D.~A., {Stetson} P.~B., 2004, \pasp, 116, 997

\bibitem[{{Wang} {et~al}\mbox{.}(2014){Wang}, {Chen}, {Lin}, {Pandey}, {Huang},
  {Panwar}, {Lee}, {Tsai}, {Tang}, {Goldman}, {Burgett}, {Chambers}, {Draper},
  {Flewelling}, {Grav}, {Heasley}, {Hodapp}, {Huber}, {Jedicke}, {Kaiser},
  {Kudritzki}, {Luppino}, {Lupton}, {Magnier}, {Metcalfe}, {Monet}, {Morgan},
  {Onaka}, {Price}, {Stubbs}, {Sweeney}, {Tonry}, {Wainscoat}, \&
  {Waters}}]{WangPF2014}
{Wang} P.~F. {et~al.}, 2014, \apj, 784, 57

\bibitem[{{Wright} {et~al}\mbox{.}(2010){Wright}, {Eisenhardt}, {Mainzer},
  {Ressler}, {Cutri}, {Jarrett}, {Kirkpatrick}, {Padgett}, {McMillan},
  {Skrutskie}, {Stanford}, {Cohen}, {Walker}, {Mather}, {Leisawitz}, {Gautier},
  {McLean}, {Benford}, {Lonsdale}, {Blain}, {Mendez}, {Irace}, {Duval}, {Liu},
  {Royer}, {Heinrichsen}, {Howard}, {Shannon}, {Kendall}, {Walsh}, {Larsen},
  {Cardon}, {Schick}, {Schwalm}, {Abid}, {Fabinsky}, {Naes}, \& {Tsai}}]{wise}
{Wright} E.~L. {et~al.}, 2010, \aj, 140, 1868

\bibitem[{{Yadav} {et~al}\mbox{.}(2008){Yadav}, {Bedin}, {Piotto}, {Anderson},
  {Cassisi}, {Villanova}, {Platais}, {Pasquini}, {Momany}, \&
  {Sagar}}]{yadav_etal08}
{Yadav} R.~K.~S. {et~al.}, 2008, \aap, 484, 609

\bibitem[{{Zasowski} {et~al}\mbox{.}(2013){Zasowski}, {Johnson}, {Frinchaboy},
  {Majewski}, {Nidever}, {Rocha Pinto}, {Girardi}, {Andrews}, {Chojnowski},
  {Cudworth}, {Jackson}, {Munn}, {Skrutskie}, {Beaton}, {Blake}, {Covey},
  {Deshpande}, {Epstein}, {Fabbian}, {Fleming}, {Garcia Hernandez}, {Herrero},
  {Mahadevan}, {M{\'e}sz{\'a}ros}, {Schultheis}, {Sellgren}, {Terrien}, {van
  Saders}, {Allende Prieto}, {Bizyaev}, {Burton}, {Cunha}, {da Costa},
  {Hasselquist}, {Hearty}, {Holtzman}, {Garc{\'{\i}}a P{\'e}rez}, {Maia},
  {O'Connell}, {O'Donnell}, {Pinsonneault}, {Santiago}, {Schiavon}, {Shetrone},
  {Smith}, \& {Wilson}}]{Zasowski_etal13}
{Zasowski} G. {et~al.}, 2013, \aj, 146, 81

\bibitem[{{Zhou} {et~al}\mbox{.}(2014){Zhou}, {Bayliss}, {Hartman}, {Bakos},
  {Penev}, {Csubry}, {Tan}, {Jord{\'a}n}, {Mancini}, {Rabus}, {Brahm},
  {Espinoza}, {Mohler-Fischer}, {Ciceri}, {Suc}, {Cs{\'a}k}, {Henning}, \&
  {Schmidt}}]{Zhou2014}
{Zhou} G. {et~al.}, 2014, \mnras, 437, 2831

\bibitem[{{Zola} {et~al}\mbox{.}(2014){Zola}, {{\c S}enavc{\i}}, {Liakos},
  {Nelson}, \& {Zakrzewski}}]{Zola2014}
{Zola} S., {{\c S}enavc{\i}} H.~V., {Liakos} A., {Nelson} R.~H., {Zakrzewski}
  B., 2014, \mnras, 437, 3718

\end{thebibliography}
